\documentclass[10pt,letterpaper]{article}
\usepackage[margin=2cm]{geometry}

\usepackage{amsmath,amssymb}

\usepackage{changepage}

\usepackage[utf8x]{inputenc}

\usepackage{textcomp,marvosym}

\usepackage{cite}

\usepackage{nameref,hyperref}

\usepackage{microtype}
\DisableLigatures[f]{encoding = *, family = * }

\usepackage[table]{xcolor}

\usepackage{array}

\newcolumntype{+}{!{\vrule width 2pt}}

\newlength\savedwidth

\usepackage[aboveskip=1pt,labelfont=bf,labelsep=period,justification=raggedright,singlelinecheck=off]{caption}

\makeatletter
\renewcommand{\@biblabel}[1]{\quad#1.}
\makeatother

\usepackage{lastpage,fancyhdr,graphicx}
\usepackage{epstopdf}
\fancyheadoffset[L]{2.25in}
\fancyfootoffset[L]{2.25in}
  \graphicspath{{./figure/}}

\usepackage{epstopdf}
\usepackage{enumitem}

\begin{document}
\vspace*{0.2in}

\begin{flushleft}
{\Large
\textbf\newline{Revealing evolutionary constraints on proteins through sequence analysis} 
}
\newline
\\
Shou-Wen Wang\textsuperscript{1,2,3\Yinyang\textcurrency},
Anne-Florence Bitbol\textsuperscript{4\Yinyang*},
Ned S. Wingreen\textsuperscript{3,5*}
\\
\bigskip
\textbf{1} Department of Engineering Physics, Tsinghua University, Beijing, 100086, China
\\
\textbf{2} Beijing Computational Science Research Center, Beijing, 100094, China
\\
\textbf{3} Lewis-Sigler Institute for Integrative Genomics, Princeton University, Princeton, NJ 08544, USA
\\
\textbf{4} Sorbonne Universit{\'e}, CNRS, Laboratoire Jean Perrin (UMR 8237), F-75005 Paris, France
\\
\textbf{5} Department of Molecular Biology, Princeton University, Princeton, NJ 08544, USA
\\
\bigskip

%
%
\Yinyang These authors contributed equally to this work.

\textcurrency Current Address: Department of Systems Biology, Harvard Medical School, Boston, MA 02115, USA. 

* anne-florence.bitbol@sorbonne-universite.fr (A.-F. B.), wingreen@princeton.edu (N.~S.~W.)

\end{flushleft}
\section*{Abstract}
Statistical analysis of alignments of large numbers of protein sequences has revealed ``sectors'' of collectively coevolving amino acids in several protein families. Here, we show that selection acting on any functional property of a protein, represented by an additive trait, can give rise to such a sector. As an illustration of a selected trait, we consider the elastic energy of an important conformational change within an elastic network model, and we show that selection acting on this energy leads to correlations among residues. For this concrete example and more generally, we demonstrate that the main signature of functional sectors lies in the small-eigenvalue modes of the covariance matrix of the selected sequences. However, secondary signatures of these functional sectors also exist in the extensively-studied large-eigenvalue modes. Our simple, general model leads us to propose a principled method to identify functional sectors, along with the magnitudes of mutational effects, from sequence data. We further demonstrate the robustness of these functional sectors to various forms of selection, and the robustness of our approach to the identification of multiple selected traits.

\section*{Author summary}
Proteins play crucial parts in all cellular processes, and their functions are encoded in their amino-acid sequences. Recently, statistical analyses of protein sequence alignments have demonstrated the existence of “sectors” of collectively correlated amino acids. What is the origin of these sectors? Here, we propose a simple underlying origin of protein sectors: they can arise from selection acting on any collective protein property. We find that the main signature of these functional sectors lies in the low-eigenvalue modes of the covariance matrix of the selected sequences. A better understanding of protein sectors will make it possible to discern collective protein properties directly from sequences, as well as to design new functional sequences, with far-reaching applications in synthetic biology.


\section*{Introduction}
Proteins play crucial roles in all cellular processes, acting as enzymes, motors, receptors, regulators, and more. The function of a protein is encoded in its amino-acid sequence. In evolution, random mutations affect the sequence, while natural selection acts at the level of function, however our ability to predict a protein's function directly from its sequence has been very limited. Recently, the explosion of available sequences has inspired new data-driven approaches to uncover the principles of protein operation. At the root of these new approaches is the observation that amino-acid residues which possess related functional roles often evolve in a correlated way. In particular, analyses of large alignments of protein sequences have identified ``sectors'' of collectively correlated amino acids~\cite{Lockless99, Suel03, socolich2005evolutionary, halabi2009protein, Dahirel11, mclaughlin2012spatial}, which has enabled successful design of new functional sequences~\cite{socolich2005evolutionary}. Sectors are spatially contiguous in the protein structure, and in the case of multiple sectors, each one may be associated with a distinct role~\cite{halabi2009protein, Rivoire16}. What is the origin of these sectors, and can we identify them from sequence data in a principled way?

To address these questions, we developed a general physical model that naturally gives rise to sectors. Specifically, motivated by the observation that many protein properties reflect additive contributions from individual amino acids~\cite{DePristo05,Starr16,Otwinowski18}, we consider any additive trait subject to natural selection. As a concrete example, we study a simple elastic-network model that quantifies the energetic cost of protein deformations~\cite{bahar2010global}, which we show to be an additive trait. We then demonstrate that selection acting on any such additive trait automatically yields collective correlation modes in sequence data. We show that the main signature of the selection process lies in the small-eigenvalue modes of the covariance matrix of the selected sequences, but we find that some signatures also exist in the widely-studied large-eigenvalue modes. Finally, we demonstrate a principled method to identify sectors and to quantify mutational effects from sequence data alone.

\section*{Results}

\subsection*{Selection on an additive trait}
We focus on selection on an additive scalar trait 
\begin{equation}
T (\vec{\alpha})=\sum_{l=1}^L\Delta_l(\alpha_l)\,,
\label{eq:trait}
\end{equation} 
where $\vec{\alpha}=(\alpha_1,\dots,\alpha_L)$ is the amino-acid sequence considered, $L$ is its length, and $\Delta_l(\alpha_l)$ is the mutational effect on the trait $T$ of a mutation to amino acid $\alpha_l$ at site $l$. Mutational effects can be measured with respect to a reference sequence $\vec{\alpha}^0$, satisfying $\Delta_l(\alpha_l^0)=0$ for all $l$. 

Eq.~\ref{eq:trait} is very general as it amounts to saying that, to lowest order, mutations have an additive effect on the trait $T$, which can be any relevant physical property of the protein, say its binding affinity, catalytic activity, or thermal stability~\cite{Cunningham17}. System-specific details are encoded by the single-site mutational effects $\Delta_l(\alpha_l)$, which can be measured experimentally. The assumption of additivity is experimentally validated in many cases. For instance, protein thermal stability, measured through folding free energy, is approximately additive~\cite{DePristo05,Wylie11}. Importantly, we allow selection to act on a phenotype that is a nonlinear function of $T$. Permitting a phenotypic nonlinearity on top of our additive trait model is motivated by the fact that actual phenotype data from recent high-throughput mutagenesis experiments were accurately modeled via a nonlinear mapping of an underlying additive trait~\cite{Otwinowski18}. 

Protein sectors are usually defined operationally as collective modes of correlations in amino-acid sequences. However, the general sequence-function relation in Eq.~\ref{eq:trait} suggests an operational definition of a \textit{functional} protein sector, namely as the set of sites with dominant mutational effects on a trait under selection. Selection can take multiple forms. To be concrete, we first consider a simple model of selection, assuming a favored value $T^*$ of the trait $T$, and using a Gaussian selection window. We subsequently show that the conclusions obtained within this simple model are robust to different forms of selection. Our Gaussian selection model amounts to selecting sequences according to the following Boltzmann distribution: 
\begin{equation}
P(\vec{\alpha})=\frac{\exp(w(\vec{\alpha}))}{ \sum_{\vec{\alpha}} \exp(w(\vec{\alpha}))}\,, 
\label{eq:distr}
\end{equation}
where the fitness $w(\vec{\alpha})$ of a sequence is given by
\begin{equation}
w(\vec{\alpha})=-\frac{\kappa}{2}\left(T(\vec{\alpha})-T^*\right)^2=-\frac{\kappa}{2}\left(\sum_{l=1}^L\Delta_l(\alpha_l)-T^*\right)^2\,.
\label{eq:quadr}
\end{equation}
The selection strength $\kappa$ sets the width of the selection window.

Such selection for intermediate values of a trait can be realistic, e.g. for protein stability~\cite{DePristo05}. However, the form of selection can vary, for example selection can be for a nonlinear transform of a trait to be above a certain threshold~\cite{Otwinowski18}, and several relevant selection variants are investigated below. Crucially, while the trait is additive (Eq.~\ref{eq:trait}), the fact that fitness (Eq.~\ref{eq:quadr}) and selection (Eq.~\ref{eq:distr}) are nonlinear functions of the trait leads to coupling between mutations. This phenomenon is known as global~\cite{Kryazhimskiy14,Otwinowski18} or nonspecific~\cite{Starr16} epistasis, and its relevance has been shown in evolution experiments~\cite{Kryazhimskiy14}, over and above contributions from specific epistasis~\cite{Starr16,Posfai18}. The focus of this paper is on global epistasis, and we do not include specific epistasis. Studying the interplay of these two types of epistasis will be an interesting future direction.

\subsection*{A toy model yielding a concrete example of an additive trait} 

\subsubsection*{Elastic-network model} 

\begin{figure}[h!]
	\centering
	\includegraphics[width=0.8\textwidth]{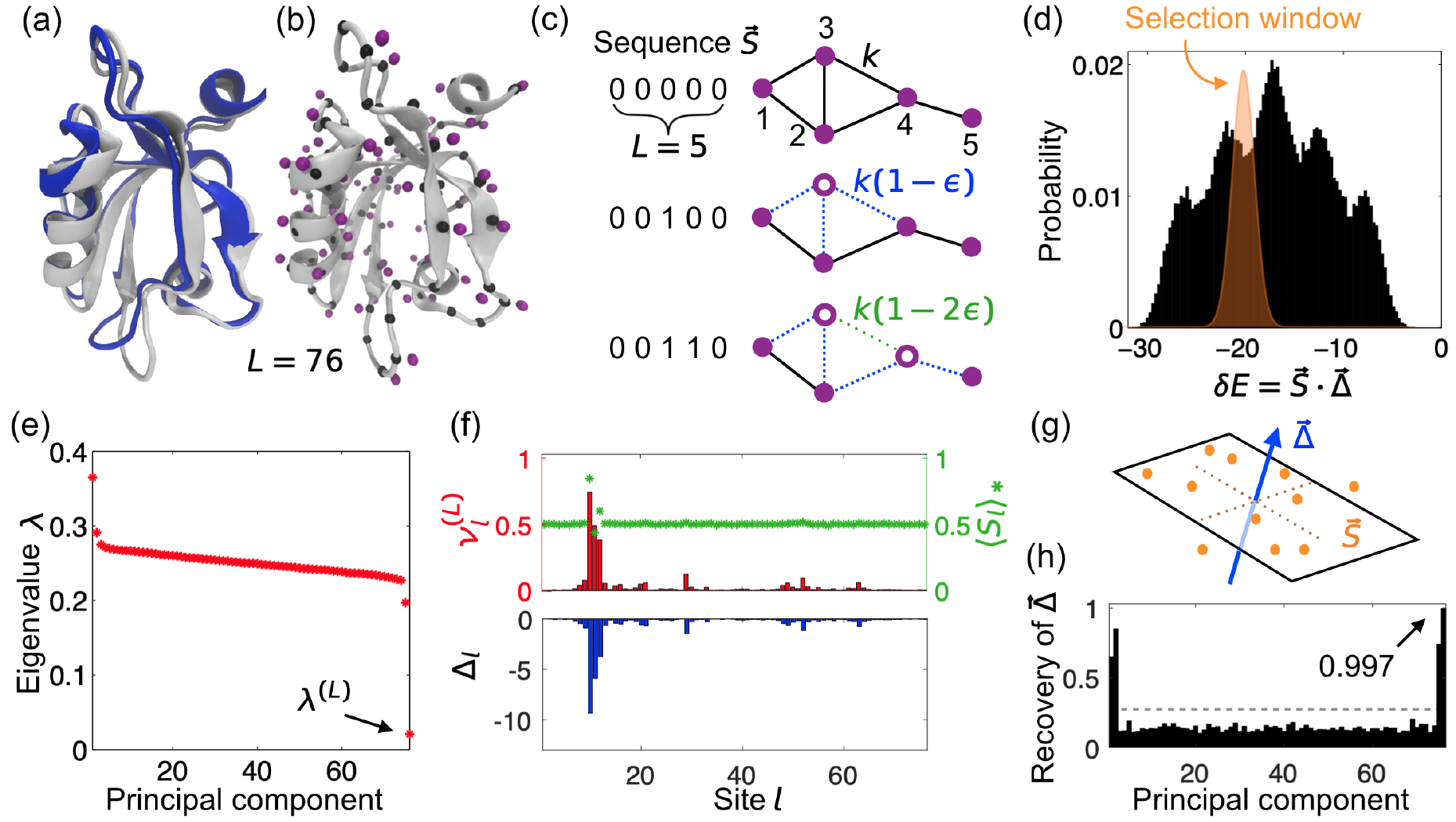}
	\vspace{0.2cm}
	\caption{{\bf Selection applied to an elastic protein model leads to a statistical signature among sequences.} (a) Cartoon representation of the third  PDZ domain of the rat postsynaptic density protein 95 from the RSCB PDB~\cite{Berman00}, (gray: ligand free, 1BFE; blue: ligand bound, 1BE9 (ligand not shown)).  (b)  Elastic network model for 1BFE, where each amino-acid residue is represented by its alpha carbon (C$\alpha$, black node) and beta carbon (C$\beta$, purple node). Nearby nodes interact through a harmonic spring~\cite{de2005functional} (\nameref{S1_Appendix}). (c) Relation between protein sequence $\vec{S}$ and elastic network: 0 denotes the reference state, while 1 denotes a mutated residue, which weakens interactions of the corresponding C$\beta$ with all its neighbors by $\epsilon$. (d) Histogram of the energy $\delta E$ required to deform the domain from its ligand-free to its ligand-bound conformation, for randomly sampled sequences where 0 and 1 are equally likely at each site. Sequences are selectively weighted using a narrow Gaussian window (orange) around $\delta E^*$.  (e) Eigenvalues of the covariance matrix $C$ for the selectively weighted protein sequences. (f) Upper panel: last principal component $\nu_l^{(L)}$ of $C$ (red) and average mutant fraction $\langle S_l\rangle_*$ (green)  at site $l$ after selection; lower panel: effect $\Delta_l$ of a single mutation at site $l$ on $\delta E$.  (g) Schematic representation of the selected ensemble in sequence space, where each dot is a highly-weighted sequence; thus dots are restricted to a narrow region around a plane perpendicular to $\vec{\Delta}$.  (h) Recovery of $\vec{\Delta}$ for all principal components $\vec{\nu}^{(j)}$, with maximum Recovery=1 (Eq.~\ref{eq:recovery-measure}). Gray dashed line: random expectation of Recovery (\nameref{S1_Appendix}).}
	\label{Fig1}
\end{figure}

To illustrate how additive traits naturally arise, we consider the elastic energy associated with a functionally important protein deformation. We explicitly derive the additivity of this trait in the regime of small deformations and weak mutational effects. This concrete example is relevant since functional deformation modes are under selection in proteins~\cite{Zheng06,Lukman09,Saldano16}, and dynamical domains possess a signature in sequence data~\cite{granata2017patterns}. Moreover, elastic-network models have elucidated a variety of protein properties~\cite{de2005functional,Delarue02,Zheng03,bahar2010global}, including the emergence of allostery~\cite{yan2017architecture,Tlusty17,Flechsig17,dutta2018green,rocks2017designing,Yan18,BraviPre}. Thus motivated, we begin by building an elastic-network model~\cite{bahar2010global,de2005functional} for a well-studied PDZ protein domain (Fig.~\ref{Fig1}(a,b))~\cite{doyle1996crystal,Hung02} and computing the relationship between its ``sequence'' and the energetic cost of a functionally-relevant conformational change. 

\clearpage

To build the elastic-network model of the PDZ domain,  we replace each of the $L=76$ amino-acid residues by its corresponding alpha carbon  C$\alpha$ and beta carbon C$\beta$, as shown in  Fig.~\ref{Fig1}(b).  Every pair of carbons within a cutoff distance $d_c$ is then connected with a harmonic spring~\cite{bahar2010global}. Following a previous analysis of the same PDZ domain~\cite{de2005functional},  we set $d_c=7.5\,\text{\normalfont\AA}$ and assign spring constants as follows:  a) 2  for C$\alpha$-C$\alpha$ pairs if adjacent along the backbone, 1 otherwise;  b) 1 for C$\alpha$-C$\beta$ pairs;  c) 0.5 for C$\beta$-C$\beta$ pairs. 

Within our elastic model, the energetic cost of a small deformation from the equilibrium structure is 
 \begin{equation}
 E=\frac{1}{2}\sum_{i,j} \left(\boldsymbol{r}_i-\boldsymbol{r}_i^0\right)M_{ij}\left(\boldsymbol{r}_j-\boldsymbol{r}_j^0\right)= \frac{1}{2}\,\delta \boldsymbol{r}^TM \delta \boldsymbol{r},
 \label{eq:energy_cost}
 \end{equation}
where $\boldsymbol{r}_i$ is the position of the $i$th carbon atom, $\boldsymbol{r}_i^0$ is its equilibrium position, and the Hessian matrix $M$ contains the second derivatives of the elastic energy with respect to atomic coordinates. Here,  we take $\delta \boldsymbol{r} $ to be the conformational change from a ligand-free state (1BFE) to a ligand-bound state (1BE9) of the same PDZ domain (Fig.~\ref{Fig1}(a)). This conformational change is central to PDZ function, so its energetic cost has presumably been under selection during evolution. Any other coherent conformational change would also be suitable for our analysis. Note that our aim is not to analyze conformational changes in all their richness, but to provide a minimal concrete example of a relevant additive trait, and to analyze the impact of selection acting on this trait on the associated family of sequences.  

To mimic simply the effect of mutation and selection within our toy model, we introduce ``mutations'' of residues that weaken the spring constants involving their beta carbons by a small fraction $\epsilon$. In practice, we take $\epsilon = 0.2$. We represent mutations using a sequence $\vec{S}$  with $S_l\in \{0,1\}$,  where $l$ is the residue index:   $S_l=0$ denotes the reference state, while $S_l=1$ implies a mutation (Fig.~\ref{Fig1}(c)). The sequence $\vec{S}$ and the spring network fully determine the Hessian matrix $M$, and thus the energy cost $E$ of a conformational change (Eq.~\ref{eq:energy_cost}). Note that here $\vec{S}$ is a binary sequence, which represents a simplification compared to real protein sequences $\vec{\alpha}$ where each site can feature 21 states (20 amino acids, plus the alignment gap). We start with the binary model for simplicity, and we then extend our results to a more realistic 21-state model. Note that binary representations of actual protein sequences, with a consensus residue state and a ``mutant'' state, have proved useful in sector analysis~\cite{halabi2009protein}, although more recent approaches for diverse protein families have employed full 21-state models~\cite{Rivoire16}. Binary representations are also appropriate to analyze sets of sufficiently close sequences, notably HIV proteins, allowing identification of their sectors~\cite{Dahirel11} and predictions of their fitness landscapes~\cite{Mann14}.

\subsubsection*{Deformation energy as an additive trait} 

Focusing on mutations that weakly perturb the elastic properties of a protein, we perform first-order perturbation analysis: 
 $M=M^{(0)}+\epsilon M^{(1)}+o(\epsilon)$. Using Eq.~\ref{eq:energy_cost} yields $E=E^{(0)}+\epsilon E^{(1)}+o(\epsilon)$, with $E^{(1)}=\delta \boldsymbol{r}^T M^{(1)} \delta \boldsymbol{r}/2$. Both $M^{(1)}$ and $E^{(1)}$ can be expressed as sums of contributions from individual mutations. We define $\Delta_l$ as the first-order energy cost $\epsilon E^{(1)}$ of a single mutation at site $l$ of the sequence. To leading order, the effect of mutations on the energy cost of a deformation reads
 \begin{equation}
 \delta E=E-E^{(0)}=\sum_{l=1}^L S_l\Delta_l.
\label{eq:generic} 
 \end{equation}
This equation corresponds to the binary-sequence case of the general additive trait defined in Eq.~\ref{eq:trait}. Hence, the deformation energy in our toy model of a protein as a sequence-dependent elastic network constitutes a practical example of an additive trait. 
 
Within our functional definition, a protein sector is the set of sites with dominant mutational effects on the trait under selection. The vector $\vec{\Delta}$ of mutational effects for our elastic-network model of the PDZ domain is shown in Fig.~\ref{Fig1}(f). The magnitudes of mutational effects are strongly heterogeneous (\nameref{S1_Appendix}, Fig.~\ref{fig:PDZ_Delta_histogram}). Here, the amino acids with largest effects, which constitute the sector, correspond to those that move most upon ligand binding. (Note that the ligand-binding deformation of PDZ is well-described by one low-frequency normal mode of the elastic network~\cite{de2005functional}: hence, our sector significantly overlaps with the sites that are most involved in this mode.)

How is such a functionally-defined sector reflected in the statistical properties of the sequences that survive evolution? To answer this question, we next analyze sequences obtained by selecting on the trait $\delta E$. While for concreteness, we use the mutational effects obtained from our elastic model, the analysis is general and applies to any additive trait. Indeed, we later present some examples using synthetically-generated random mutational effect vectors, both binary and more realistic 21-state ones (see Figs. 3, 4, and \nameref{S1_Appendix}).

\subsection*{Signature of selection in sequences} For our elastic model of the PDZ domain, the distribution of the additive trait $\delta E$ for random sequences is shown in Fig.~\ref{Fig1}(d). We use the selection process introduced in Eqs.~\ref{eq:distr}-\ref{eq:quadr} to limit sequences to a narrower distribution of $\delta E$s, corresponding, e.g., to a preferred ligand-binding affinity. The fitness of a binary sequence $\vec{S}$, a particular case of Eq.~\ref{eq:quadr}, reads:
\begin{equation}
w(\vec{S})=-\frac{\kappa}{2} \left( \sum_{l=1}^L \Delta_l S_l-\delta E^*\right)^2.
\label{eq:fitness}
\end{equation}
Here, the selection strength $\kappa$ sets the width of the selection window, and $\delta E^*$ is its center. For all selections, we take $\kappa=10/(\sum_l  \Delta_l^2)$, so that the width of the selection window scales with that of the unselected distribution. We have confirmed that our conclusions are robust to varying selection strength, provided $\kappa \sum_l\Delta_l^2\gg 1$ (see Fig.~\ref{fig:Supp_effect_kappa}).

Although mutations have additive effects on the trait $\delta E$, the nonlinearities involved in fitness and selection give rise to correlations among sites. For instance, if $\delta E^*=0$ and if $\Delta_l<0$ for all $l$, as in Fig.~\ref{Fig1}, a mutation at a site with large $|\Delta_l|$ will decrease the likelihood of additional mutations at all other sites with large $|\Delta_l|$. 

Previous approaches to identifying sectors from real protein sequences have relied on modified forms of Principal Component Analysis (PCA). So we begin by asking: can PCA identify sectors in our physical model? PCA corresponds to diagonalizing the covariance matrix $C$ of sequences: it identifies the principal components (eigenvectors)  $\vec{\nu}^{(j)}$  associated with progressively smaller variances (eigenvalues) $\lambda^{(j)}$. We introduce $\langle \cdot\rangle_*$ to denote ensemble averages over the selectively weighted sequences, reserving $\langle \cdot\rangle$ for averages over the unselected ensemble. The mutant fraction at site $l$ in the selected ensemble is $\langle S_l\rangle_*= \sum_{\vec{S}}S_l P(\vec{S})$, and the covariance matrix $C$ reads 
\begin{equation}
C_{ll'}=\Big\langle (S_l-\langle S_l\rangle_*)\cdot (S_{l'}-\langle S_{l'}\rangle_*) \Big\rangle_*.
\end{equation}

To test the ability of PCA to identify a functional sector, we employed the selection window shown in orange in Fig.~\ref{Fig1}(d). The resulting eigenvalues are shown in Fig.~\ref{Fig1}(e). One sees outliers. In particular, why is the last eigenvalue so low? Due to the narrow selection window, according to Eq.~\ref{eq:fitness} the highly-weighted sequences satisfy $\sum_l S_l \Delta_l=\vec{S} \cdot \vec{\Delta}\approx \delta E^*$. This means that in the $L$-dimensional sequence space, the data points for the highly-weighted sequences lie in a narrow region around a plane perpendicular to $\vec{\Delta}$ (Fig.~\ref{Fig1}(g)). Hence, the data has exceptionally small variance in this direction, leading to a particularly small eigenvalue of $C$. Moreover, the corresponding last principal component $\vec{\nu}^{(L)}$ points in the direction with the smallest variance and is consequently  parallel to $\vec{\Delta}$ (Fig.~\ref{Fig1}(f)). Formally, in Eq.~\ref{eq:fitness}, $\vec{\Delta}$  appears in a quadratic coupling term where it plays the part of a repulsive pattern in a generalized Hopfield model~\cite{Cocco11,cocco2013principal}: alone, such a term would penalize sequences aligned with $\vec{\Delta}$. But here, $\vec{\Delta}$ also appears in a term linear in $\vec{S}$ and as a result Eq.~\ref{eq:fitness} penalizes sequences that fail to have the selected projection onto $\vec{\Delta}$. 
 
In this example, the last principal component accurately recovers the functional sector corresponding to the largest elements of the mutational-effect vector $\vec{\Delta}$. More generally, to quantify the recovery of $\vec{\Delta}$ by a given vector $\vec{\nu}$,  we introduce
  \begin{equation}
\mathrm{Recovery}=\frac{\sum_l |\nu_l  \Delta_l |}{\sqrt{\sum_l \nu_l^2 }\sqrt{\sum_l \Delta_l^2 }},
\label{eq:recovery-measure}
 \end{equation} 
which is nonnegative, has a random expectation of  $(\sqrt{2/\pi L})\sum_l |\Delta_l|/\sqrt{\sum_l \Delta_l^2}$ for $L\gg1$ (\nameref{S1_Appendix}), and saturates at 1 (including the case of parallel vectors). For our test case, Fig.~\ref{Fig1}(h) shows Recovery for all principal components. The last one features the highest Recovery, almost 1, confirming that it carries substantial information about $\vec{\Delta}$. The second-to-last principal component and the first two also provide a value of Recovery substantially above random expectation. Outlier eigenvalues arise from the sector, and accordingly, we find that the number of modes with high Recovery often corresponds to the number of sites with strong mutational effects. A more formal analysis of this effect will be an interesting topic for further study.
	
In our model, $\vec{\Delta}$ is fundamentally a direction of \textit{small variance}. So why do the first principal components also carry information about $\vec{\Delta}$? Qualitatively, when variance is decreased in one direction due to a repulsive pattern $\vec{\Delta}$, variance tends to increase in orthogonal directions involving the same sites. To illustrate this effect, let $L=3$ and $\vec{\Delta}=(-1,1,0)$, and consider the sequences $\vec{S}$ satisfying $\vec{\Delta}\cdot\vec{S}=0$ (namely $(0,0,0)$; $(1,1,0)$; $(0,0,1)$; $(1,1,1)$). The last principal component is $\vec{\Delta}$, with zero variance, and the first principal component is $(1,1,0)$: Recovery is 1 for both of them. This selection conserves the trace of the covariance matrix (i.e. the total variance), so that decreasing the variance along $\vec{\Delta}=(-1,1,0)$ necessarily increases it along $(1,1,0)$. This simple example provides an intuitive understanding of why the large-eigenvalue modes of the covariance matrix also carry information about $\vec{\Delta}$.

It is worth remarking that Eq.~\ref{eq:fitness} is a particular case of a general fitness function with one- and two-body terms (known as fields and couplings in Ising or Potts models in physics). Here, the values of these one- and two-body terms are constrained by their expressions in terms of $\vec{\Delta}$. In practice, several traits might be selected simultaneously (see below), yielding more independent terms among the fields and couplings. More generally, such one- and two-body descriptions have been very successfully employed via Direct Coupling Analysis (DCA) to identify strongly coupled residues that are in contact within a folded protein~\cite{Weigt09,morcos2011direct,Marks11}, to investigate folding~\cite{Morcos14}, and to predict fitness~\cite{Dwyer13,Cheng14,Cheng16,Mann14,Figliuzzi16,Barton16,Hopf17} and conformational changes~\cite{Morcos11,Malinverni15}, as well as protein-protein interactions~\cite{Bitbol16,Gueudre16}. A complete model of protein covariation in nature should necessarily incorporate both the collective modes described here and the strongly coupled residue pairs which are the focus of DCA.

\subsection*{ICOD method} 
\begin{figure}[h!]
\centering
\includegraphics[width=8.5cm]{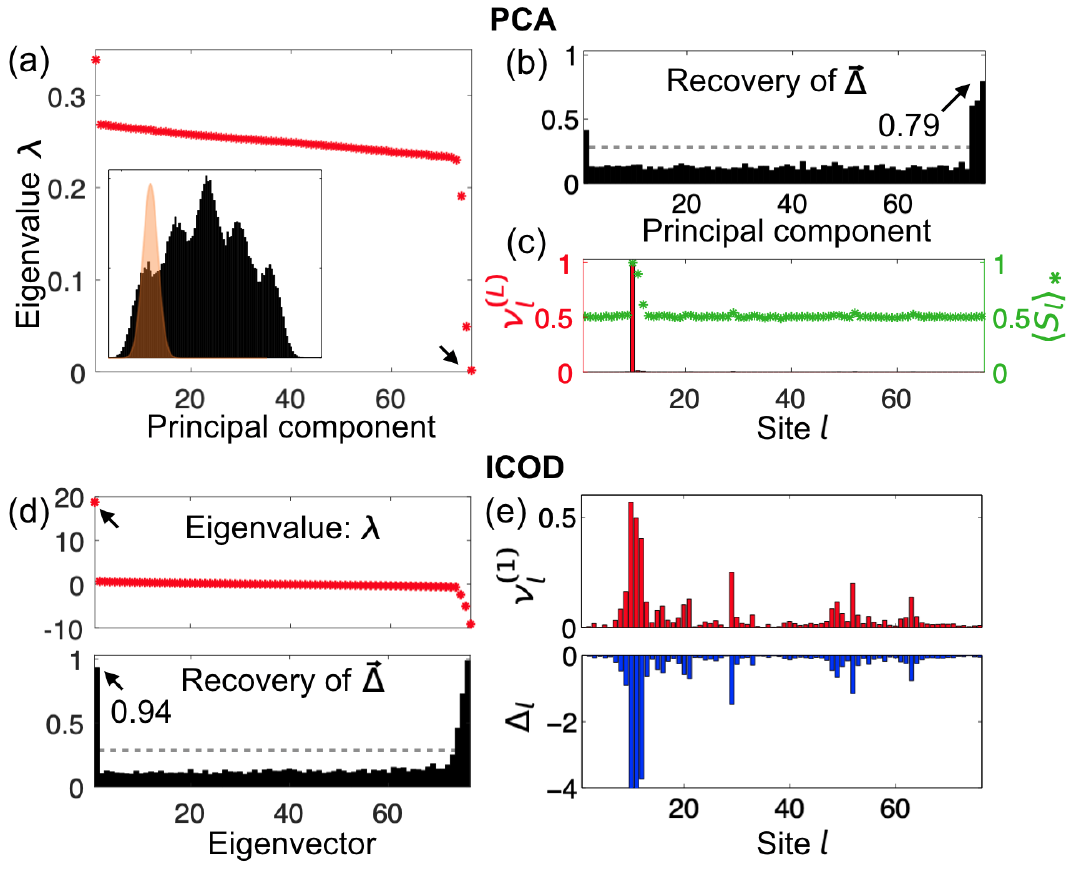}
\vspace{0.2cm}
\caption{{\bf Recovery of mutational-effect vector $\vec{\Delta}$ from sequence analysis in the case of strongly biased selection.} (a-c) Principal Component Analysis (PCA)  performs poorly due to strong conservation at some sites of large mutational effect. (a) Eigenvalues of covariance matrix obtained for strongly biased selection around $\delta E_{\rm biased}^*$  (inset, orange window) for same model proteins as in Fig. 1. (b) Recovery of $\vec{\Delta}$  for all principal components.  (c) Last principal component $\nu_l^{(L)}$ (red) and average mutant fraction $\langle S_l\rangle_*$ (green) at site $l$. (d-e) The ICOD method performs robustly. (d) Eigenvalues of $\tilde{C}^{-1}_{ll'}$ (Eq.~\ref{eq:ICODmatrix}) (upper) and Recovery of $\vec{\Delta}$ for all eigenvectors (lower). (e) Leading eigenvector $\nu_l^{(1)}$ (upper) and mutational effect $\Delta_l$ at site $l$ (lower, same as in Fig. 1(f)). Gray dashed lines in (b,d): random expectation of Recovery (\nameref{S1_Appendix}).  }
\label{Fig2}
\end{figure}

An important concern is whether the last principal component is robust to small and/or noisy datasets. Indeed, other directions of small variance can appear in the data. As a second example, we applied a different selection window, centered in the tail of the distribution of $\delta E$s from our elastic model of the PDZ domain (Fig.~\ref{Fig2}(a), inset). This biased selection generates strong conservation, $\langle S_l\rangle_*\approx 1$, for some sites with significant mutational effects. Extreme conservation at one site now dictates the last principal component, and disrupts PCA-based recovery of $\vec{\Delta}$ (Fig.~\ref{Fig2}(a,b)).

To overcome this difficulty, we developed a more robust approach that relies on inverting the covariance matrix. Previously, the inverse covariance matrix was successfully employed in Direct Coupling Analysis (DCA) to identify strongly coupled residues that are in contact within a folded protein~\cite{Weigt09,morcos2011direct,Marks11}. The fitness in our model (Eq.~\ref{eq:fitness}) involves one and two-body interaction terms, and constitutes a particular case of the DCA Hamiltonian (\nameref{S1_Appendix}). A small-coupling approximation ~\cite{Plefka82,morcos2011direct,Marks11,bitbol2016inferring} (\nameref{S1_Appendix}) gives
\begin{equation}
C^{-1}_{ll'}\approx  \left(1-\delta_{ll'}\right)\,\kappa \Delta_l\Delta_{l'}+\delta_{ll'}\left(\frac{1}{P_l}+\frac{1}{1-P_l}\right),
\label{eq:inverse-PCA-q=2}
\end{equation}
 where $P_l$ denotes the probability that site $l$ is mutated. Since we are interested in extracting $\vec{\Delta}$, we can simply set to zero the diagonal elements of $C^{-1}$, which are dominated by conservation effects, to obtain a new matrix  
\begin{equation}
\tilde{C}^{-1}_{ll'}\approx (1-\delta_{ll'}) \kappa \Delta_l\Delta_{l'}.
\label{eq:ICODmatrix}
\end{equation}
The first eigenvector of $\tilde{C}^{-1}$ (associated with its largest eigenvalue) should accurately report $\vec{\Delta}$ since, except for its zero diagonal, $\tilde{C}^{-1}$ is proportional to the outer product $\vec{\Delta} \otimes \vec{\Delta}$.  We call this approach the  \emph{Inverse Covariance Off-Diagonal} (ICOD) method. As shown in Fig.~\ref{Fig2}(d-e), ICOD overcomes the difficulty experienced by PCA for biased selection, while performing equally well as PCA for unbiased selection (Fig.~\ref{eq:supp-inverse-Fig1}, \nameref{S1_Appendix}). Removing the diagonal elements of $C^{-1}$ before diagonalizing is crucial: otherwise, the first eigenvector of $C^{-1}$ is the same as the last eigenvector of $C$ and suffers from the same shortcomings for strong conservation. Here too, eigenvectors associated to both small and large eigenvalues contain information about $\vec{\Delta}$ (Figs.~\ref{Fig2}(b,d)).

\subsection*{Selection on multiple traits} An important challenge in sector analysis is distinguishing multiple, independently evolving sectors~\cite{halabi2009protein,Tesileanu15,Rivoire16}. We can readily generalize our fitness function (Eqs.~\ref{eq:quadr},~\ref{eq:fitness}) to allow for selection on multiple additive traits:  
\begin{equation}
w(\vec{S})=-\sum_{i=1}^N \frac{\kappa_i}{2} \left( \sum_{l=1}^L \Delta_{i,l} S_l-T^*_{i}\right)^2,
\label{eq:newfitness}
\end{equation}
where $N$ is the number of distinct additive traits $T_i(\vec{S})=\sum_l \Delta_{i,l} S_l$ under selection, $\vec{\Delta}_i$ is the vector of mutational effects on trait $T_i$, $\kappa_i$ is the strength of selection on this trait, and $T^*_{i}$ is the associated selection bias.  For example, 
 $\vec{\Delta}_1$ might measure how mutations change a protein's binding affinity, while $\vec{\Delta}_2$ might be related to its thermal stability, etc.  
 
In \nameref{S1_Appendix} Fig.~\ref{fig:two-sector-fig3}, we consider selection on two distinct additive traits, using synthetically-generated random mutational-effect vectors $\vec{\Delta}_1$ and $\vec{\Delta}_2$ (\nameref{S1_Appendix}). Note that these mutational effects are thus unrelated to our toy model of protein elastic deformations: as stated above, our approach holds for any additive trait under selection. ICOD then yields \emph{two} large outlier eigenvalues of the modified inverse covariance matrix $\tilde{C}^{-1}$. The associated eigenvectors accurately recover both $\vec{\Delta}_1$ and $\vec{\Delta}_2$, after a final step of Independent Component Analysis (ICA)~\cite{Hyvarinen, Hansen01, Rivoire16} that successfully disentangles the contributions coming from the two constraints (see \nameref{S1_Appendix}).

\subsection*{Performance in sector recovery} 

\begin{figure}[h!]
\centering
\includegraphics[width=8.5cm]{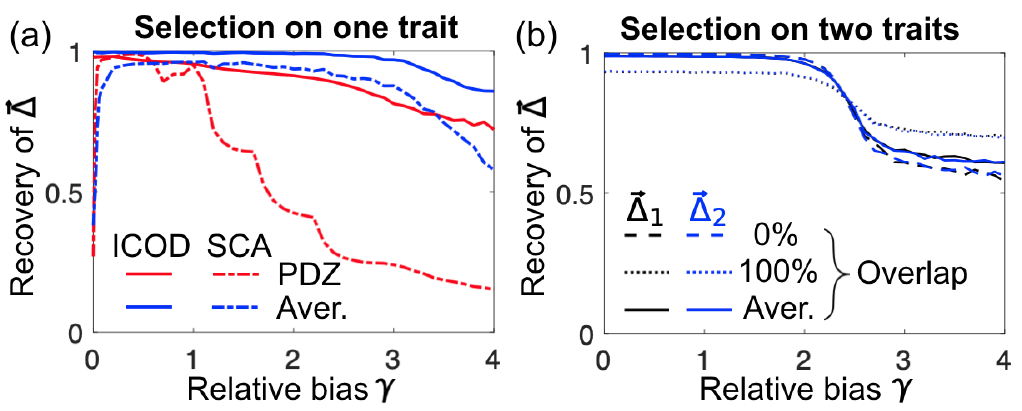}
\vspace{0.2cm}
\caption{{\bf Average recovery of mutational-effect vectors $\vec{\Delta}$ as a function of relative selection bias $\gamma\equiv (T^*-\langle T\rangle)/\sqrt{\langle (T-\langle T\rangle)^2 \rangle}$ on the selected additive trait $T$.}  (a) Selection on a single trait.  Different $\vec{\Delta}$s are used to generate sequence ensembles: the elastic-network  $\vec{\Delta}$ from Fig. 1 (red); synthetic $\vec{\Delta}$s (\nameref{S1_Appendix}) with number of sites of large mutational effect (sector sites) ranging from 1 to 100, for sequences of length $L=100$  (blue). Recovery is shown for ICOD (solid curves) and for SCA~\cite{halabi2009protein,Rivoire16} (dashed curves).  (b) Selection on two distinct traits. Different pairs of synthetic $\vec{\Delta}$s (\nameref{S1_Appendix}) are used to generate sequence ensembles (with $L=100$):  ``0$\%$'' indicates two non-overlapping sectors, each with 20 sites; ``100$\%$'' indicates two fully overlapping sectors, each with 100 sites;  ``Aver.'' indicates average Recovery over 100 cases of double selection, where the single-sector size increases from 1 to 100, and the overlap correspondingly increases from 0 to 100.  ICA was applied to improve Recovery (\nameref{S1_Appendix}).  }
\label{Fig3}
\end{figure}

We further tested the performance of ICOD by systematically varying the selection bias, both for our toy model of PDZ elastic deformations and for more general synthetically-generated random mutational-effect vectors (Fig.~\ref{Fig3}). ICOD achieves high Recovery of these various mutational-effect vectors for both single and double selection over a broad range of selection biases $T^*$, albeit performance falls off in the limit of extreme bias.

How does ICOD compare with other approaches to identifying sectors? We compared the performance of ICOD with Statistical Coupling Analysis (SCA), the original PCA-based method~\cite{halabi2009protein,Rivoire16}. In SCA, the covariance matrix $C$ is reweighted by a site-specific conservation factor $\phi_l$, the absolute value is taken, $\tilde{C}_{ll'}^{\mathrm{(SCA)}}=|\phi_l C_{ll'}\phi_{l'}|$,
and sectors are identified from the leading eigenvectors of $\tilde{C}^{\mathrm{(SCA)}}$. We therefore tested the ability of the first eigenvector of $\tilde{C}^{\mathrm{(SCA)}}$ to recover $\vec{\Delta}$ for a single selection.  We found that the square root of the elements of the first SCA eigenvector can provide high Recovery of $\vec{\Delta}$ (Figs.~\ref{Fig3},~\ref{fig:Covariance_structure},~\ref{fig:reweighting-method}) (\nameref{S1_Appendix}). However, the performance of SCA relies on conservation through  $\phi_l$, and it has been shown that residue conservation actually dominates sector identification by SCA in certain proteins~\cite{Tesileanu15}. Consequently, for unbiased selection, SCA breaks down (Fig.~\ref{Fig3}(a), dashed curves) and cannot identify sector sites (\nameref{S1_Appendix} Fig.~\ref{fig:SCA_ICOD}). ICOD does not suffer from such shortcomings, and performs well over a large range of selection biases. Note that in SCA, only the top eigenvectors of $\tilde{C}^{\mathrm{(SCA)}}$ convey information about sectors (Figs.~\ref{fig:Covariance_structure}, \ref{fig:SCAsquareroot}). 

We also compared ICOD with another PCA-based approach~\cite{Cocco11}, which employs an inference method specific to the generalized Hopfield model, and should thus be well adapted to identifying sectors within our physical model (Eq.~\ref{eq:fitness}). Overall, this specialized approach performs similarly to ICOD, being slightly better for very localized sectors, but less robust than ICOD for strong selective biases and small datasets (\nameref{S1_Appendix}). Exactly as for PCA and ICOD, within this method, the top Recovery is obtained for the bottom eigenvector of the (modified) covariance matrix, consistent with $\vec{\Delta}$ in our model being a repulsive pattern~\cite{Cocco11}, but large Recoveries are also obtained for the top eigenvectors (Fig.~\ref{fig:Cocco_performance}).

\subsection*{Robustness to different forms of selection} 

\begin{figure}[h!]
\centering
\includegraphics[width=8.5cm]{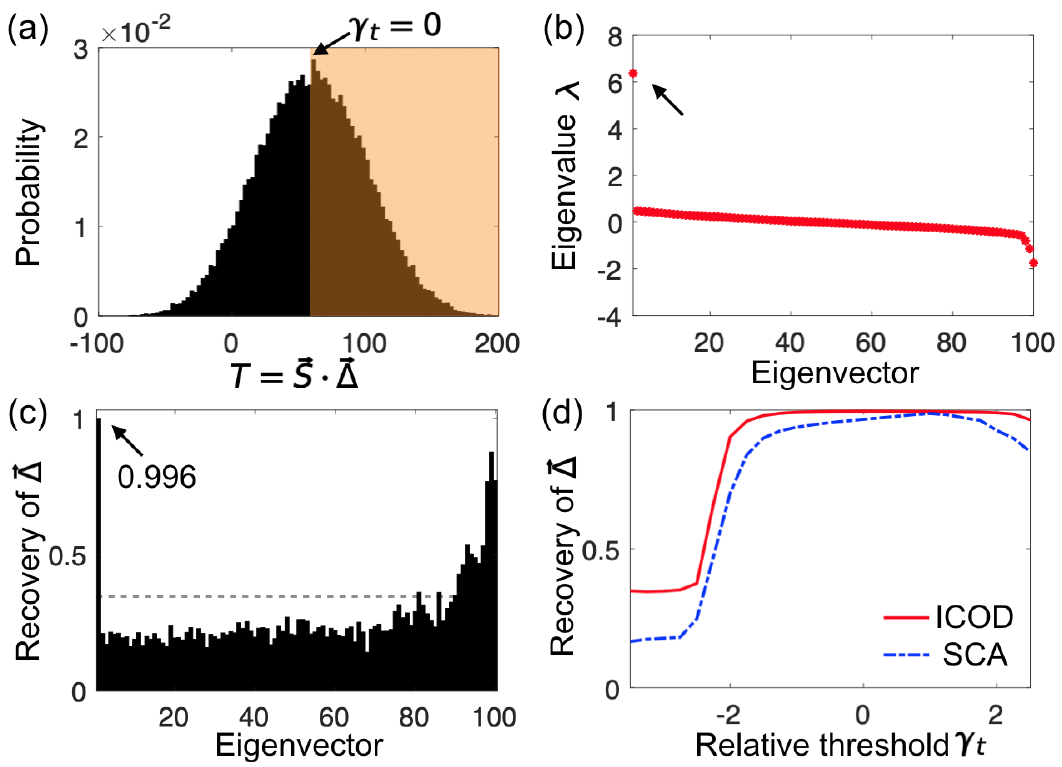}
\vspace{0.2cm}
\caption{{\bf  Identification of sectors that result from threshold-based selection. } (a)  Histogram of the additive trait $T(\vec{S})=\vec{S}\cdot\vec{\Delta}$ for randomly sampled sequences where 0 and 1 are equally likely at each site. Sequence length is $L=100$, mutational effects are synthetically generated with 20 sector sites (see \nameref{S1_Appendix}). Sequences are selected if they have a trait value $T(\vec{S})>T_t$ (orange shaded region). Selection is shown for $T_t=\langle T\rangle$, or equivalently $\gamma_t=0$, in terms of the relative threshold $\gamma_t\equiv (T_t-\langle T\rangle)/\sqrt{\langle (T-\langle T\rangle)^2 \rangle}$.  (b) Eigenvalues of the ICOD-modified inverse covariance matrix $\tilde{C}^{-1}$ (Eq.~\ref{eq:ICODmatrix}) of the selected sequences for $\gamma_t=0$. (c) Recovery of $\vec{\Delta}$ for all eigenvectors of $\tilde{C}^{-1}$ for $\gamma_t=0$. Gray dashed line: random expectation of Recovery. (d) Recovery of $\vec{\Delta}$ for ICOD and for SCA as functions of the relative selection threshold $\gamma_t$. The data in (d) is averaged over 100 realizations of $\vec \Delta$.   
}
\label{Fig4}
\end{figure}

To assess the robustness of functional sectors to selections different from the simple Gaussian selection window of Eqs.~\ref{eq:distr}-\ref{eq:quadr}, we selected sequences with an additive trait $T$ above a threshold $T_t$, and varied this threshold. For instance, a fluorescent protein might be selected to be fluorescent enough, which could be modeled by requiring that (a nonlinear transform of) an additive trait be sufficiently large~\cite{Otwinowski18}. As shown in Fig.~\ref{Fig4}, the corresponding sectors are identified by ICOD as well as those resulting from our initial Gaussian selection window. In Fig.~\ref{Fig4}(d), we show the performance of both ICOD and SCA at recovering sectors arising from selection with a threshold. Consistent with previous results (see Fig.~\ref{Fig3}), we find that ICOD is more robust than SCA to extreme selections. We also successfully applied ICOD to other forms of selection: Fig.~\ref{fig:quartic_selection} shows the case of a quartic fitness function replacing the initial quadratic one (Eq.~\ref{eq:quadr}) in the Boltzmann distribution (Eq.~\ref{eq:distr}) and Fig.~\ref{fig:square_selection} shows the case of a rectangular selection window (\nameref{S1_Appendix}). These results demonstrate the robustness of functional sectors, and of ICOD, to different plausible forms of selection. 

\newpage

\subsection*{Extension to 21-state sequences and to natural sequences} 

So far, we have considered binary sequences, with only one type of mutation with respect to the reference state. In the \nameref{S1_Appendix}, we demonstrate that our formalism, including the ICOD method, extends to mutations among $q$ different states. The case $q=21$, which includes the 20 different amino-acid types plus the alignment gap is the relevant one for real proteins. The single-site mutational effects $\Delta_l$ are then replaced by state-specific mutational effects $\Delta_l(\alpha_l)$ with $\alpha_l\in\{1,\dots,21\}$ (see Eq.~\ref{eq:trait}). Fig.~\ref{fig:protein-data-no-pseudocount} (\nameref{S1_Appendix}) shows that the generalized version of ICOD performs very well on synthetic data generated for the case $q=21$.  
We further demonstrate that sector identification is robust to gauge changes (reference changes) and to the use of pseudocounts (\nameref{S1_Appendix}). 

\begin{figure}[h!]
	\centering
	\includegraphics[width=8.5cm]{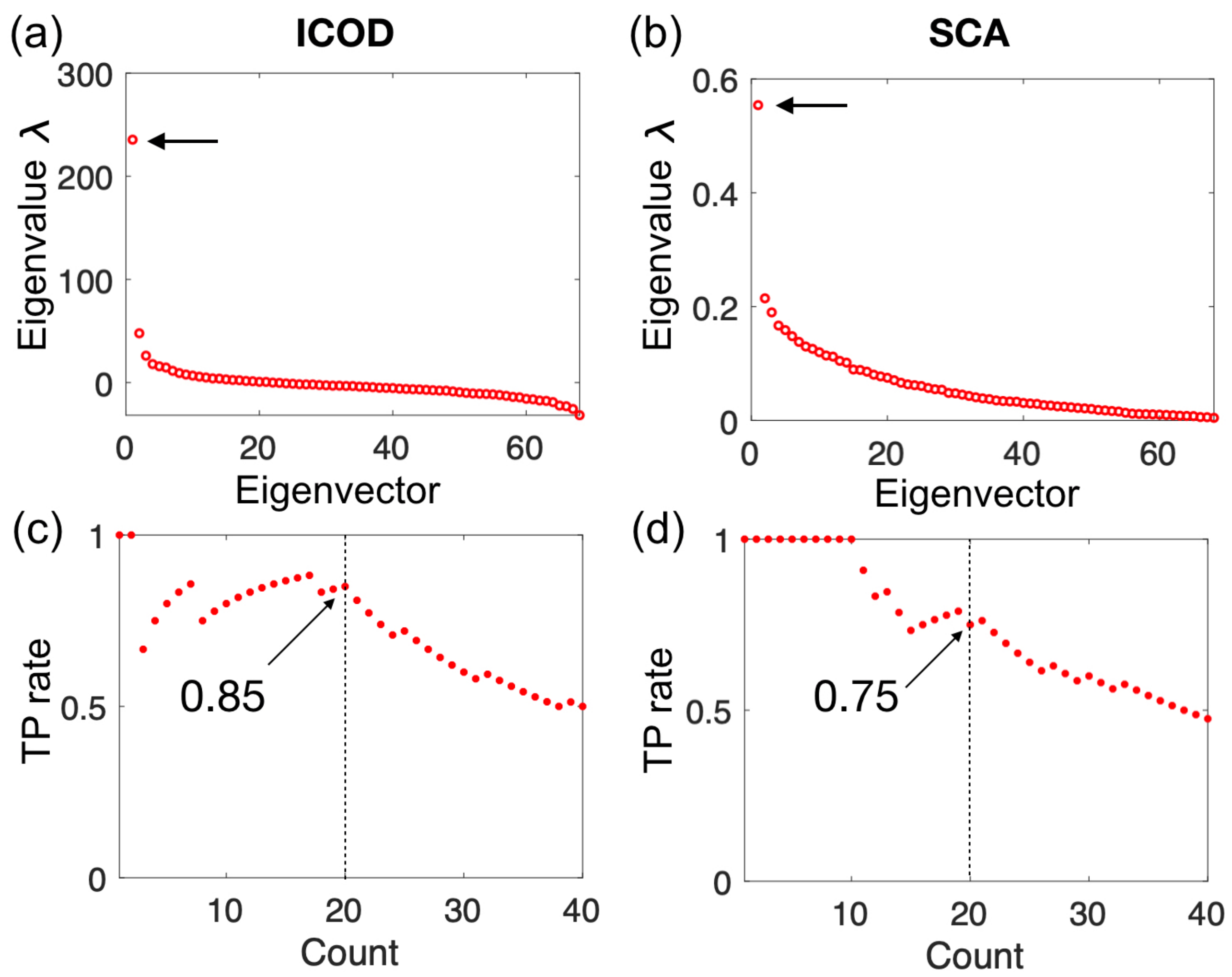}
	\vspace{0.2cm}
	\caption{ 
		{\bf Performance of ICOD and SCA at predicting the 20 sites with largest experimentally-determined mutational effects in a PDZ domain.}  (a) Eigenvalues of the compressed ICOD-modified inverse covariance matrix $\tilde{C}^{-1}$ (\nameref{S1_Appendix}). (b) Eigenvalues of the SCA matrix. (c) True Positive (TP) rates obtained by taking the first eigenvector $\vec{\nu}^{(1)}$ from the compressed ICOD-modified inverse covariance matrix,  generating a ranked list of sites of descending magnitudes of the components $||\nu_l^{(1)}||$ of this eigenvector at each site $l$ (\nameref{S1_Appendix}), and computing the fraction of the top sites in this predicted ordering that are also among the 20 experimentally most important sites~\cite{mclaughlin2012spatial}. Results are shown versus the number of top predicted sites (``count''). (d) TP rates from SCA, computed as in panel (c). In panels (c) and (d), the TP rate values obtained for the top 20 predicted sites are indicated by arrows. In all panels, a pseudocount ratio $\Lambda=0.02$ was used, and sites with more than $15\%$ gap state were discarded (see \nameref{S1_Appendix} for details).
}
	\label{Fig5}
\end{figure}

While the main purpose of this article is to propose an operational definition of functional protein sectors and to understand how they can arise, an interesting next question will be to investigate what ICOD can teach us about real data. As a first step in this direction, we applied ICOD to a multiple sequence alignment of PDZ domains. In this analysis, we employed a complete description with $q=21$, but we compressed the ICOD-modified inverse matrix using the Frobenius norm to focus on overall (and not residue-specific) mutational effects (see \nameref{S1_Appendix} for details). As shown in Figs.~\ref{Fig5}(a) and (b), both ICOD and SCA identify one strong outlying large eigenvalue, thus confirming that PDZ has only one sector~\cite{mclaughlin2012spatial}. Recall that due to the inversion step, the largest eigenvalue in ICOD is related to the mode with smallest variance, whose importance was demonstrated above. Furthermore, as seen in Figs.~\ref{Fig5}(c) and (d), both methods correctly predict the majority of residues found experimentally to have important mutational effects on ligand binding to the PDZ domain shown in Fig.~\ref{Fig1}(a)~\cite{mclaughlin2012spatial}. For instance, over the 20 top sites identified by ICOD (resp. SCA), we find that 85\% (resp. 75\%) of them are also among the 20 experimentally most important sites. Note that for SCA, we recover the result from Ref.~\cite{mclaughlin2012spatial}. The performance of ICOD is robust to varying the cutoff for removal of sites with a large proportion of gaps (see Fig.~\ref{fig:Supp_Effect_Cut_gap}), but notably less robust than SCA to pseudocount variation (see Fig.~\ref{fig:Supp_PseudoCount_ICOD_SCA}). 

Importantly, both ICOD and SCA perform much better than random expectation, which is $29\%$. Hence, both of these methods can be useful to identify functionally important sites. The slightly greater robustness of SCA to pseudocounts on this particular dataset (see Fig.~\ref{fig:Supp_PseudoCount_ICOD_SCA}) might come from the fact that many of the experimentally-identified functionally important sites in the PDZ domain are strongly conserved~\cite{Tesileanu15}, which makes the conservation reweighting step in SCA advantageous. Since residue conservation alone is able to predict most of the experimentally important PDZ sites~\cite{Tesileanu15}, we also compared conservation to SCA and ICOD: ranking sites by conservation (employing the conservation score of Ref.~\cite{Rivoire16}, see \nameref{S1_Appendix}) indeed identifies 70\% of the top 20 experimentally-determined sites with important mutational effects. Interestingly, ICOD scores are slightly more strongly correlated with conservation than SCA scores are correlated with conservation (see Fig.~\ref{fig:Supp_Similarity_Conservation_ICOD_SCA}), despite the fact that conservation is explicitly used in SCA and not in ICOD. 

Overall, this preliminary application to real data highlights the ability of ICOD to identify functionally related amino acids in a principled way that only relies on covariance. We emphasize that the main goal of this paper is to provide insight into the possible physical origins of sectors, and into the statistical signatures of these physical sectors in sequence data. A more extensive application of ICOD and related methods to real sequence data will be the subject of future work.

\section*{Discussion}

We have demonstrated how sectors of collectively correlated amino acids can arise from evolutionary constraints on functional properties of proteins. Our model is very general, as it only relies on the functional property under any of various forms of selection being described by an underlying additive trait, which has proven to be valid in many relevant situations~\cite{DePristo05,Wylie11,Starr16,Otwinowski18}. 

We showed that the primary signature of functional selection acting on sequences lies in the small-eigenvalue modes of the covariance matrix. In contrast, sectors are usually identified from the large-eigenvalue modes of the SCA matrix~\cite{halabi2009protein,Rivoire16}. This is not in contradiction with our results because, as we showed, signatures of our functional sectors are often also found in large-eigenvalue modes of the covariance matrix. Besides, the construction of the SCA matrix from the covariance matrix involves reweighting by conservation and taking an absolute value or a norm~\cite{halabi2009protein,Rivoire16}, which can substantially modify its eigenvectors, eigenvalues, and their order. Conservation is certainly important in real proteins, especially in the presence of phylogeny; indeed, the SCA matrix, which includes both conservation and covariance, was recently found to capture well experimentally-measured epistasis with respect to the free energy of PDZ ligand binding~\cite{Salinas18}. However, the fundamental link we propose between functional sectors and small-eigenvalue modes of the covariance matrix is important, since large-eigenvalue modes of the covariance matrix also contain confounding information about subfamily-specific residues~\cite{Casari95} and phylogeny~\cite{Qin18}, and consistently, some sectors identified by SCA have been found to reflect evolutionary history rather than function~\cite{halabi2009protein}. Interestingly, the small-eigenvalue modes are also the ones that contain most information about structural contacts in real proteins~\cite{cocco2013principal}. Hence, our results help explain previously observed correlations between sectors and contacts, e.g. the fact that contacts are overrepresented within a sector but not across sectors~\cite{rivoire2013elements}. 

We introduced a principled method to detect functional sectors from sequence data, based on the primary signature of these sectors in the small-eigenvalue modes of the covariance matrix. We further demonstrated the robustness of our approach to the existence of multiple traits simultaneously under selection, to various forms of selection, and to data-specific questions such as reference choices and pseudocounts. 

Importantly, our modeling approach allowed us to focus on functional selection alone, in the absence of historical contingency and of specific structural constraints, thus yielding insights complementary to purely data-driven methods. The collective modes investigated here are just one source of residue-residue correlations. Next, it will be interesting to study the intriguing interplay between functional sectors, phylogeny, and contacts, and to apply our methods to multiple protein families. Our results shed light on an aspect of the protein sequence-function relationship and open new directions in protein sequence analysis, with implications in synthetic biology, building toward function-driven protein design.

\section*{Supporting information}


\paragraph*{S1~Appendix.}
\label{S1_Appendix}
{\bf Methodological details and further results.} In \nameref{S1_Appendix}, we present additional details about our model and methods, as well as additional results.

\section*{Acknowledgments}
S.-W. W. and N. S. W. acknowledge the Center for the Physics of Biological Function under NSF Grant PHY-1734030. A.-F. B. and N. S. W. acknowledge the Aspen Center for Physics, which is supported by NSF Grant PHY-1607611. S.-W. W. was  supported by  the NSFC under Grants No. U1430237 and 11635002. S.-W. W. also acknowledges Tsinghua University for supporting a half-year visit in Princeton University. N. S. W. was supported by NSF Grant MCB-1344191 and by NIH Grant R01 GM082938.

\newpage


\noindent{\huge
\textbf{ S1~Appendix} 
}

\bigskip

\renewcommand{\thefigure}{S\arabic{figure}}
\setcounter{figure}{0} 
\renewcommand{\theequation}{S\arabic{equation}}
\setcounter{equation}{0} 
\renewcommand{\thetable}{S\arabic{table}}
\setcounter{table}{0} 

\def\theequation{S\arabic{equation}}
\setcounter{equation}{0}
\def\thefigure{S\arabic{figure}}
\setcounter{figure}{0}
\newcommand{\wh}[1]{\widehat{#1}}

\tableofcontents

\newpage

\section{Supplemental results for elastic network model of PDZ domain}

\begin{figure}[htb]
\centering
\includegraphics[width=8.5cm]{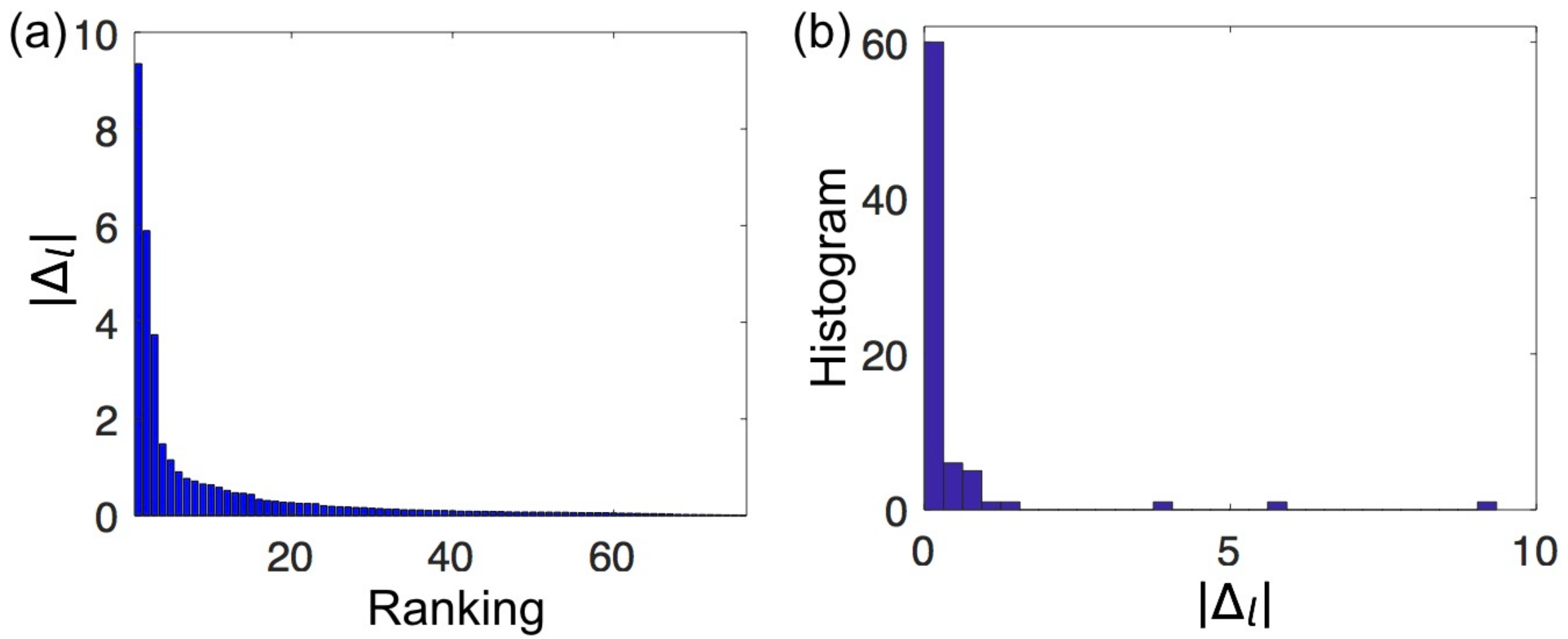}
\vspace{0.2cm}
\caption{{\bf Magnitude of single-site mutational effects $\Delta_l$ for the PDZ domain conformational change from Fig.~1.}  (a)  Magnitudes by rank. (b) Histogram of magnitudes. According to our definition, the sites of large magnitude constitute ``sector'' sites with respect to selection on the energy cost of this conformational change, while all others are ``non-sector'' sites.  
}
\label{fig:PDZ_Delta_histogram}
\end{figure}

\begin{figure}[htb]
\centering
\includegraphics[width=8.5cm]{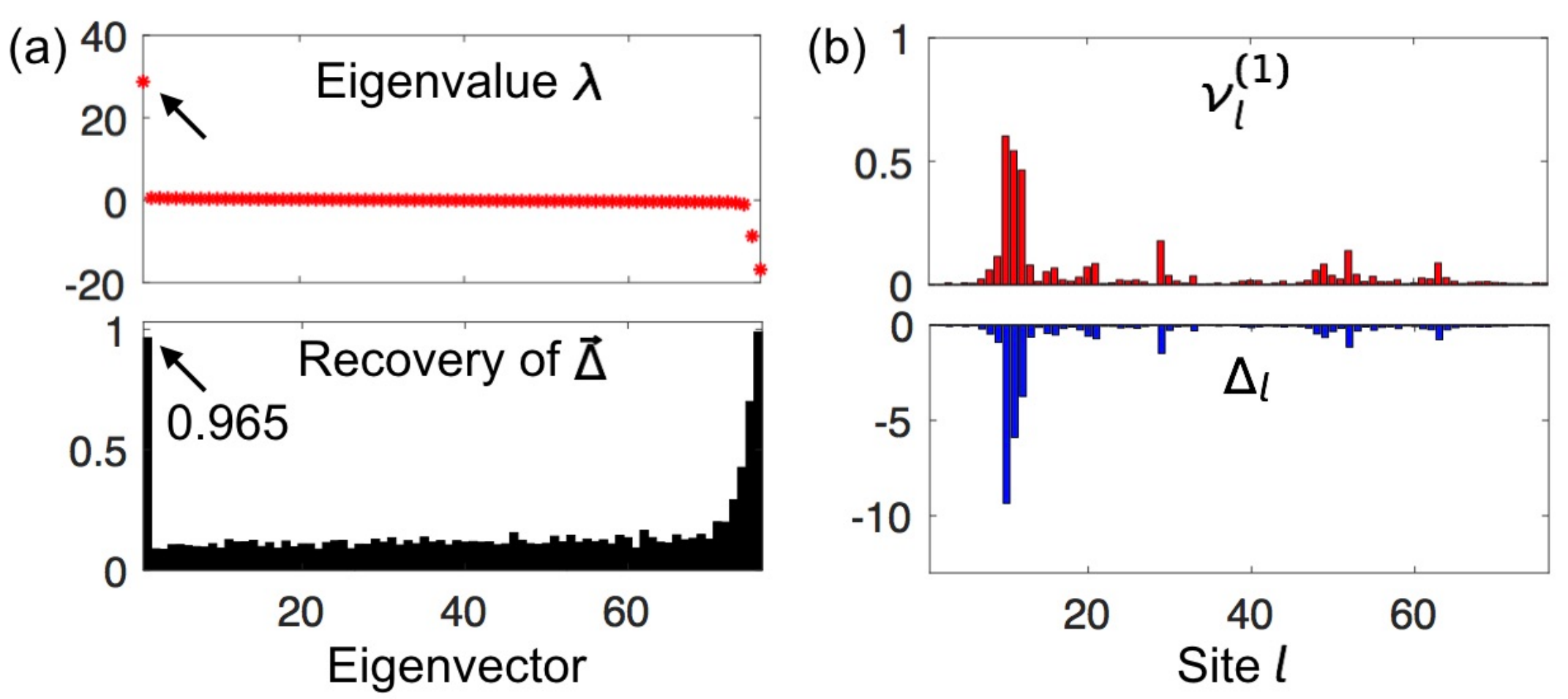}
\vspace{0.2cm}
\caption{{\bf Performance of ICOD for the selected sequence ensemble from Fig.~1.} (a) Eigenvalues for ICOD method (upper) and Recovery of $\vec{\Delta}$ for all eigenvectors (lower). (b) Leading eigenvector $\nu_l^{(1)}$ (upper) and mutational effect $\Delta_l$ at site $l$ (lower, same as in Fig. 1(f)). The excellent performance of ICOD on this unbiased ensemble of sequences supports the general applicability of the ICOD method to both biased and unbiased sequence ensembles.
}
\label{eq:supp-inverse-Fig1}
\end{figure}

\section{Recovery by a random vector}
Here, we calculate the random expectation of the Recovery of the mutational-effect vector $\vec{\Delta}$ by a generic other vector $\vec{\nu}$, in order to establish a null model to which to compare. For a binary sequence, Recovery, as defined in Eq.~\ref{eq:recovery-measure}, can be expressed as
\begin{equation}
\mathrm{Recovery}= \vec{\Delta}'\cdot \vec{\nu'}=\sum_{l=1}^L \Delta'_l\,\nu'_l,
\label{rec_new}
\end{equation}
with $\Delta'_l=|\Delta_l|/\sqrt{\sum_l \Delta_l^2}$ and $\nu'_l=|\nu_l|/\sqrt{\sum_l \nu_l^2}$. As before, $L$ denotes the length of the sequence and hence the number of components of $\vec{\Delta}$ and $\vec{\nu}$. As $\vec{\nu'}$ is a normalized $L$-dimensional vector, its components can be expressed in $L$-dimensional spherical coordinates using $L-1$ angles $\theta_i$:
\begin{align}
\nu'_l&=\left(\prod_{i=1}^{l-1}\sin\theta_i \right)\cos\theta_{l}\,\,\,\,\,\,\,\,\forall l\in\{1,\dots,L-1\}\,,\label{nuprime}\\
\nu'_L&=\prod_{i=1}^{L-1}\sin\theta_i\,,
\label{nuprimeL}
\end{align}
where $\theta_i\in [0,\pi/2]$ for all $i\in\{1,\cdots,L\}$, because all components of $\vec{\nu'}$ are nonnegative. Note that we employ the usual convention that empty products are equal to one: Eq.~\ref{nuprime} yields $\nu'_1=\cos\theta_1$.

The average Recovery for a random vector $\vec{\nu'}$ with an orientation uniformly distributed in the $L$-dimensional sphere reads:
\begin{equation}
\langle  \mathrm{Recovery}\rangle= \frac{\int_\Omega d\Omega\,\, \sum_l \Delta'_l \nu'_l}{\int_\Omega d\Omega}=\frac{ \sum_l \Delta'_l\,I_l}{\int_\Omega d\Omega}\,,
\label{recov_expr_int}
\end{equation} 
where the angular element is $d\Omega=\prod_{i=1}^{L-1}d\theta_{i}\,\sin^{L-i-1}(\theta_i)$, the integration domain is $\Omega=\left[0,\pi/2\right]^{L-1}$, and we have introduced $I_l=\int_\Omega d\Omega\,\nu'_l$. Using Eq.~\ref{nuprime}, we obtain for $1\leq l\leq L-1$
\begin{equation}
I_l=\int_\Omega d\Omega\,\nu'_l=\left(\prod_{i=1}^{l-1}\int_0^{\pi/2}d\theta_i\,\sin^{L-i}(\theta_i)\right)\left(\int_0^{\pi/2}d\theta_l\,\sin^{L-l-1}(\theta_l)\cos(\theta_l)\right)\left(\prod_{i=l+1}^{L-1}\int_0^{\pi/2}d\theta_i\,\sin^{L-i-1}(\theta_i)\right)\,,
\end{equation} 
and similarly, Eq.~\ref{nuprimeL} yields
\begin{equation}
I_L=\int_\Omega d\Omega\,\nu'_L=\prod_{i=1}^{L-1}\int_0^{\pi/2}d\theta_i\,\sin^{L-i}(\theta_i)\,.
\end{equation} 
Using the following results valid for $n>-1$:
\begin{equation}
\int_0^{\pi/2} d\theta\,\sin^n(\theta)=\frac{\sqrt{\pi}}{2}\,\frac{\Gamma \left(\frac{1+n}{2}\right)}{\Gamma\left(\frac{n+2}{2}\right)}\,;\quad \int_0^{\pi/2}d\theta\, \sin^n(\theta) \cos(\theta) =\frac{1}{n+1}\,,
\end{equation}
where $\Gamma$ denotes the Euler Gamma function, which satisfies $\Gamma(x+1)=x\,\Gamma(x)$ for all $x$, we obtain for $1\leq l\leq L$:
\begin{equation}
I_l=\frac{\pi^{(L-1)/2}}{2^{L-1}\,\Gamma\left(\frac{L+1}{2}\right)},
\label{I_l}
\end{equation}
which is independent of $l$. Besides,
\begin{equation}
\int_ \Omega d\Omega=\frac{\pi^{L/2}}{2^{L-1}\,\Gamma(L/2)}.
\label{Om}
\end{equation}
Combining Eq.~\ref{recov_expr_int} with Eqs.~\ref{I_l} and~\ref{Om} finally yields
\begin{equation}
\langle  \mathrm{Recovery}\rangle=\frac{ \sum_l \Delta'_l\,I_l}{\int_\Omega d\Omega}= \frac{\Gamma \left(L/2\right)}{\sqrt{\pi}\,\,\Gamma\left(\frac{L+1}{2}\right)}\sum_l \Delta'_l=\frac{\Gamma \left(L/2\right)}{\sqrt{\pi}\,\,\Gamma\left(\frac{L+1}{2}\right)}\,\frac{\sum_l |\Delta_l|}{\sqrt{\sum_l \Delta_l^2}}.
\label{eq:exact-average-Recovery}
\end{equation}

In particular, in the relevant regime $L\gg 1$, an asymptotic expansion of $\Gamma$ yields: 
\begin{equation}
\langle  \mathrm{Recovery}\rangle \approx  \sqrt{\frac{2}{\pi L}} \,\,\frac{\sum_l |\Delta_l|}{\sqrt{\sum_l \Delta_l^2}}.
\label{eq:random-Recovery}
\end{equation}
The maximum expectation of Recovery is obtained when all components of $\vec{\Delta}$, i.e. all mutational effects, are identical: 
\begin{equation}
\langle  \mathrm{Recovery}\rangle_\textrm{max}= \sqrt{\frac{2}{\pi}}\approx 0.798.
\end{equation}
Conversely, the average Recovery becomes minimal when only one component of $\vec{\Delta}$ is nonzero, which constitutes the limit of the case where the mutational effect at one site is dominant: 
\begin{equation}
\langle  \mathrm{Recovery}\rangle_\textrm{min}= \sqrt{\frac{2}{\pi L}},
\end{equation}
which approaches zero in the limit $L\to \infty$.

 \section{Inverse covariance matrix of our sequence ensembles}
 
 Here, we present a derivation of the small-coupling approximation of the inverse covariance matrix for our artificially-generated sequence ensembles. In this small-coupling limit, the inverse covariance matrix provides an estimate of the energetic couplings used to generate the data. More generally, deducing energetic parameters from observed statistics is a well-known inference problem, also known as an inverse problem. Two-body energetic couplings can be inferred from the one and two-body frequencies observed in the data, using a standard maximum entropy approach. However, the exact calculation of the energetic terms is difficult, and various approximations have been developed. Following Refs~\cite{Marks11, morcos2011direct}, we use the mean-field or small-coupling approximation, which was introduced in Ref.~\cite{Plefka82} for the Ising spin-glass model. For the sake of completeness, we now review the main steps of the calculation, which follow Ref.~\cite{morcos2011direct}. Note that we do not use inference methods specific to low-rank coupling matrices~\cite{Cocco11,cocco2013principal} because we wish to retain generality, with the application to real sequence data in mind.

We begin with the case of binary sequences, which is discussed in the main text. Following that, we generalize to cases where more than two states are allowed at each site, such as the 21 possible states for real protein sequence (20 amino acids plus gap).

\subsection{Binary sequences}

We begin by deriving Eq.~\ref{eq:inverse-PCA-q=2} from the main text, which provides an approximation for the inverse covariance matrix of the ensembles of our binary artificial sequences. Each sequence $\vec{S}$ is such that $S_l\in\{0,1\}$ for each site $l$ with $1\leq l\leq L$, where $L$ is the length of the sequence.

\subsubsection{From a sector model for binary sequences to an Ising model}
\label{Sec_Ising}

Recall the fitness $w$ of a binary sequence $\vec{S}$ under selection for trait $T$ to be close to $T^*$ (Eq.~\ref{eq:fitness}):
\begin{equation}
w(\vec{S})=-\frac{\kappa}{2}\left(T(\vec{S})-T^* \right)^2=-\frac{\kappa}{2}\left(\sum_l \Delta_l S_l-T^* \right)^2\,.
\label{Hini}
\end{equation}
We introduce $s_l=2 S_l-1$: it is an ``Ising spin'' variable ($S_l=0\Leftrightarrow s_l=-1$ and $S_l=1\Leftrightarrow s_l=1$). The fitness in Eq.~\ref{Hini} can be rewritten as
\begin{equation}
w(\vec{s})=-\frac{\kappa}{2}\left(\sum_l D_l s_l-\alpha \right)^2\,,
\label{HIs}
\end{equation}
with $D_l=\Delta_l/2$ and $\alpha=T^*-\sum_l D_l$. Expanding yields
\begin{equation}
w(\vec{s})=-\frac{\kappa}{2}\left(\sum_{l\neq p} D_l D_p s_l s_p +\sum_l D_l^2 - 2\alpha\sum_l D_l s_l +\alpha^2  \right)\,,
\label{HIs2}
\end{equation}
where we have used the fact that $s_l^2=1$. The second term and the last term in Eq.~\ref{HIs2} are both constants, and therefore our fitness is equivalent to
\begin{equation}
w(\vec{s})=-\frac{\kappa}{2}\left(\sum_{l\neq p} D_l D_p s_l s_p - 2\alpha\sum_l D_l s_l \right)\,.
\label{SecToIs}
\end{equation}
This fitness has the form of a standard Ising Hamiltonian with inter-spin couplings and local fields, albeit with the convention difference in overall sign between fitness and energy.

\subsubsection{First-order small-coupling expansion} 

We next consider the general Ising Hamiltonian with inter-spin couplings and local fields
\begin{equation}
H(\vec{s})=-\frac{1}{2}\epsilon \sum_{i\neq j} J_{ij} s_i s_j -\sum_i h_i s_i\,,
\label{HIs_SC}
\end{equation}
where  $\epsilon$ is a constant to be employed in a small-coupling expansion. With this Hamiltonian, taking thermal energy $k_{\rm B}T = 1$, the equilibrium probability of  finding a particular sequence  $\vec{s}$ is
\begin{equation}
P(\vec{s})=\frac{1}{Z}e^{-H\left(\vec{s}\right)},
\end{equation}
where $Z=\sum_{\vec{s}}e^{-H\left(\vec{s}\right)}$.

Introducing $F=-\log Z$, we have
\begin{align}
\frac{\partial F}{\partial h_i}&=-\langle s_i\rangle= -m_i\,,\nonumber\\
\frac{\partial^2 F}{\partial h_i\partial h_j}&=-\frac{\partial m_i}{\partial h_j}=\langle s_i\rangle \langle s_j\rangle-\langle s_i s_j\rangle=-C'_{ij}\,,
\end{align}
where, following the Ising terminology, $m_i$ denotes the average magnetization at site $i$, while $C'$ denotes the covariance matrix in the Ising convention. Note that, using the identity $m_i=2P_i-1$, where $P_i$ denotes the probability that $s_i=1$, we obtain
\begin{equation}
C'_{ij}=\langle s_i s_j\rangle-\langle s_i\rangle \langle s_j\rangle=4\left(P_{ij}-P_i P_j\right)=4\, C_{ij}\,,
\label{Cconv}
\end{equation}
where $P_{ij}$ is the probability that $s_i=s_j=1$, and $C$ denotes the covariance matrix in the Potts convention, which is used in the main text because it allows straightforward generalization to the case where more than two states are possible at each site.

Performing a Legendre transform, we introduce $G=F+\sum_i m_i h_i$, yielding
\begin{align}
\frac{\partial G}{\partial m_i}&=h_i\,,
\label{eq1}\\
\frac{\partial^2 G}{\partial m_i\partial m_j}&=\frac{\partial h_i}{\partial m_j}={C'_{ij}}^{-1}\,.
\label{eq2}
\end{align}

We now perform a small-coupling expansion and express $G$ to first order in $\epsilon$ (see Eq.~\ref{HIs_SC}): $G(\epsilon)\approx G(0)+\epsilon G'(0)$. Since sites are independent for $\epsilon=0$, it is straightforward to express $G(0)$ and $G'(0)$ as a function of the one-body expectations, represented by $m_i$, and of the couplings. We obtain
\begin{equation}
G(0)=\sum_i \frac{m_i+1}{2}\log\left(\frac{m_i+1}{2}\right)+\frac{1-m_i}{2}\log\left(\frac{1-m_i}{2}\right)\,,
\end{equation}
and
\begin{equation}
G'(0)=\frac{\partial G}{\partial \epsilon}(0)=-\frac{1}{2}\sum_{i\neq j} J_{ij} m_i m_j\,.
\end{equation}
Using these expressions, and taking $\epsilon=1$ in the expansion, we obtain the following approximation for $G$:
\begin{equation}
G\approx \sum_i \frac{m_i+1}{2}\log\left(\frac{m_i+1}{2}\right)+\frac{1-m_i}{2}\log\left(\frac{1-m_i}{2}\right)-\frac{1}{2}\sum_{i\neq j} J_{ij} m_i m_j\,.
\label{Gappx}
\end{equation}
Using Eqs.~\ref{eq1} and~\ref{eq2}, we obtain the elements of the inverse  covariance matrix from Eq.~\ref{Gappx}:
\begin{align}
{C'_{kl}}^{-1}&=-J_{kl}\,,\,\,\forall l\neq k\,,\nonumber\\
{C'_{ll}}^{-1}&= \frac{1}{2}\left(\frac{1}{1+m_l}+\frac{1}{1-m_l}\right)= \frac{1}{4}\left(\frac{1}{P_l}+\frac{1}{1-P_l}\right)\,,
\label{Corr_Ising}
\end{align}
where $P_l$ denotes the probability that $s_l=1$.

Note that Eq.~\ref{Gappx} is a first-order small-coupling (or mean-field) approximation. The expansion can be extended to higher order, and the second-order expansion is known as the Thouless, Anderson, and Palmer (TAP) free energy~\cite{Thouless77,Plefka82}. 

\subsubsection{Application to our sector model}
Comparing Eqs.~\ref{SecToIs} and~\ref{HIs_SC} (with $\epsilon=1$) allows us to identify the couplings in our sector model as
\begin{equation}
J_{kl}=-\kappa\, D_k D_l=- \kappa\, \Delta_k \Delta_l /4\,,\,\,\forall k\neq l\,.
\end{equation}
Note that this expression is in the Ising gauge (also known as the zero-sum gauge). Recall also that the link to the Potts convention is made through $C'=4\,C$ (Eq.~\ref{Cconv}), which implies ${C'}^{-1}=C^{-1}/4$. Finally, recall that fitness and energy have opposite signs.

Hence, in the Potts convention, Eq.~\ref{Corr_Ising} yields for our sector model:
\begin{align}
C_{kl}^{-1}&=\kappa \Delta_k \Delta_l\,,\,\,\forall l\neq k\,,\nonumber\\
C_{ll}^{-1}&=\frac{1}{P_l}+\frac{1}{1-P_l}\,. \label{eqn2}
\end{align}
This corresponds to Eq.~\ref{eq:inverse-PCA-q=2} in the main text.

\subsection{Sequences with $q$ possible states at each site}
\label{Sec:q-state}

\subsubsection{From a sector model to a Potts model for sequences}

Motivated by the fact that a real protein sequence has 21 possible states at each site (20 different amino acids plus gap), we now generalize the above result to the case where $q$ states are possible at each site. We denote these states by $\alpha$ with $\alpha\in\{1,..,q\}$. Our sector model can then be mapped to a $q$-state Potts model. The length-$L$ vector $\vec{\Delta}$ of single-site mutational effects introduced in the two-state case in the main text is replaced by a $(q-1)\times L$ matrix of mutational effects, each being denoted by $\Delta_l(\alpha_l)$. These mutational effects can be measured with respect to a reference sequence $\vec{\alpha}^0$ satisfying $\Delta_l(\alpha_l^0)=0,\,\,\forall l\in\{1,\dots,L\}$: at each site $l$, the state present in the reference sequence $\vec{\alpha}^0$ serves as the reference with respect to which the mutational effects at that site are measured. For the sake of simplicity, we will take state $q$ as reference state at all sites. This does not lead to any loss of generality, since it is possible to reorder the states for each $l$.

The generalization of the fitness function Eq.~\ref{eq:fitness} (Eq.~\ref{Hini}) to our $q$-state model can be written as
\begin{equation}
w(\vec{\alpha})=-\frac{\kappa}{2}\left(T(\vec{\alpha})-T^*\right)^2=-\frac{\kappa}{2}\left(\sum_{l=1}^L\Delta_l(\alpha_l)-T^*\right)^2\,, 
\label{Ham0}
\end{equation}
 (see Eq.~\ref{eq:quadr} in the main text). Expanding this expression, discarding a constant term, and using the fact that there can only be one state at each site, we find that the fitness of sequences can be expressed as
\begin{equation}
w(\vec{\alpha})=-\frac{\kappa}{2}\sum_{l\neq k}\Delta_l(\alpha_l)\Delta_k(\alpha_k)
-\frac{\kappa}{2}\sum_{l=1}^L\Delta_l(\alpha_l)\left(\Delta_l(\alpha_l)-2\,T^*\right)\,.
\label{Ham2}
\end{equation}
This is a particular case of the more general Potts Hamiltonian
\begin{equation}
H(\vec{\alpha})=-\frac{1}{2}\sum_{l\neq k} e_{lk}(\alpha_l,\alpha_k)-\sum_{l=1}^{L}h_{l}(\alpha_l)\,,
\label{Potts1}
\end{equation}
which is the one usually considered in Direct Coupling Analysis (DCA)~\cite{morcos2011direct,Marks11}.

In order to identify Eq.~\ref{Ham2} and Eq.~\ref{Potts1}, one must deal with the degeneracies present in Eq.~\ref{Potts1}, where the number of independent parameters is $L(q-1)+L(L-1) (q-1)^2/2$~\cite{Ekeberg14}. To lift this degeneracy, we choose the gauge usually taken in mean-field DCA~\cite{morcos2011direct}: $e_{lk}(\alpha_l,q)=e_{lk}(q,\alpha_k)=h_l(q)=0$ for all $l, k, \alpha_l, \alpha_k$. This choice is consistent with taking state $q$ as the reference state for mutational effects (see above), and we will refer to it as the reference-sequence gauge. This gauge choice enables us to identify the couplings between Eq.~\ref{Ham2} and Eq.~\ref{Potts1}:
\begin{align}
e_{lk}(\alpha_l,\alpha_k)&=-\kappa\Delta_l(\alpha_l)\Delta_k(\alpha_k)\,,
\label{identify}
\end{align}
 for all $l \neq k$, and all $\alpha_l, \alpha_k$, with $\Delta_l(q)=0$ for all $l$ (recalling that fitness and energy have opposite signs). 
 
 \subsubsection{First-order small-coupling expansion} 

The derivation of the first-order mean-field or small-coupling approximation for $q$-state models is very similar to the Ising case presented above. Hence, we will simply review the main results (see Ref.~\cite{morcos2011direct}).

We start with the Hamiltonian
\begin{equation}
H(\vec{\alpha})=-\frac{\epsilon}{2}\sum_{l\neq k} e_{lk}(\alpha_l,\alpha_k)-\sum_{l=1}^{L}h_{l}(\alpha_l)\,,
\label{Potts1b}
\end{equation}
where $\epsilon$ has been introduced to perform the small-coupling expansion. Eq.~\ref{Potts1b} coincides with Eq.~\ref{Potts1} for $\epsilon=1$. Considering $F=-\log(Z)$ with $Z=\sum_{\vec{\alpha}}e^{-H\left(\vec{\alpha}\right)}$, where $H\left(\vec{\alpha}\right)$ is the Potts Hamiltonian in Eq.~\ref{Potts1b}, we have for all $k$ and all $\alpha_k<q$:
\begin{equation}
\frac{\partial F}{\partial h_k(\alpha_k)}=-P_k(\alpha_k)\,,
\end{equation}
where $P_k(\alpha_k)$ is the one-body probability. Similarly, we have for all $k, l$ and all $\alpha_k<q$ and $\alpha_l<q$:
\begin{equation}
\frac{\partial^2 F}{\partial h_l(\alpha_l)\partial h_k(\alpha_k)}=-\frac{\partial P_k(\alpha_k)}{\partial h_l(\alpha_l)}=-C_{kl}(\alpha_k,\alpha_l)\,,
\label{Fseconde}
\end{equation}
where we have introduced the covariance $C_{kl}(\alpha_k,\alpha_l)=P_{kl}(\alpha_k,\alpha_l)-P_k(\alpha_k)P_l(\alpha_l)$. 

We perform a Legendre transform and introduce $G=F-\sum_i\sum_{\alpha_i}h_i(\alpha_i)P_i(\alpha_i)$, yielding
\begin{align}
\frac{\partial G}{\partial P_k(\alpha_k)}&=h_k(\alpha_k)\,,\label{field}\\
\frac{\partial^2 G}{\partial P_l(\alpha_l)\partial P_k(\alpha_k)}&=\frac{\partial h_l(\alpha_l)}{\partial P_k(\alpha_k)}=C^{-1}_{kl}(\alpha_k,\alpha_l)\,,
\label{invC}
\end{align}
for all $k, l$ and all $\alpha_k<q$ and $\alpha_l<q$. Note that, in the latter equation, $C^{-1}_{kl}(\alpha,\beta)$ is shorthand for $A^{-1}_{ij}$, where $A$ is the $(q-1)L\times (q-1)L$  covariance matrix where terms involving the reference state $q$ have been excluded: $A_{ij}=C_{kl}(\alpha,\beta)$, where $i=(q-1)(k-1)+\alpha$ and $j=(q-1)(l-1)+\beta$, with $\alpha\in\{1,\dots,q-1\}$ and $\beta\in\{1,\dots,q-1\}$~\cite{Ekeberg13}.

We next perform a first-order expansion of $G$ in $\epsilon$, and take $\epsilon=1$, yielding:
\begin{equation}
G\approx \sum_l\sum_{\alpha_l}P_l(\alpha_l)\log\left(P_l(\alpha_l)\right)-\frac{1}{2}\sum_{l\neq k}\sum_{\alpha_l,\alpha_k}e_{lk}(\alpha_l,\alpha_k)P_l(\alpha_l)P_k(\alpha_k)\,.
\label{G1}
\end{equation}
Applying Eqs.~\ref{field},~\ref{invC} to Eq.~\ref{G1}, and using $P_l(q)=1-\sum_{\alpha_l<q}P_l(\alpha_l)$ gives
\begin{align}
C^{-1}_{kl}(\alpha_k,\alpha_l)&=-e_{kl}(\alpha_k,\alpha_l),\,\,\,\forall \,l\neq k\,,\nonumber\\
C^{-1}_{ll}(\alpha_k,\alpha_l)&=\frac{1}{P_k(q)}+\frac{\delta_{\alpha_k\alpha_l}}{P_k(\alpha_k)}\,.
\label{mf_coupl}
\end{align}
This result is the standard one found in DCA ~\cite{morcos2011direct}.

\subsubsection{Application to our sector model}

Combining Eqs.~\ref{identify} and \ref{mf_coupl}, we obtain for our sector model:
\begin{align}
C^{-1}_{kl}(\alpha_k,\alpha_l)&=\kappa \Delta_{k}(\alpha_k)\Delta_l(\alpha_l),\,\,\,\forall \,l\neq k\,,\nonumber\\
C^{-1}_{ll}(\alpha_k,\alpha_l)&=\frac{1}{P_k(q)}+\frac{\delta_{\alpha_k\alpha_l}}{P_k(\alpha_k)}\,.
\label{invcorrPotts}
\end{align}
For $q=2$,  Eq.~\ref{invcorrPotts} reduces to Eq.~\ref{Corr_Ising} (Eq.~\ref{eq:inverse-PCA-q=2} in the main text), using $1-P_l=P_{l}(q)$. 

\subsubsection{Selection on multiple traits}

So far, we have mainly discussed the case where there selection on only one trait (yielding one sector). However, real proteins face various selection pressures. The generalization of the fitness in Eq.~\ref{Ham0} to $N$ simultaneous selection on different traits reads
 \begin{equation}
 w(\vec{S})=-\sum_{i=1}^N \frac{\kappa_i}{2} \left(T_i-T^*_i\right)^2=-\sum_{i=1}^N \frac{\kappa_i}{2} \left(\sum_{l=1}^L \Delta_{i,l}(\alpha_l)-T^*_i\right)^2\,,
 \label{eq:landscape-1}
 \end{equation}
which corresponds to Eq.~\ref{eq:newfitness} in the main text. We choose the reference-state gauge, assuming again for simplicity that the reference state is $q$ at each site. The identification to the general Potts Hamiltonian Eq.~\ref{Potts1} (recalling that fitnesses and energies have opposite signs) then yields
 \begin{equation}
e_{lk}(\alpha_l,\alpha_k)=-\sum_{i=1}^N \kappa_i \Delta_{i,l}(\alpha_l)\Delta_{i,k}(\alpha_k)\,,
 \end{equation}
which generalizes Eq.~\ref{identify} to the multiple selection case. Using the small-coupling expansion result in Eq.~\ref{mf_coupl}, we obtain the following approximation for the inverse covariance matrix:
 \begin{align}
 C^{-1}_{kl}(\alpha_k,\alpha_l)&=\sum_{i=1}^N \kappa_i \Delta_{i,k}(\alpha_k)\Delta_{i,l}(\alpha_l),\,\,\,\forall \,l\neq k\,,\nonumber\\
 C^{-1}_{ll}(\alpha_k,\alpha_l)&=\frac{1}{P_k(q)}+\frac{\delta_{\alpha_k\alpha_l}}{P_k(\alpha_k)}\,.
 \label{invcorrPottsMulti}
 \end{align}
This generalizes Eq.~\ref{invcorrPotts} to the case of simultaneous selection on multiple traits.

\section{Robustness of functional sectors and of ICOD}

In the main text, we introduced the Inverse Covariance Off-Diagonal (ICOD) method to identify protein sectors from sequence data. The ICOD method exploits the approximate expression derived above for the inverse covariance matrix (Eq.~\ref{invcorrPotts}); in particular, ICOD makes use of the fact that the off-diagonal elements of $C^{-1}$ are simply related to the elements of the mutational effect vector $\vec\Delta$. In this section, we first describe our comparison of ICOD to SCA for single selection, and detail our test of ICOD for double selection, using synthetic binary sequences.  Next, we confirm the robustness of the ICOD method to different forms of selection and then show how ICOD can be extended to sequences with more than two states per site, and finally demonstrate its robustness to gauge choice and pseudocounts. 

\subsection{Robustness of ICOD to selection bias, selection strength, and multiple selections}

To quantify the performance of ICOD and to compare to SCA over a range of selection biases we focused on binary sequences. To obtain the average curve for single selections in Fig.~3(a), we first generated 100 distinct synthetic $\vec{\Delta}$s, one for each sector size from $n=1$ to 100, where sector sites are defined as those with large mutational effects. To this end, the mutational effects of the sector sites and the non-sector sites were sampled, respectively, from zero-mean Gaussian distributions with standard deviations 20 and 1. For each sector size and each selection bias we generated a sequence ensemble of 50,000 random sequences and weighted each sequence according to the distribution 
\begin{equation}
P(\vec{S})=\frac{\exp(w(\vec{S}))}{ \sum_{\vec{S}} \exp(w(\vec{S}))}\,,
\label{eq:distrb}
\end{equation}
where $w(\vec{S})$ is the fitness of sequence $\vec{S}$, given by the single selection formula Eq.~\ref{eq:fitness}. In general, we wish to employ a selection window whose width in energy (or any other selected variable) scales with the overall width of the unselected distribution. Hence, as mentioned in the main text, we perform all selections with a strength
\begin{equation}
\kappa=\frac{10}{\sum_{l}\Delta_l^2}\,.
\label{eq:kappa}
\end{equation}
Then, for each method (ICOD or SCA), performance as measured by Recovery of $\vec{\Delta}$ by the first eigenvector was averaged over the 100 different sector sizes.

\begin{figure}[htb]
	\centering
	\includegraphics[width=5cm]{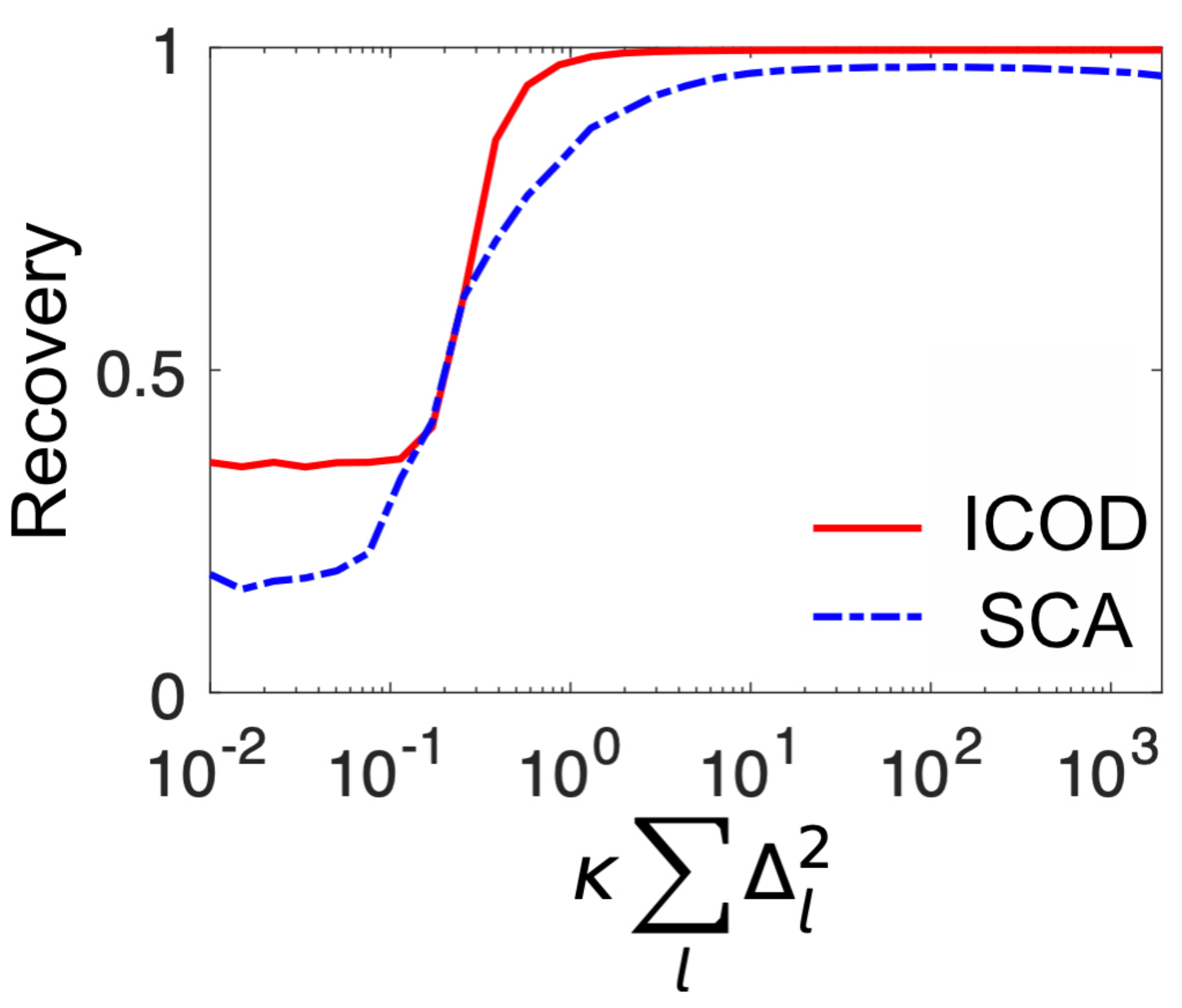}
	\vspace{0.2cm}
	\caption{ {\bf Impact of selection strength $\kappa$ on the performance of ICOD and SCA on synthetic data.}  Results obtained on binary synthetic sequences with $L=100$, selected using a synthetic $\vec{\Delta}$ where the first 20 and the other 80 mutational effects are, respectively, sampled from Gaussian distributions with variances of 20 and 1. Selection is performed on ensembles of 50,000 random sequences, and each data point is obtained by averaging over 100 realizations. The relative bias is $\gamma=0.5$.  }
	\label{fig:Supp_effect_kappa}
\end{figure}

 As an aside, Fig.~\ref{fig:Supp_effect_kappa} demonstrates that the performance of ICOD and SCA is robust to varying selection strength $\kappa$, as long as $\kappa \sum_l\Delta_l^2\gg 1$. (A small value of $\kappa \sum_l\Delta_l^2$ implies weak selection, where most random sequences pass selection and the resulting ensemble does not significantly reflect the constraint.)

\begin{figure}[htb]
\centering
\includegraphics[width=9cm]{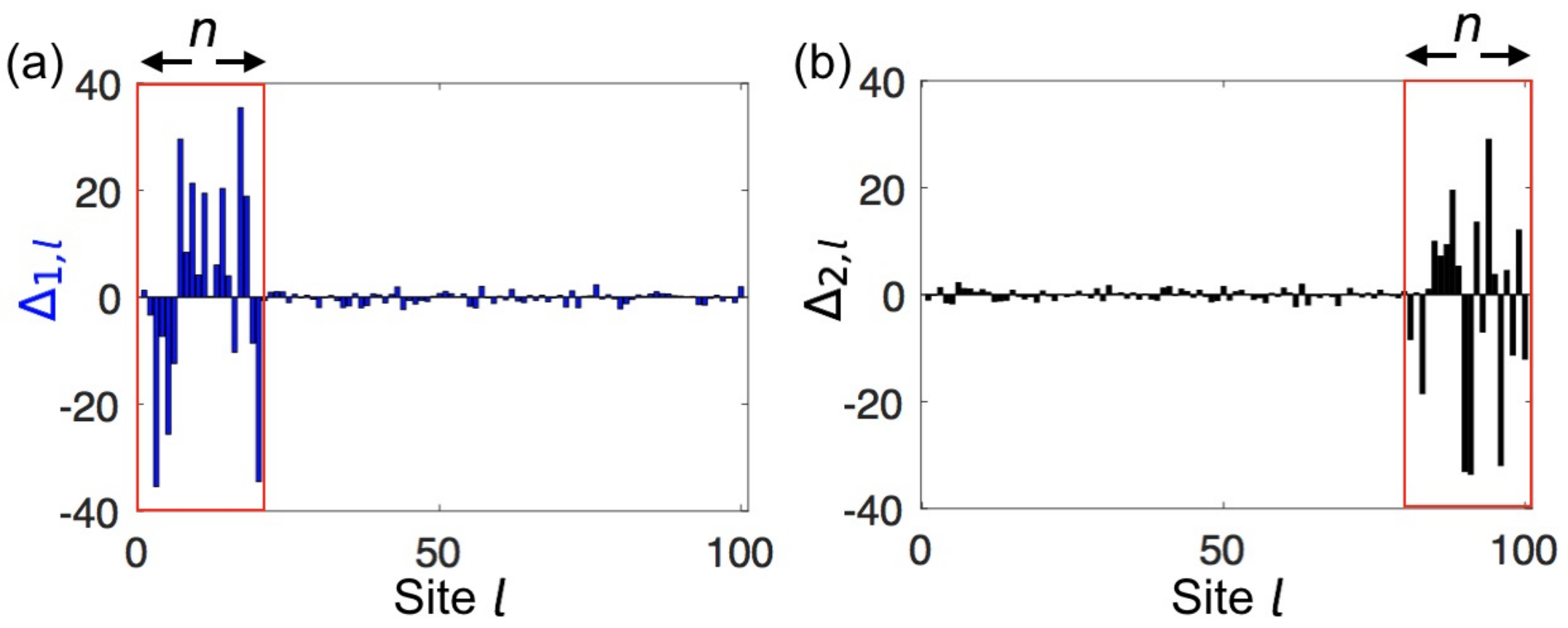}
\vspace{0.2cm}
\caption{{\bf Example of two synthetic $\vec{\Delta}$s generated for the double selection in Fig.~3(b).} (a) Generation of $\vec{\Delta}_1$, where the mutational effects for the first $20$ sites and for the last $80$ sites are sampled, respectively, from zero-mean Gaussian distributions with a standard deviation of 20 and 1.  (b) Generation of  $\vec{\Delta}_2$, where the mutational effects for the last $20$ sites and for the first $80$ sites are sampled, respectively, from zero-mean Gaussian distributions with a standard deviation of 20 and 1.   
}
\label{fig:synthetic_delta_explain}
\end{figure}

Similarly, to obtain the average curve for double selection in Fig.~3(b), we generated 100 distinct pairs of $\vec{\Delta}_1$s and $\vec{\Delta}_2$s, one pair for each sector size from $n=1$ to 100. Specifically, the sector for $\vec{\Delta}_1$ consisted of the first $n$ sites, while the sector for $\vec{\Delta}_2$ corresponded to the last $n$ sites, so that the two sectors overlap for  $n>50$. As for the single selections, the mutational effects of the sector sites and the non-sector sites were sampled, respectively, from Gaussian distributions with standard deviations 20 and  1.  As an example, two synthetic $\vec{\Delta}$s for $n=20$ are shown in Fig.~\ref{fig:synthetic_delta_explain}. Again, for each sector size and each selection bias, we generated an ensemble of 50,000 random sequences and weighted them according to Eq.~\ref{eq:distrb} along with the double selection formula Eq.~\ref{eq:landscape-1} (i.e. Eq.~\ref{eq:newfitness} in the main text). The performance of ICOD as measured by Recovery of $\vec{\Delta}_1$ and $\vec{\Delta}_2$ by the first two eigenvectors was averaged over the 100 different sector sizes.  In Fig.~3(b) we also reported the performance of ICOD for two non-overlapping sectors, each with 20 sites, and for two fully overlapping sectors, each with 100 sites. We followed a protocol similar to that described above, but in each of these cases, we averaged Recovery over 100 realizations using distinct pairs of $\vec{\Delta}_1$ and $\vec{\Delta}_2$.

 Unless otherwise stated, data for other plots were generated in the same way, i.e. using 50,000 random sequences, sequence length $L=100$,  selection strength $\kappa$ in Eq.~\ref{eq:kappa}, and standard deviation 20/1 of $\Delta_l$ in the sector/non-sector sites.  

\begin{figure}[htb]
		\centering
		\includegraphics[width=8.5cm]{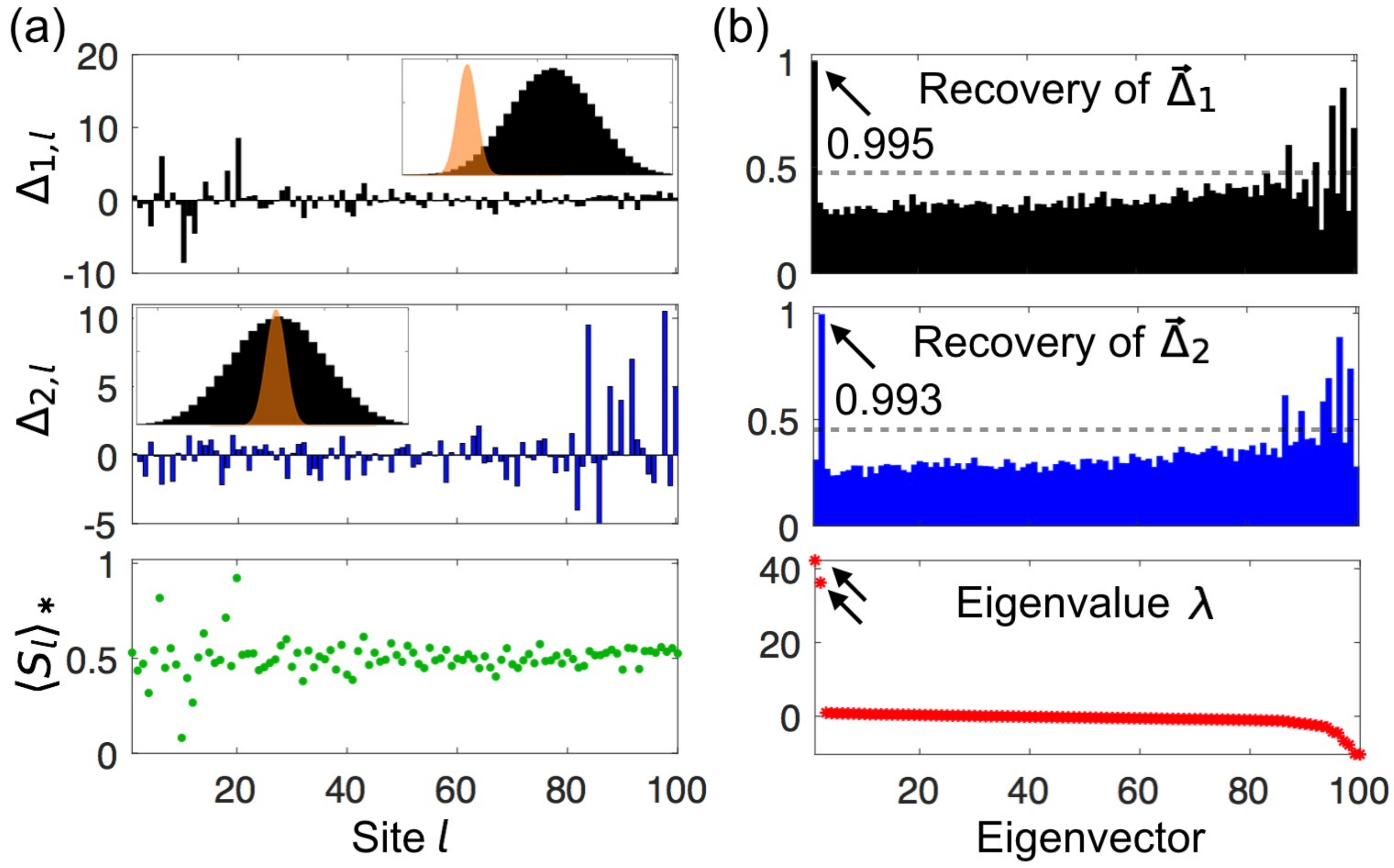}
		\vspace{0.2cm}
		\caption{{\bf ICOD method for simultaneous selection on two traits.}  (a)  Upper panels: Components at each site $l$ of two synthetically generated mutational-effect vectors, with insets showing biased selection around $T_1^*$ for $\vec{\Delta}_1$ and neutral selection around $T_2^*$ for $\vec{\Delta}_2$. Lower panel: average mutant fraction $\langle S_l\rangle_*$ at site $l$ after selection on both traits.  (b)  Performance of ICOD method.   Recovery of $\vec{\Delta}_1$ and $\vec{\Delta}_2$ for all eigenvectors (upper) and corresponding eigenvalues (lower). The gray dashed line indicates the random expectation of Recovery (Eq.~\ref{eq:random-Recovery}).  
		} 
		\label{fig:two-sector-fig3}
	\end{figure}

\begin{figure}[htb]
		\centering
		\includegraphics[width=8.5cm]{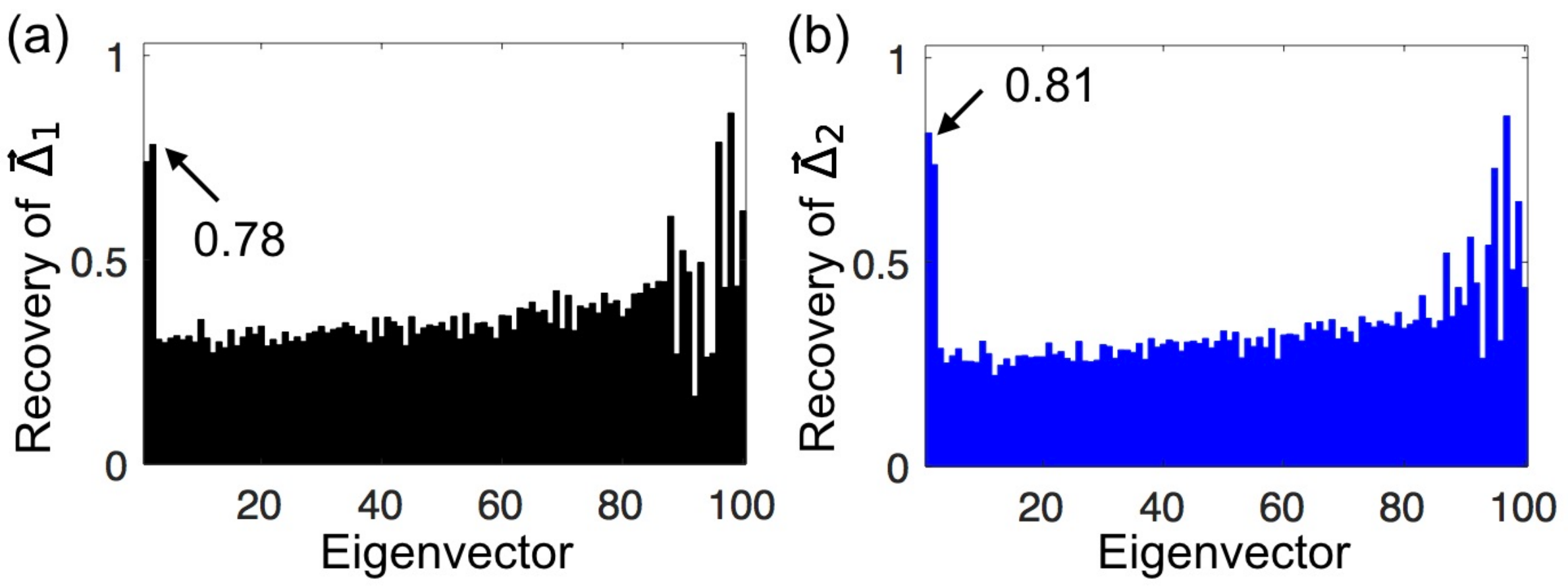}
		\vspace{0.2cm}
		\caption{{\bf Performance of ICOD for the two-sector case in Fig.~\ref{fig:two-sector-fig3}, without applying ICA.} }
		\label{fig:beforeICA-inverse}
	\end{figure}

Note that to improve Recovery in the case of double selection, we applied Independent Component Analysis (ICA)~\cite{Hyvarinen, Hansen01, Rivoire16} to the first two eigenvectors in order to disentangle the contributions coming from the two constraints. In general, we expect that the first $N$ eigenvectors of the ICOD matrix $\tilde{C}^{-1}$ will report $N$ constraints. However, each of these $N$ eigenvectors is likely to include a mixture of contributions from different constraints.  Applying ICA to the first $N$ eigenvectors to recover the individual constraints amounts to assuming that all the constraints are statistically independent. As an example, in Fig.~\ref{fig:two-sector-fig3}, we consider the case of two selections targeting a different set of sites and with different selection windows (one biased, one non-biased). In this case, ICOD plus ICA yields excellent Recovery (Fig.~\ref{fig:two-sector-fig3}). Without ICA, the results are noticeably worse (Fig.~\ref{fig:beforeICA-inverse}). Moreover, Fig.~3(b) demonstrates that   ICOD plus ICA can achieve a high Recovery for a broad range of overlaps between two sectors. 
	
\begin{figure}[htb]
\centering
\includegraphics[width=8.5cm]{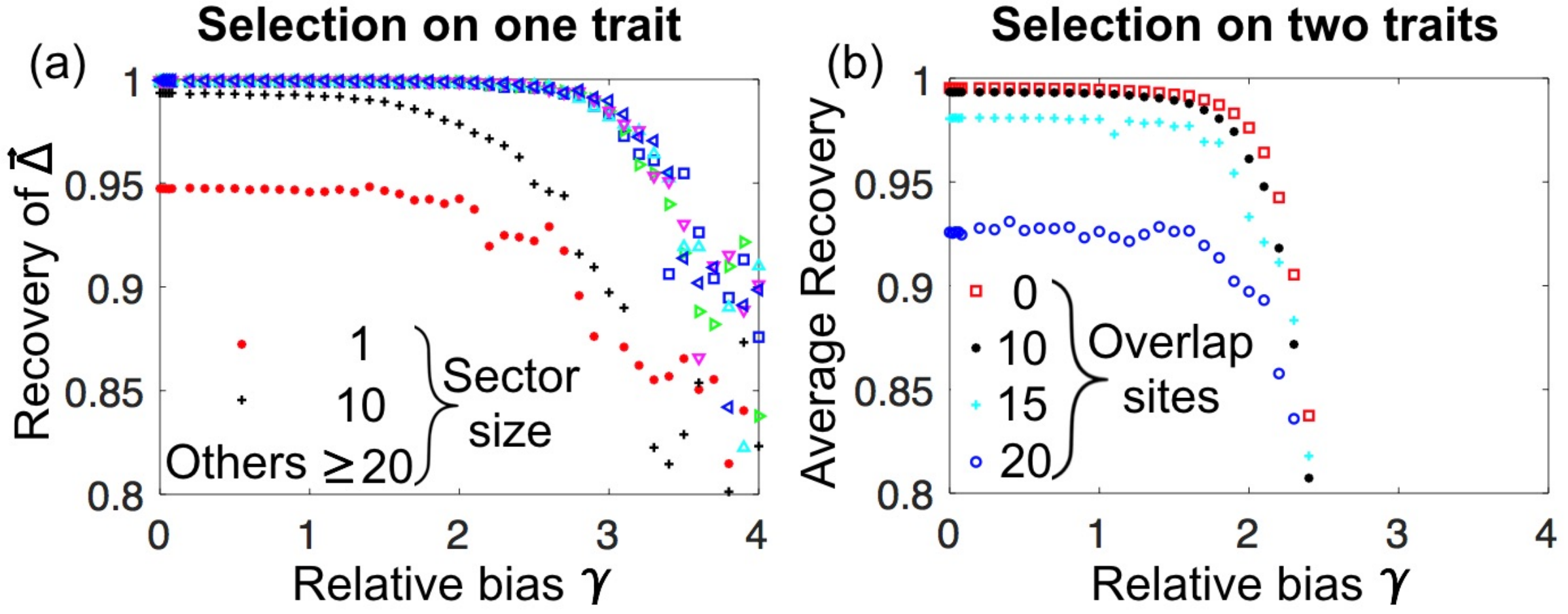}
\vspace{0.2cm}
\caption{ {\bf Performance of ICOD for different  sector sizes and sector overlaps.} (a)  Selection on a single trait with varying sector size. Recovery is shown as a function of relative selection bias $\gamma\equiv (T^*-\langle T\rangle)/\sqrt{\langle (T-\langle T\rangle)^2 \rangle}$ for sectors of size 1, 10, 20, 40, 60, 80, and 100 out of 100 sequence sites ({\it cf.} Fig.~3(a)).  Recovery is almost perfect for sectors of size larger than 10, but is substantially lower for sector size 1, which violates the criteria $\Delta_l\ll \sqrt{\sum_{l'}\Delta_{l'}^2}$. (b) Simultaneous selection on two traits with different degrees of sector overlap.  For each selection, the sector size is 20 out of 100 sequence sites, and the overlap varies from 0 to 20 sites.  The average Recovery for $\vec{\Delta}_1$ and $\vec{\Delta}_2$  is shown as a function of relative selection bias. The data in (b) is averaged over 20 realizations of $\vec \Delta$s.    
}
\label{fig:Supp_overlap_sites}
\end{figure}

In Fig.~3(b), one observes a slight decrease of performance of ICOD plus ICA for double selection with overlapping sectors. Does this arise from  increasing sector size or  from increasing overlap?   As expected from Eqs.~\ref{eq:recovery-measure} and~\ref{eq:inverse-PCA-q=2}, Fig.~\ref{fig:Supp_overlap_sites}(a) shows that Recovery does not fall off with increased sector size. Thus, we tested whether larger sector overlaps could reduce Recovery. Fig.~\ref{fig:Supp_overlap_sites}(b) shows that this is indeed the case for sequence ensembles subject to two selections each with a fixed sector size of 20, but with different numbers of overlapping sites.   However, the reduction of Recovery is quite modest, as even for 100\% overlap, Recovery remains above 0.9. It is interesting to note that, independent of sector size and overlap, Recovery decreases faster for double selection than for single selection at large relative biases (see Figs.~3 and~\ref{fig:Supp_overlap_sites}).

\subsection{Robustness of functional sectors to different forms of selection}

To assess the robustness of physical sectors to forms of selection other than the simple Gaussian selection window of Eqs.~\ref{eq:distr}-\ref{eq:quadr}, we generated ensembles of 50,000 random binary sequences as above, and used synthetically generated mutational effects, with 20 sector sites out of $L=100$ total sites. As before, the mutational effects of the sector sites and the non-sector sites were sampled, respectively, from zero-mean Gaussian distributions with standard deviations 20 and 1. 

We first addressed selection for sequences with an additive trait $T$ above a threshold $T_t$. We thus considered the selected ensembles of sequences such that the value of the trait $T(\vec{S})=\vec{S}\cdot\vec{\Delta}$ is larger than a threshold $T_t$, and we varied this threshold. Fig.~\ref{Fig4} in the main text demonstrates that the corresponding sectors are identified by ICOD just as well as those resulting from our initial Gaussian selection window. 

\begin{figure}[h!]
\centering
\includegraphics[width=8.5cm]{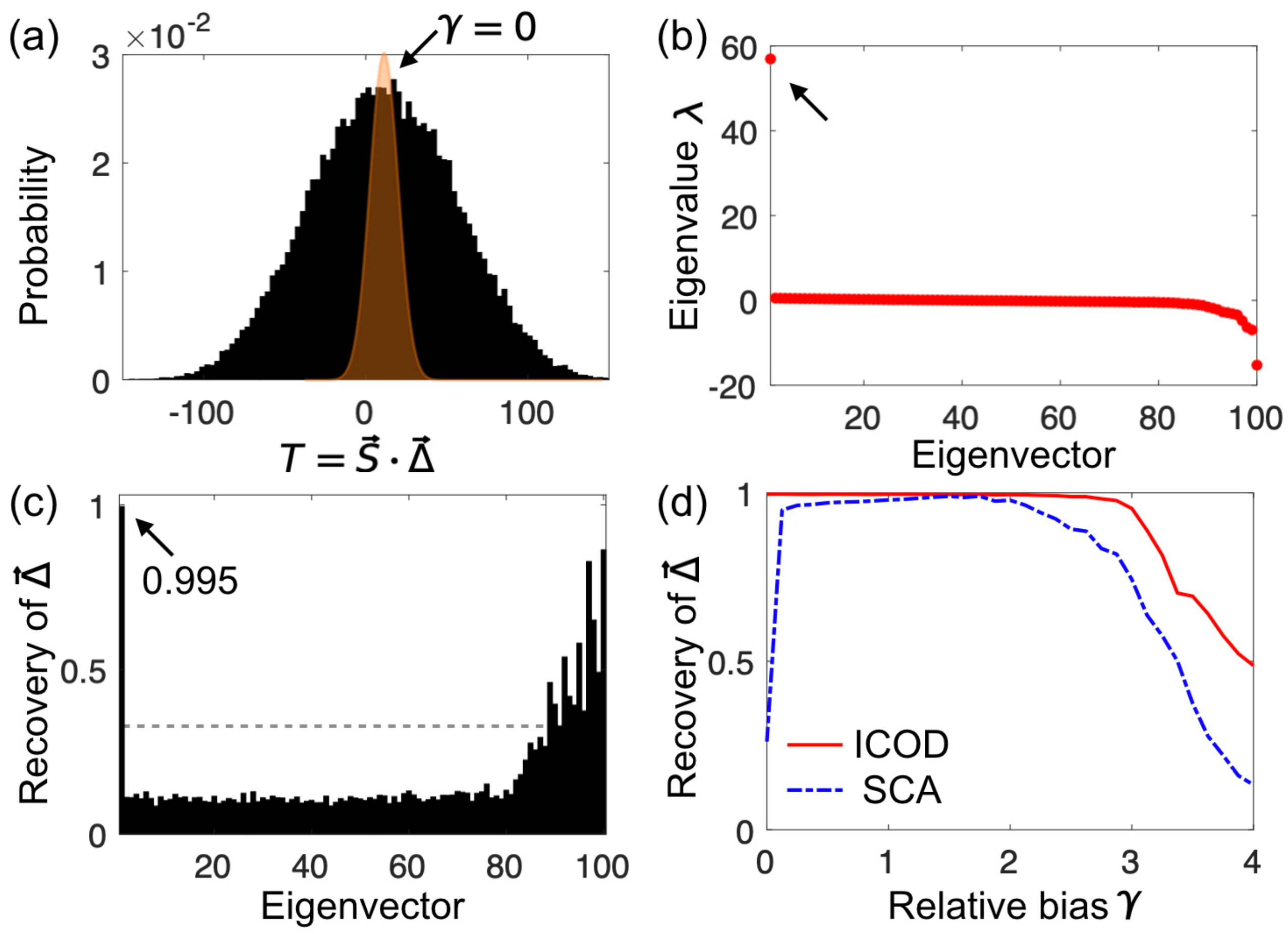}
\vspace{0.2cm}
\caption{{\bf  Identification of sectors that result from quartic selection. } (a)  Histogram of the additive trait $T(\vec{S})=\vec{S}\cdot\vec{\Delta}$ for randomly sampled sequences where 0 and 1 are equally likely at each site. Sequence length is $L=100$, mutational effects are synthetically generated with 20 sector sites. Sequences are selectively weighted using a quartic window (orange) around $T^*$. Selection is shown for $T^*=\langle T\rangle$, or equivalently $\gamma=0$, in terms of the relative selection bias $\gamma\equiv (T^*-\langle T\rangle)/\sqrt{\langle (T-\langle T\rangle)^2 \rangle}$.  (b) Eigenvalues of the ICOD-modified inverse covariance matrix $\tilde{C}^{-1}$ (Eq.~\ref{eq:ICODmatrix}) of the selected sequences for $\gamma=0$. (c) Recovery of $\vec{\Delta}$ for all eigenvectors of $\tilde{C}^{-1}$ for $\gamma=0$. Gray dashed line: random expectation of Recovery. (d) Recovery of $\vec{\Delta}$ for ICOD and for SCA as functions of the relative selection bias $\gamma$. The data in (d) is averaged over 100 realizations of $\vec \Delta$.   
}
\label{fig:quartic_selection}
\end{figure}

\begin{figure}[h!]
\centering
\includegraphics[width=8.5cm]{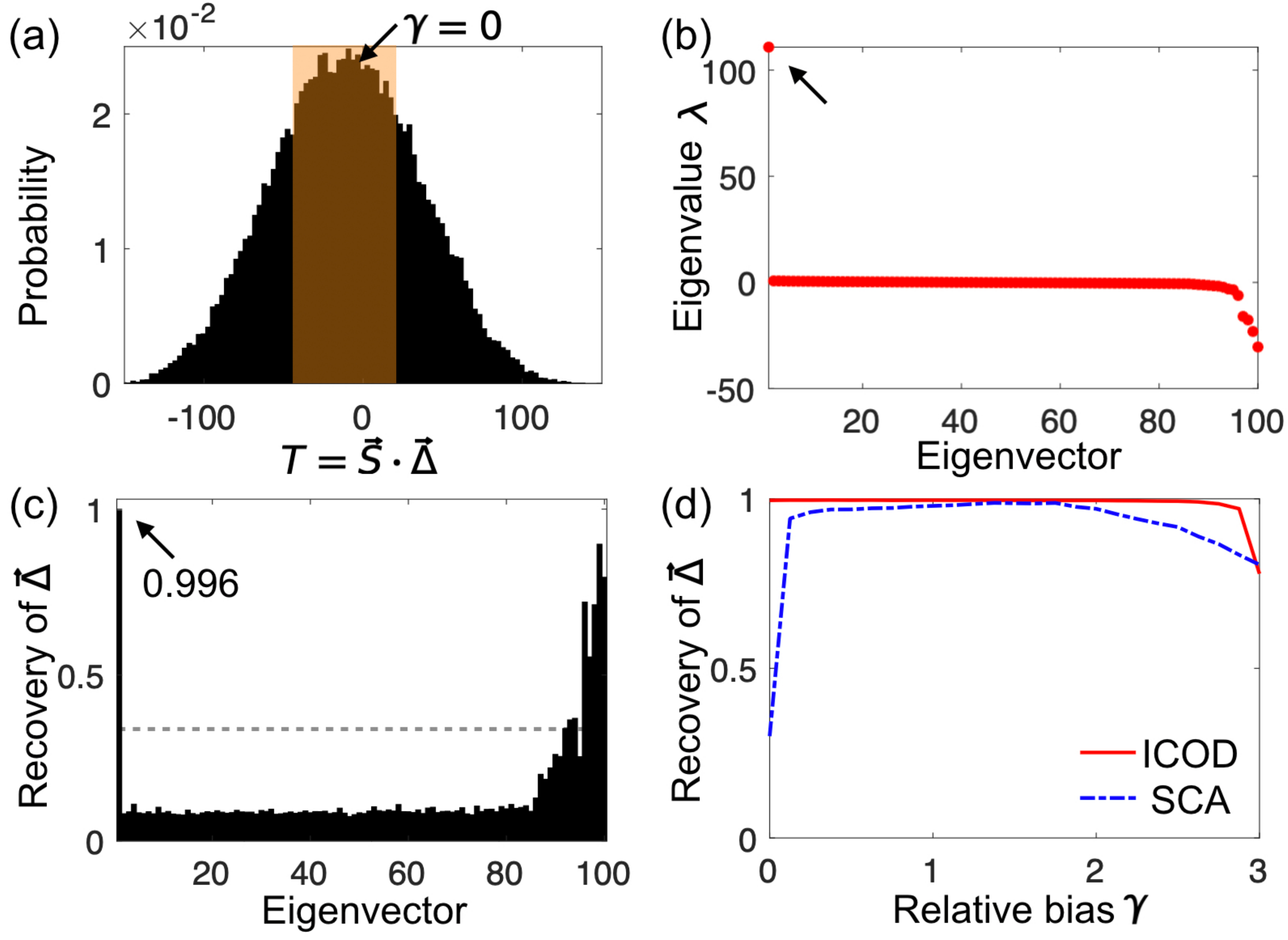}
\vspace{0.2cm}
\caption{{\bf  Identification of sectors that result from rectangular-window selection. } (a)  Histogram of the additive trait $T(\vec{S})=\vec{S}\cdot\vec{\Delta}$ for randomly sampled sequences where 0 and 1 are equally likely at each site. Sequence length is $L=100$, mutational effects are synthetically generated with 20 sector sites. Sequences are selected if they have a trait value $T^*-\eta/2<T(\vec{S})<T^*+\eta/2$ (orange shaded region). Selection is shown for $T^*=\langle T\rangle$, or equivalently $\gamma=0$, in terms of the relative selection bias $\gamma\equiv (T^*-\langle T\rangle)/\sqrt{\langle (T-\langle T\rangle)^2 \rangle}$.  (b) Eigenvalues of the ICOD-modified inverse covariance matrix $\tilde{C}^{-1}$ (Eq.~\ref{eq:ICODmatrix}) of the selected sequences for $\gamma=0$. (c) Recovery of $\vec{\Delta}$ for all eigenvectors of $\tilde{C}^{-1}$ for $\gamma=0$. Gray dashed line: random expectation of Recovery. (d) Recovery of $\vec{\Delta}$ for ICOD and for SCA as functions of the relative selection threshold $\gamma$. The data in (d) is averaged over 100 realizations of $\vec \Delta$.  
}
\label{fig:square_selection}
\end{figure}

We also successfully applied ICOD to various other forms of selection. In Fig.~\ref{fig:quartic_selection}, we used the quartic fitness function:
\begin{equation}
w(\vec{S})=-\frac{\kappa_1}{4}\left(\sum_{l=1}^L \Delta_lS_l-T^* \right)^4\,,
\label{eq:quartic}
\end{equation}
with $\kappa_1=(10/ \sum_l \Delta_l^2)^{2}$,  instead of our initial quadratic fitness function (see Eq.~\ref{Hini} and Eq.~\ref{eq:quadr} in the main text) and we weighted sequences using the Boltzmann distribution in Eq.~\ref{eq:distr}. Finally, in Fig.~\ref{fig:square_selection}, we considered the selected ensembles of sequences such that the value of the trait $T(\vec{S})=\vec{S}\cdot\vec{\Delta}$ is between $T^*-\eta/2$ and $T^*+\eta/2$, where $\eta$ is the width of the selection window. In Fig.~\ref{fig:square_selection}, we used $\eta=0.6\sqrt{\sum_l\Delta_l^2}$.

These results confirm the robustness of our approach to different plausible forms of selection.

\clearpage

\subsection{Multiple states per site and alternative gauge choice}
\label{secMulti}

\begin{figure}[htb]
	\centering
	\includegraphics[width=8cm]{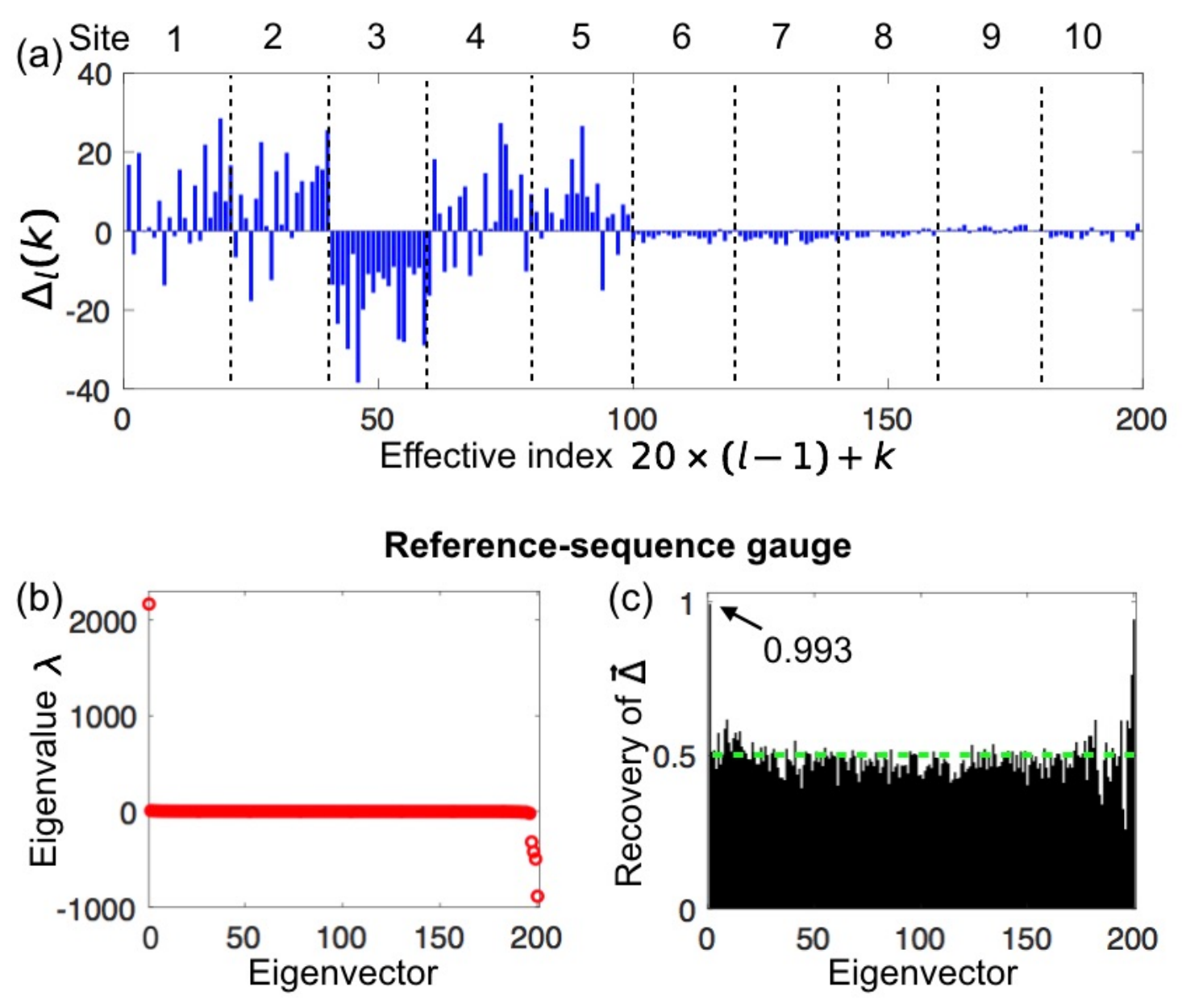}
	\vspace{0.2cm}
	\caption{ {\bf Performance of ICOD on synthetic sequence data with $q=21$ possible states at each site.}    (a) Mutational effects $\Delta_l(k)$  with respect to a reference sequence, chosen to be state $21$  at every site.  The mutational effect at $q=21$ is not shown. Note that while mutational effects are initially generated from a Gaussian distribution, \emph{relative} mutational effects (calculated with respect to the reference sequence) can have a systematic bias at each site $l$.   (b) Eigenvalues of the ICOD-modified inverse covariance matrix $\tilde C^{-1}$ defined in Eq.~\ref{ICOD_Potts}.  (c) Recovery of $\vec{\Delta}$ (see Eq.~\ref{eq:qstaterecovery}). The green dashed line indicates the random expectation of Recovery (Eq.~\ref{eq:random-Recovery}). 
 }
	\label{fig:protein-data-no-pseudocount}
\end{figure}

In Section~\ref{Sec:q-state} above, we described how to generalize from binary sequences to sequences with $q$ possible states at each site.
Correspondingly, we now generalize the ICOD method to higher values of $q$. Since we are interested in extracting the single-site mutational effects $\Delta_l(\alpha_l)$ with respect to a reference state at each site, we can simply set to zero the diagonal blocks of $C^{-1}$ in Eq.~\ref{invcorrPottsMulti}, yielding the modified inverse covariance matrix
 \begin{equation}
 \tilde{C}^{-1}_{kl}(\alpha_k,\alpha_l)=(1-\delta_{kl})\sum_{i=1}^N \kappa_i \Delta_{i,k}(\alpha_k)\Delta_{i,l}(\alpha_l)\,,
 \end{equation}
for the case of multiple selections, or more simply for a single selection
\begin{equation}
\tilde{C}^{-1}_{kl}(\alpha_k,\alpha_l)=(1-\delta_{lk})\,\kappa \Delta_{k}(\alpha_k)\Delta_l(\alpha_l).
\label{ICOD_Potts}
\end{equation}
This equation generalizes Eq.~\ref{eq:ICODmatrix} obtained for $q=2$ in the main text. As in that case, the first eigenvector of $\tilde{C}^{-1}$ (associated with the largest eigenvalue) should accurately report the single-site mutational effects $\Delta_k(\alpha_k)$. Indeed, Fig.~\ref{fig:protein-data-no-pseudocount} in the main text shows that this generalized version of ICOD performs very well on synthetic data generated for the case $q=21$ relevant to real protein sequences. Note that in the reference-sequence gauge, Recovery generalizes naturally to the $q$-state model as
  \begin{equation}
\mathrm{Recovery}=\frac{\sum_{l,\alpha_l} |\nu_l(\alpha_l)  \Delta_l(\alpha_l) |}{\sqrt{\sum_{l,\alpha_l} \nu_l(\alpha_l)^2 }\sqrt{\sum_{l,\alpha_l} \Delta_l(\alpha_l)^2 }},
\label{eq:qstaterecovery}
 \end{equation} 
where the sums over states $\alpha_l$ do not include the reference state at each site.

While the reference-sequence gauge is convenient and allows a clear interpretation of the mutational effects, other gauge choices are possible. For instance, in the DCA literature, the zero-sum (or Ising) gauge is often employed~\cite{Ekeberg13,Baldassi14}. In this gauge, the couplings satisfy
\begin{equation}
\sum_{\alpha}e_{ij}(\alpha,\beta)=\sum_{\beta}e_{ij}(\alpha,\beta)=0\,,
\label{gauge2}
\end{equation}
Qualitatively, the gauge degree of freedom means that contributions to the Hamiltonian in Eq.~\ref{Potts1} can be shifted between the fields and the couplings~\cite{Weigt09}. In DCA, the focus is on identifying the dominant two-body interactions, so one does not want the couplings to include contributions that can be accounted for by the one-body fields~\cite{Ekeberg14}. The zero-sum gauge satisfies this condition because it minimizes the Frobenius norms of the couplings
\begin{equation}
\left\Vert e_{ij}\right\Vert=\sqrt{\sum_{\alpha,\beta=1}^q \left[e_{ij}(\alpha,\beta)\right]^2}\,.
\label{Frob}
\end{equation}
Hence, the zero-sum gauge attributes the smallest possible fraction of the energy in Eq.~\ref{Potts1} to the couplings, and the largest possible fraction to the fields~\cite{Weigt09,Ekeberg13}. In order to transform to the zero-sum gauge defined in Eq.~\ref{gauge2}, each coupling $e_{ij}(\alpha,\beta)$ is replaced by
\begin{equation}
\tilde{e}_{ij}(\alpha,\beta)=e_{ij}(\alpha,\beta)-\langle e_{ij}(\zeta,\beta)\rangle_\zeta-\langle e_{ij}(\alpha,\eta)\rangle_\eta+\langle e_{ij}(\zeta,\eta)\rangle_{\zeta,\eta}\,,
\label{gc}
\end{equation}
where $\langle .\rangle_\zeta$ denotes an average over $\zeta\in\{1,...,q\}$~\cite{Ekeberg13}.

Shifting from the reference-sequence gauge where one state (in our derivations, state $q$) is taken as a reference at each site to the zero-sum gauge requires the replacement
\begin{equation}
\tilde{\Delta}_{l}(\alpha)= \Delta_{l}(\alpha)-\frac{1}{q}\sum_{\beta=1}^q \Delta_{l}(\beta),
\label{gc_Delta}
\end{equation}
The new reference-state-free mutational effects satisfy $\sum_{\beta=1}^q \tilde{\Delta}_{l}(\beta)=0$, and the associated couplings $\tilde{e}_{lk}(\alpha_l,\alpha_k)=-\kappa \tilde{\Delta}_l(\alpha_l)\tilde{\Delta}_k(\alpha_k)$ (see Eq.~\ref{identify}) are related to the initial ones $e_{lk}(\alpha_l,\alpha_k)$ through Eq.~\ref{gc}. 

\begin{figure}[htb]
	\centering
	\includegraphics[width=9cm]{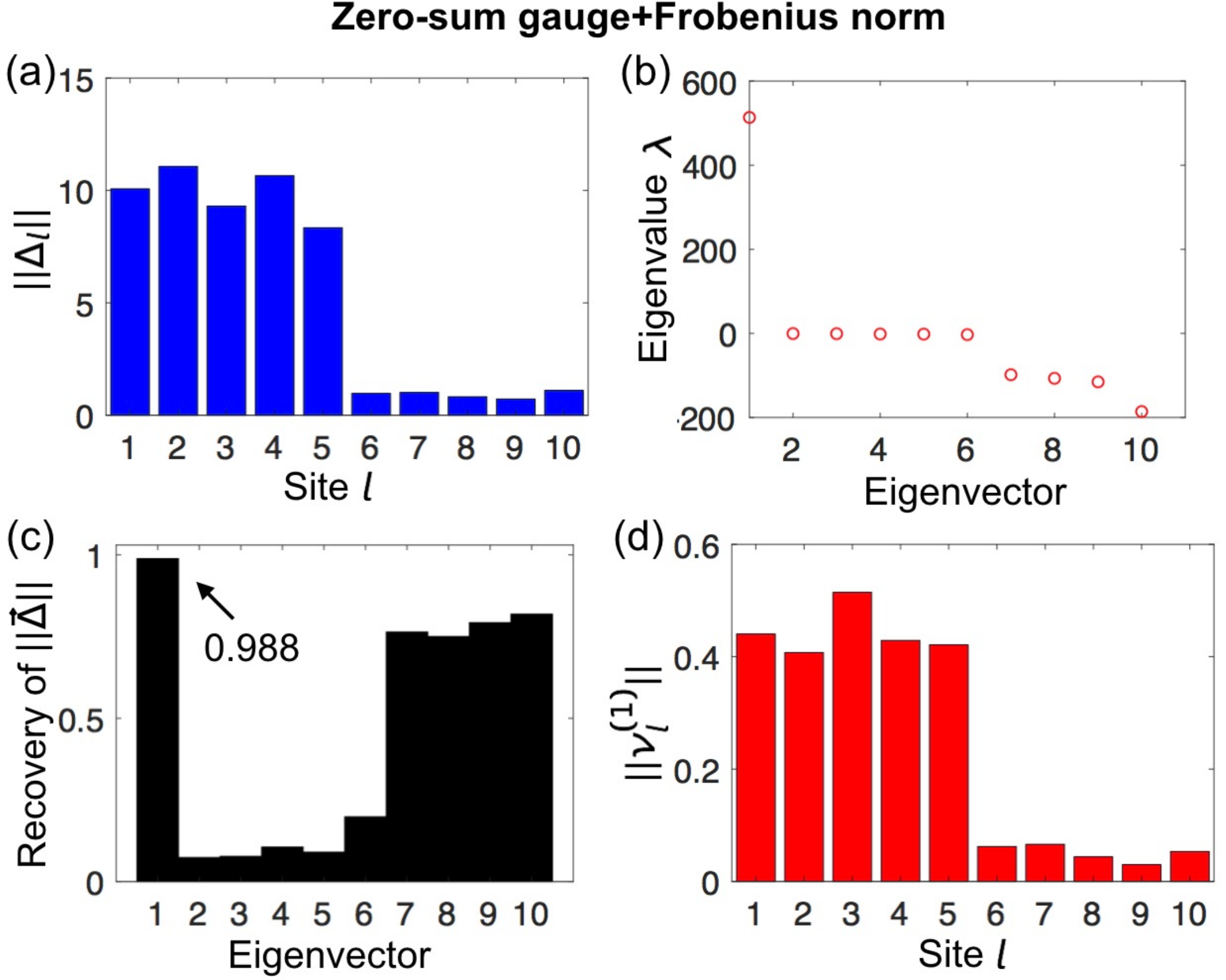}
	\vspace{0.2cm}
	\caption{{\bf Assessing site significance for synthetic sequence data.}  The same synthetic data as in Fig.~\ref{fig:protein-data-no-pseudocount} (with $q=21$ possible states at each site) is used. (a) Significance $||\Delta_l||$ of each site $l$, computed directly by applying Eqs.~\ref{gc_Delta} and~\ref{eq:fn_Delta} to the mutational effects $\Delta_l(k)$ shown in Fig.~\ref{fig:protein-data-no-pseudocount}(a). (b) Eigenvalues of the compressed ($L\times L$) ICOD-modified inverse covariance matrix, calculated by applying the zero-sum gauge to the ICOD-modified inverse covariance matrix (see Eq.~\ref{ICOD_Potts}), and by computing the Frobenius norm of each $20\times 20$ block associated to each pair of sites $(i,j)$ according to Eq.~\ref{Frob}.  (c) Recovery of site significances $|| \vec{\Delta}||$ from each eigenvector of the compressed ICOD-modified inverse covariance matrix (see panel (a) and Eq.~\ref{eq:qstaterecovery}) (d) Estimated site significances computed from the first eigenvector $\vec{\nu}^{(1)}$ of the compressed ICOD-modified inverse covariance matrix.
 }
	\label{fig:Supp_protein_data_no_pseudocount_zero_sum_gauge}
\end{figure}

Importantly, these reference-state-free mutational effects can be used to assess the overall importance of mutations at each particular site in the sequence. To this end, let us introduce the Frobenius norm of the reference-state-free mutational effects: 
\begin{equation}
||\Delta_l||= \sqrt{\sum_{\beta=1}^q \left(\tilde{\Delta}_{l}(\beta)\right)^2}.
\label{eq:fn_Delta}
\end{equation}
This quantity, which we refer to as the ``site significance'', measures the overall importance of mutational effects at site $l$. In order to assess site significances from an ensemble of sequences, without investigating the impact of each particular mutation at each site, one can apply the zero-sum gauge to the ICOD-modified inverse covariance matrix (see Eq.~\ref{ICOD_Potts}), and compute the Frobenius norm of each $20\times 20$ block associated to each pair of sites $(i,j)$ according to Eq.~\ref{Frob}. The first eigenvector of this compressed $L\times L$ matrix accurately reports the mutational significance of each site, as illustrated in Fig.~\ref{fig:Supp_protein_data_no_pseudocount_zero_sum_gauge}. Specifically, it yields a high Recovery of site significances as defined in Eq.~\ref{eq:fn_Delta} (see Fig.~\ref{fig:Supp_protein_data_no_pseudocount_zero_sum_gauge}(c)), and it successfully predicts the most important sites, i.e. the sector sites, in our synthetic data (see Fig.~\ref{fig:Supp_protein_data_no_pseudocount_zero_sum_gauge}(d)).

\newpage

\subsection{Pseudocounts}

As pseudocounts are often necessary to regularize real sequence data, and as a high fraction of pseudocounts is generally used in DCA, we consider here whether the ICOD method is robust to the addition of pseudocounts. 

Until now, we used only raw empirical frequencies obtained from sequence data. For instance, one-body frequencies were obtained by counting the number of sequences where a given state occured at a given site and dividing by the total number $M$ of sequences in the ensemble. Covariances were computed from the empirical single-site frequencies of occurrence of each state $\alpha$ at each site $i$, denoted by $f^e_i(\alpha)$, and the empirical two-site frequencies of occurrence of each ordered pair of states $(\alpha,\beta)$ at each ordered pair of sites $(i,j)$, denoted by $f^e_{ij}(\alpha,\beta)$. Specifically, we obtained the covariance matrix as $C_{ij}(\alpha,\beta)=f^e_{ij}(\alpha,\beta)-f^e_{i}(\alpha)f^e_{j}(\beta)$~\cite{Weigt09}. 

To avoid issues arising from limited sample size, such as states that never appear at some sites (which present mathematical difficulties, e.g. a non-invertible covariance matrix~\cite{morcos2011direct}), one can introduce pseudocounts via a parameter $\Lambda$~\cite{Weigt09,Procaccini11,Marks11,morcos2011direct}. The one-site frequencies $f_i$ then become
\begin{equation}
f_i(\alpha)=\frac{\Lambda}{q}+(1-\Lambda)f^e_i(\alpha)\,,
\label{fi}
\end{equation}
where $q$ is the number of states per site. Similarly, the two-site frequencies $f_{ij}$ become
\begin{align}
f_{ij}(\alpha,\beta)&=\frac{\Lambda}{q^2}+(1-\Lambda)f^e_{ij}(\alpha,\beta)\textrm{ if }i\neq j\,, \label{fij}\\
f_{ii}(\alpha,\beta)&=\frac{\Lambda}{q} \delta_{\alpha\beta}+(1-\Lambda)f^e_{ii}(\alpha,\beta)= f_i(\alpha)\delta_{\alpha\beta}\,. \label{fii}
\end{align}
These pseudocount corrections are uniform (i.e. they have the same weight $1/q$ for all states), and their influence relative to the raw empirical frequencies can be tuned through the parameter $\Lambda$. In DCA, a high value of f $\Lambda$ has been found to improve contact prediction: typically $\Lambda\approx 0.5$~\cite{morcos2011direct,Marks11,bitbol2016inferring}. Note that the correspondence of $\Lambda$ with the parameter $\lambda$ in Refs.~\cite{Procaccini11,Marks11,morcos2011direct} is obtained by setting $\Lambda=\lambda/(\lambda+M)$. 

\begin{figure}[htb]
\centering
\includegraphics[width=12cm]{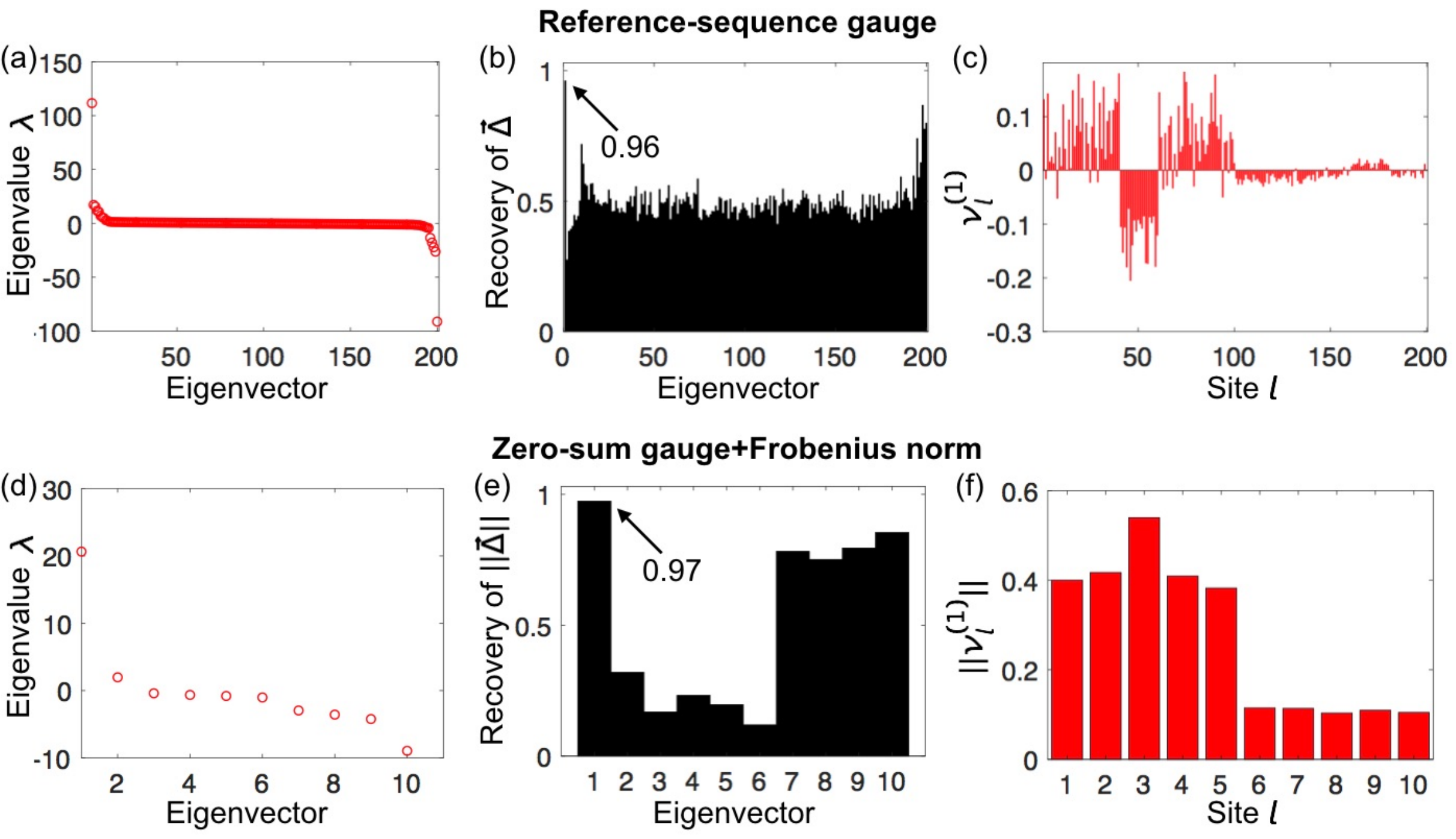}
\vspace{0.2cm}
\caption{ {\bf Effect of pseudocounts on ICOD performance on synthetic sequence data with $q=21$ possible states at each site.} The same synthetic data as in Fig.~\ref{fig:protein-data-no-pseudocount} and~\ref{fig:Supp_protein_data_no_pseudocount_zero_sum_gauge} is used, but here pseudocounts are employed, with weight $\Lambda=0.3$. (a-c) Similar analysis as in Fig.~\ref{fig:protein-data-no-pseudocount}: (a) Eigenvalues of the ICOD-modified inverse covariance matrix.  (b) Recovery of $\vec  \Delta$ from each eigenvector of the ICOD-modified inverse covariance matrix. (c) First eigenvector of the ICOD-modified inverse covariance matrix.  (d-f)   Similar analysis as in Fig.~\ref{fig:Supp_protein_data_no_pseudocount_zero_sum_gauge}: (d) Eigenvalues of the compressed ICOD-modified inverse covariance matrix.  (e) Recovery of site significances $|| \vec{\Delta}||$ from each eigenvector of the compressed ICOD-modified inverse covariance matrix. (f) Estimated site significances computed from the first eigenvector of the compressed ICOD-modified inverse covariance matrix.  } 
\label{fig:protein-data-with-pseudocount}
\end{figure}

From these quantities, we define the pseudocount-corrected covariances
\begin{equation}
C'_{ij}(\alpha,\beta)=f_{ij}(\alpha,\beta)-f_i(\alpha)f_j(\beta)\,.
\label{Cij}
\end{equation}
We show in Fig.~\ref{fig:protein-data-with-pseudocount} that adding pseudocounts as high as $\Lambda = 0.3$ still allows for accurate extraction of mutational effects (Recovery 0.96) and provides a reliable prediction of sector sites. 

\section{Performance of SCA}

\subsection{Analytical estimates for  $\langle S_l\rangle_*$ and $C_{ll'}$ for a single selection with binary sequences}
{Protein sectors were first discovered from sequence data using a PCA-based method called Statistical Coupling Analysis (SCA)~\cite{halabi2009protein,Rivoire16}. Interestingly, in SCA, sectors are found from the eigenvectors associated to the largest eigenvalues, while in ICOD they are found from the (modified) eigenvectors associated to the smallest eigenvalues. This difference stems from the fact that SCA and ICOD do not start from the same matrix. For binary sequences, SCA uses the absolute value of a conservation-weighted covariance matrix, $\tilde{C}_{ll'}^{\mathrm{(SCA)}}=|\phi_l C_{ll'}\phi_{l'}|$ (see main text and Ref.~\cite{halabi2009protein}). When all amino-acid states are accounted for, SCA compresses each block of the conservation-weighted matrix corresponding to two sites to obtain one positive value, e.g. the Frobenius norm of the block~\cite{Rivoire16}. Conversely, ICOD employs the regular covariance matrix, suppressing the diagonal blocks of its inverse at the last step before diagonalization. To better understand the performance of SCA in recovering the site-dependent mutational effects associated with a selective constraint, it is helpful to have analytical estimates for the average mutant fraction $\langle S_l\rangle_*$ at each site $l$ and the covariance matrix $C_{ll'}$ for an ensemble of {binary sequences obtained from a single selection using vector of mutational effects $\vec{\Delta}$. To this end,
 we provide the following two ansatzes: 
 \begin{equation}
\langle S_l\rangle_*-\langle S_l\rangle\approx  (T^*-\langle T\rangle) \frac{\Delta_l}{\sum_l \Delta_l^2},
\label{eq:S_ansatz}
\end{equation} 
 \begin{equation}
C_{ll'} \approx   \begin{cases} 
& - \frac{\Delta_l\Delta_{l'}\sigma_l^2\sigma_{l'}^2}{\sum_l \Delta_l^2\sigma_l^2},\quad l\neq l'\\
& \sigma_l^2 ,\quad l=l',
\end{cases}
\label{eq:Corr_ansatz}
\end{equation}
where $\sigma_l^2=\langle S_l^2\rangle_* -\langle S_l\rangle_*^2=\langle S_l\rangle_* \left(1-\langle S_l\rangle_*\right)$  represents the variance of $S_l$.
Recall that $S_l\in\{0,1\}$, where 0 is the reference state and 1 the mutant state, and that $\langle \cdot\rangle_*$ denotes ensemble averages over the selectively weighted subset of sequences, while $\langle \cdot\rangle$ denotes averages over the unselected (unweighted) ensemble. 

Although we have not proven these two ansatzes, numerical tests (Fig.~\ref{fig:Covariance_structure}) have verified these two relations for ensembles generated from a $\vec{\Delta}$  with multiple sites of comparably large mutational effects so as not to be dominated by a single site, i.e., $\Delta_{l}/\sqrt{\sum_{l'} \Delta_{l'}^2}\ll 1$ for any $l$.   As a counterexample, the $\vec{\Delta}$ from our elastic network model does not satisfy this condition.

\begin{figure}[htb]
	\centering
	\includegraphics[width=10cm]{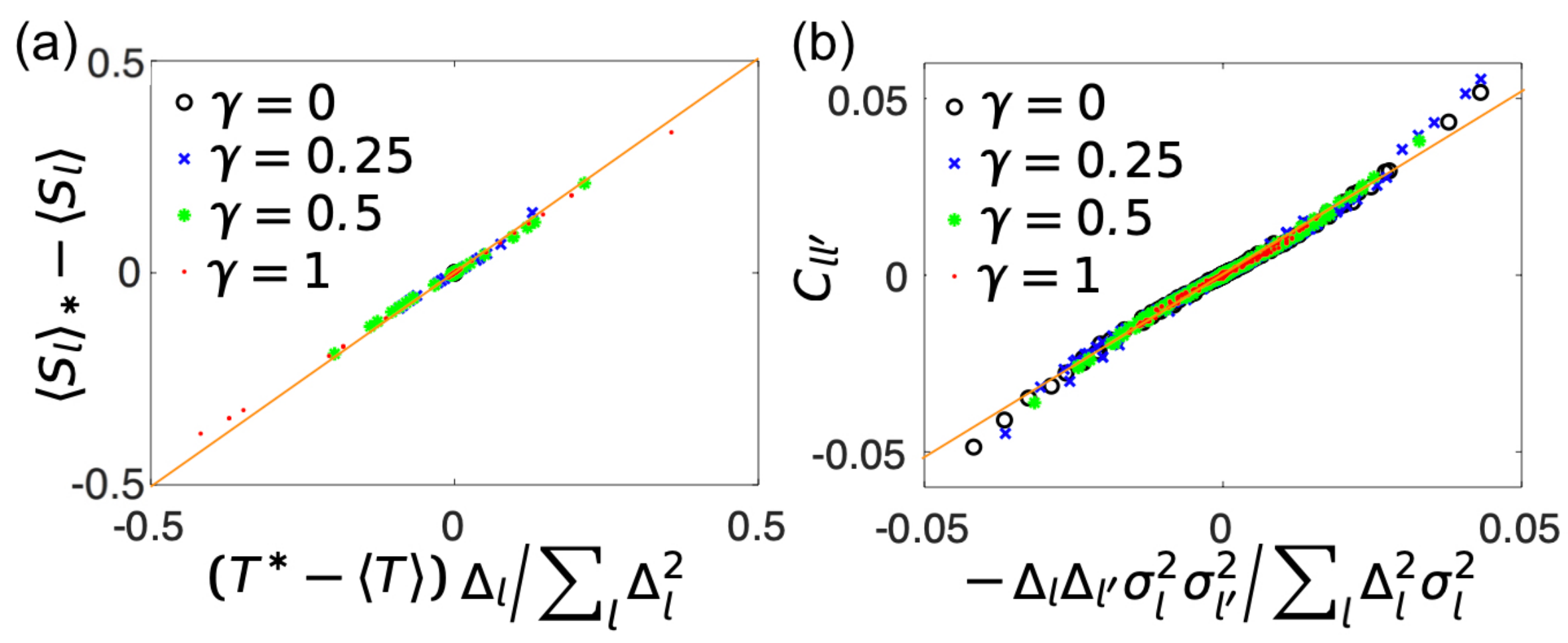}
	\vspace{0.2cm}
	\caption{ {\bf Numerical verification of the ansatzes in Eq.~\ref{eq:S_ansatz} and Eq.~\ref{eq:Corr_ansatz}.}   We generate a sequence ensemble by considering four values of relative selection bias $\gamma\equiv (T^*-\langle T\rangle)/\sqrt{\langle (T-\langle T\rangle)^2 \rangle}= 0, 0.25, 0.5, 1$ and for each case we use a synthetic $\vec{\Delta}$  with a sector size of 20.   (a)   Numerically computed average bias of the mutant fractions $\langle S_l\rangle_*-\langle S_l\rangle$. Here, $\langle S_l\rangle=0.5$ for the unselected ensemble.  (b) Numerically computed covariances $C_{ll'}$. The results in (a,b) compare well with the analytical predictions (orange lines),  provided that $\Delta_{l}/\sqrt{\sum_l' \Delta_l'^2}\ll 1$ for any $l$.  For each case,  $10^6$ random sequences  were generated   to minimize noise from sampling. }
		\label{fig:Covariance_structure}
\end{figure}
 
\subsection{Analysis of the SCA method}
\label{SCAsub}

 Here, we provide a detailed analysis of the SCA method from Refs.~\cite{halabi2009protein,Rivoire16}. Following these references, the reweighting factor is chosen to be
\begin{equation}
\phi_l=\frac{\partial D\left(\langle S_l\rangle_* ,\,\langle S_l\rangle\right)}{\partial \langle S_l\rangle_*},
\label{phil}
\end{equation}
where, for each site $l$, $D\left(\langle S_l\rangle_* ,\,\langle S_l\rangle\right)$ is the Kullback-Leibler divergence between the distribution of mutant fractions for the selected sequences and the background distribution:
\begin{equation}
D\left(\langle S_l\rangle_*,\,\langle S_l\rangle\right)=\langle S_l\rangle_* \log\frac{\langle S_l\rangle_*}{\langle S_l\rangle}+\left(1-\langle S_l\rangle_*\right) \log\frac{1-\langle S_l\rangle_*}{1-\langle S_l\rangle}.
\label{eq:KL}
\end{equation}
In our case, the background distribution is obtained from the unselected sequence ensemble, for which $\langle S_l\rangle=0.5$.  Hence,  we have 
 \begin{equation}
 \phi_l= \log\left[\frac{\langle S_l\rangle_*\left(1-\langle S_l\rangle\right)}{\langle S_l\rangle \left(1-\langle S_l\rangle_*\right)}\right],
 \end{equation}
as illustrated in Fig.~\ref{fig:reweighting-method}(a). In the regime of relatively weak conservation, i.e. 
 when  $\langle S_l\rangle$ is not close to 0 or 1, and $|\langle S_l\rangle_*-\langle S_l\rangle|\ll  \langle S_l\rangle$, a first-order expansion yields
  \begin{equation}
 \phi_l\approx \frac{\langle S_l\rangle_*-\langle S_l\rangle}{\langle S_l\rangle(1-\langle S_l\rangle)},
 \end{equation}
as shown in Fig.~\ref{fig:reweighting-method}(b).
 Employing the ansatz~(\ref{eq:S_ansatz}) in this regime, we obtain 
 \begin{equation}
 \phi_l\propto (T^*-\langle T\rangle)\Delta_l.
 \label{eq:phi-propto}
 \end{equation}
  This relation is verified in Fig.~\ref{fig:reweighting-method}(c) for a sequence ensemble generated with a synthetic $\Delta_l$.   Hence, the SCA} reweighting factor carries information about $\Delta_l$ as long as $T^*\neq \langle T\rangle$.  In this regime, information about conservation (namely $\phi_l$) should thus be sufficient to recover mutational effects and sectors. This was indeed found to be the case for some real proteins with a single sector~\cite{Tesileanu15}. However, when the selection bias, $T^*-\langle T\rangle$, is small, random noise due to finite sampling will typically swamp this relationship.
 
\begin{figure}[htb]
\centering
\includegraphics[width=14cm]{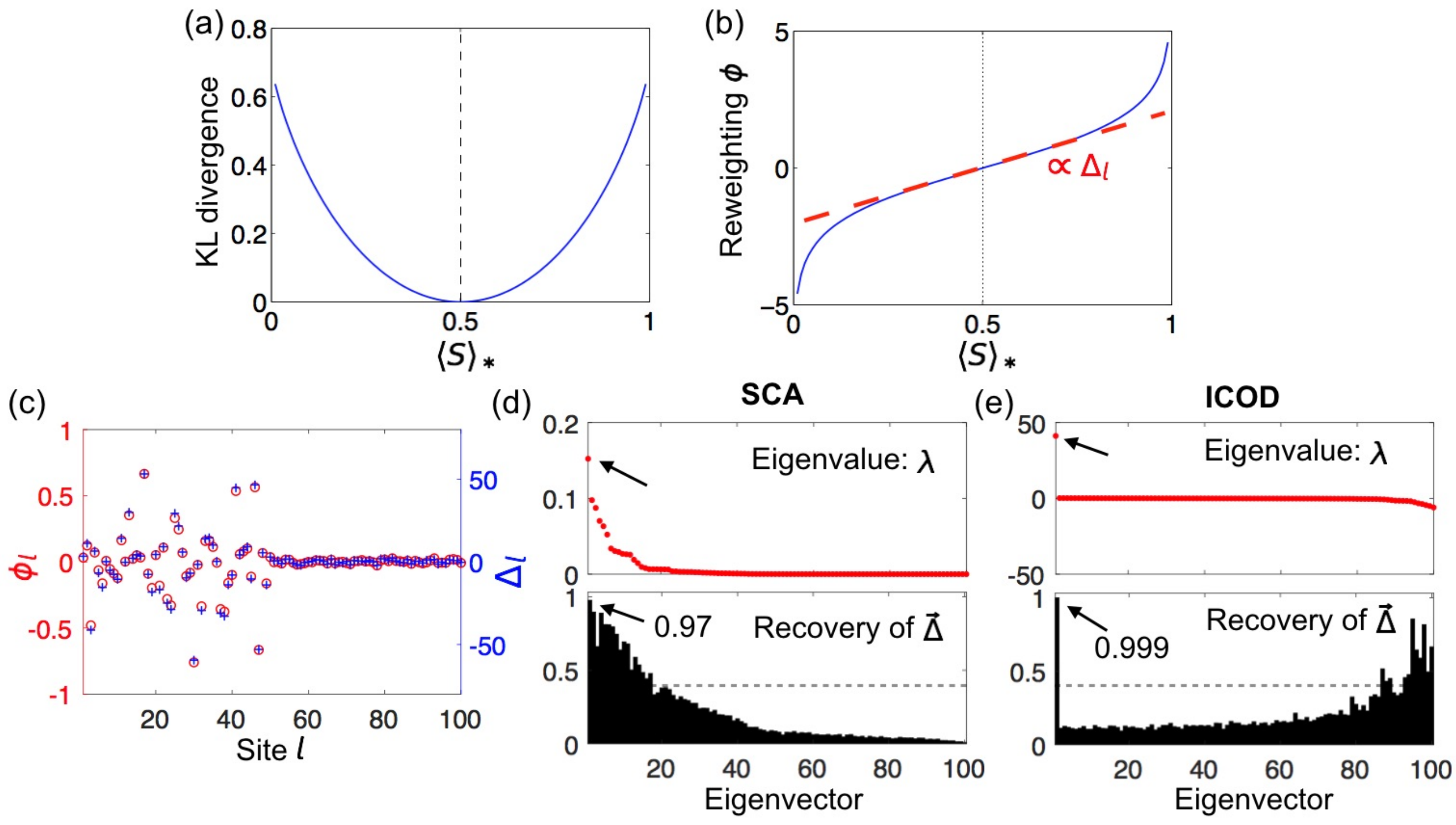}
\vspace{0.2cm}
\caption{ {\bf Underpinnings of Recovery of mutational effect vector $\vec{\Delta}$ by SCA.} (a) Kullback-Leibler divergence versus mutant fraction $\langle S\rangle_*$ for background mutant fraction $\langle S\rangle=0.5$.  (b) Reweighting factor $\phi$ as a function of mutant fraction $\langle S\rangle_*$   for background mutant fraction $\langle S\rangle=0.5$.  (c) Reweighting factor $\phi_l$ and synthetic $\Delta_l$ for an ensemble of sequences generated with a single selection at relative selection bias $\gamma=1$.  $\vec{\Delta}$ was generated with the first 50 sites as sector sites, and 50,000 sequences were employed, as in most of our examples using ICOD (see above). (d-e) Performance of  SCA and ICOD  for this ensemble, respectively.   In computing Recovery  using SCA, we use the normalized vector $\sqrt{\nu_l^{(j)}}$ to predict $\vec\Delta$. The gray dashed lines in (d) and (e) indicate the random expectation of Recovery (Eq.~\ref{eq:random-Recovery}).
}
\label{fig:reweighting-method}
\end{figure}

\clearpage

In Refs.~\cite{halabi2009protein,Rivoire16}, the first eigenvectors of the conservation-reweighted SCA covariance matrix, $\tilde{C}_{ll'}^{\mathrm{(SCA)}}=|\phi_l C_{ll'}\phi_{l'}|$, were used to find sectors from sequence data. How does the first eigenvector of  $\tilde{C}^{\mathrm{(SCA)}}$ relate to the mutational effect vector $\vec \Delta$?  Utilizing both Eq.~\ref{eq:Corr_ansatz} and Eq.~\ref{eq:phi-propto}, and assuming $T^*\neq \langle T\rangle$, we obtain 
\begin{equation}
\tilde{C}_{ll'}^{\mathrm{(SCA)} }\propto 
   \begin{cases} 
&  \Delta_l^2\Delta_{l'}^2\sigma_l^2\sigma_{l'}^2,\quad l\neq l'\\
& \Delta_l^2 \sigma_l^2 ,\quad l=l'.
\end{cases}
\end{equation}
 Apart from the diagonal, the matrix is approximately proportional to the tensor product of $\Delta_l^2\sigma_l^2$  with itself.  If we neglect the contribution from the diagonal elements of $\tilde{C}^{\mathrm{(SCA)}}$, the first eigenvector $\vec{\nu}^{(1)}$ satisfies 
\begin{equation}
\nu_l^{(1)}\propto \Delta_l^2\sigma_l^2.
\label{eq:SCA_eigenvector}
\end{equation}
Eq.~\ref{eq:SCA_eigenvector} explains why $\sqrt{\nu_l^{(1)}}$ carries information about $\Delta_l$.   In Fig.~\ref{fig:reweighting-method}(d), Recovery using SCA (and Eq.~\ref{eq:recovery-measure} with $\sqrt{\nu_l^{(1)}}$ instead of $\nu_l^{(1)}$) is 0.97,   which remains lower than  Recovery using ICOD, which  is 0.999  here.   Besides, Fig.~\ref{fig:SCAsquareroot} illustrates  that Recovery of $\vec \Delta$ by SCA is much better using $\sqrt{\nu_l^{(1)}}$ than $\nu_l^{(1)}$.

\begin{figure}[htb]
\centering
\includegraphics[width=4.5cm]{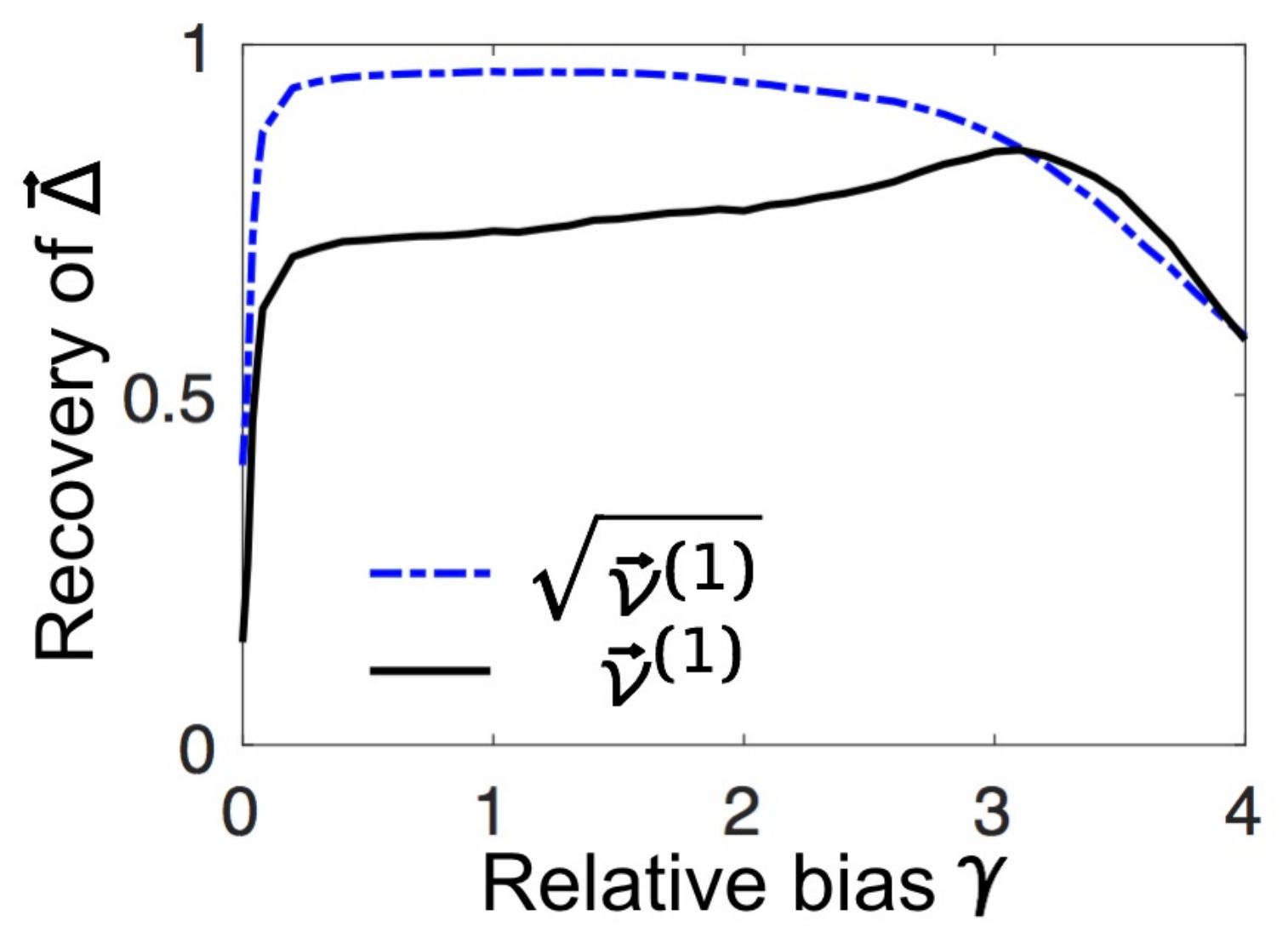}
\vspace{0.2cm}
\caption{ {\bf Recovery of $\vec{\Delta}$ from the first SCA eigenvector using $\vec{\nu}^{(1)}$ or $\sqrt{\vec{\nu}^{(1)}}$.}  The sequence data are the same as used for the blue curves in Fig.~3(a). As suggested by Eq.~\ref{eq:SCA_eigenvector}, use of the square root of $\vec{\nu}^{(1)}$ significantly improves Recovery. 
}
\label{fig:SCAsquareroot}
\end{figure}

\subsection{Comparison between ICOD and SCA}

In the main text, we compared the performance of ICOD and SCA with respect to Recovery of mutational-effect vectors $\vec\Delta$ in synthetic data (see Fig.~3). We found that ICOD performs well over a broader range of relative biases $\gamma$ than SCA. The failure of SCA at biases close to zero can be explained by the fact that the conservation weights $\phi_l$ then vanish (see Eq.~\ref{eq:phi-propto}). A further example of the failure of SCA for non-biased selections is given by the case studied in Fig.~\ref{fig:two-sector-fig3}, where we considered two selections, a biased one associated to $\vec{\Delta}_1$ and a non-biased one associated to $\vec{\Delta}_2$. Fig.~\ref{fig:Rama-two-sectors} shows that SCA recovers $\vec{\Delta}_1$ well, but performs badly for $\vec{\Delta}_2$, while ICOD recovers both of them very well (see Fig.~\ref{fig:two-sector-fig3}). 

\begin{figure}[htb]
	\centering
	\includegraphics[width=13cm]{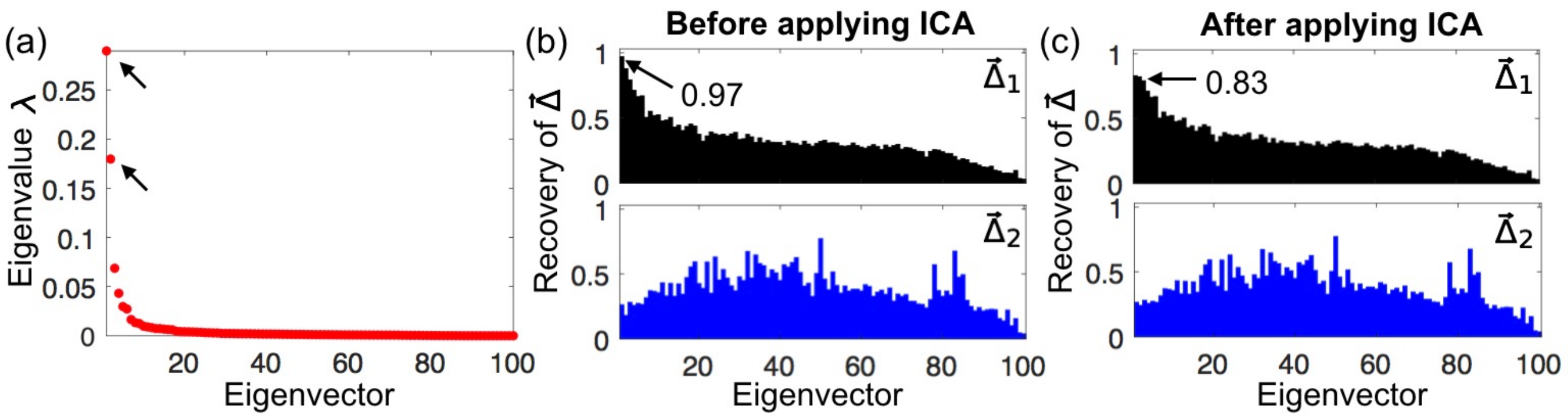}
	\vspace{0.2cm}
	\caption{{\bf Performance of SCA for the double selection from Fig.~\ref{fig:two-sector-fig3}.} (a) Eigenvalues. (b) Before applying ICA, the first eigenvector has high Recovery of $\vec{\Delta}_1$, but no eigenvector has substantial Recovery of $\vec{\Delta}_2$. This difference matches our observation that SCA performs well for selections of intermediate bias, but not for unbiased selections. (c) Applying ICA on the first two eigenvectors does not improve Recovery. 
	}
	\label{fig:Rama-two-sectors}
\end{figure}

While the comparison of Recovery favors ICOD, SCA was originally used to identify sectors (in our model, sites with important mutational effects under a given selection) rather than to recover complete mutational effect vectors $\vec \Delta$. Hence, in Fig.~\ref{fig:SCA_ICOD}, we compare the ability of ICOD and SCA to predict the $n$ sites with the largest mutational effects. Note that this comparison is independent of whether we use $\vec{\nu}^{(1)}$ or $\sqrt{\vec{\nu}^{(1)}}$ as the predictor in SCA, since the square-root function is increasing and preserves order.  Using this criterion, we again find that ICOD performs well over a broad range of relative biases $\gamma$, while SCA only works well for sequences selected under moderate biases.

\begin{figure}[htb]
\centering
\includegraphics[width=10cm]{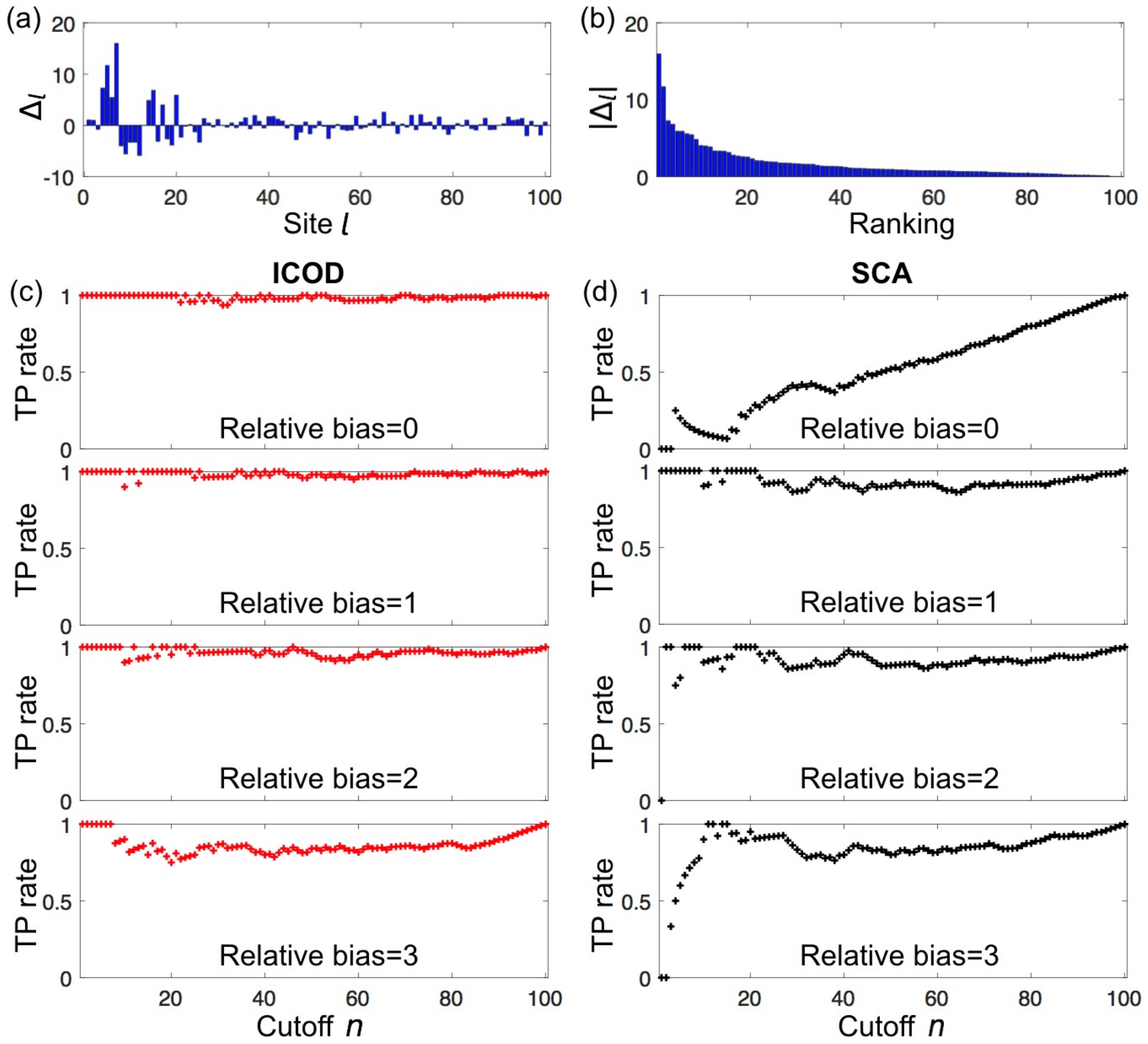}
\vspace{0.2cm}
\caption{{\bf Comparison of sector-site identification by ICOD and SCA (see also Fig.~3).} We use the synthetic $\vec{\Delta}$ in (a) to selectively weight 5,000 random sequences at four relative bias values $\gamma\equiv (T^*-\langle T\rangle)/\sqrt{\langle (T-\langle T\rangle)^2 \rangle} = 0,1,2,3$ and test the ability of ICOD or SCA to correctly predict the sites with the $n$ largest mutational effects. (b) Magnitudes of mutational effects of $\vec{\Delta}$ by rank. (c-d). True Positive (TP) rates obtained by taking the first eigenvector $\vec{\nu}^{(1)}$ from either ICOD or SCA,  generating a ranked list of sites of descending $|\nu_l^{(1)}|$ at each site $l$, and computing the fraction of the top $n$ sites in this predicted ordering that are also among the top $n$ sites of the actual ordering of mutational effect magnitudes $|\Delta_l|$. The  effect of relative bias $\gamma$ on Recovery is shown in Fig.~3. (c) As expected, the prediction of ICOD is very good under all relative biases. (d) On the other hand, SCA does not perform well at the smallest or largest relative biases.
}
\label{fig:SCA_ICOD}
\end{figure}

\section{Performance of a method based on the generalized Hopfield model}

As mentioned in the main text, we also compared ICOD with another PCA-based approach developed in Ref.~\cite{Cocco11}, which employs an inference method specific to the generalized Hopfield model. For $L$ Ising spins ($s_l\in\{-1,1\}$ for $1\leq l \leq L$), the Hamiltonian of the generalized Hopfield model reads (see Eq.~\ref{eq:fitness} in Ref.~\cite{Cocco11})
\begin{equation}
H(\vec{s})=-\sum_{l=1}^L h_l \,s_l-\frac{1}{2L}\sum_{i=1}^N\left(\sum_{l=1}^L\xi_{i,l}\,s_l\right)^2+\frac{1}{2L}\sum_{i=1}^{N'}\left(\sum_{l=1}^L\xi'_{i,l}\,s_l\right)^2\,,
\label{Hop}
\end{equation}
where $h_l$ is the local field at site $l$, while $\vec{\xi}_i=(\xi_{i,1},\dots,\xi_{i,L})$ is an attractive pattern and $\vec{\xi'}_i=(\xi'_{i,1},\dots,\xi'_{i,L})$ is a repulsive pattern. Here there are $N$ attractive patterns and $N'$ repulsive ones. In our model, in the single-selection case, the fitness of a sequence $\vec{s}$ in the Ising representation reads (see above, Sec.~\ref{Sec_Ising}, Eq.~\ref{HIs}) 
\begin{equation}
 w(\vec{s})=-\frac{\kappa}{2}\left(\sum_{l=1}^L D_l s_l-\alpha \right)^2=-\frac{\kappa}{2}\left[\left(\sum_{l=1}^L D_l s_l\right)^2-2\alpha \sum_{l=1}^L D_l s_l +\alpha^2\right]\,,
 \label{HIs_b}
 \end{equation}
 with $D_l=\Delta_l/2$ and $\alpha=T^*-\sum_l D_l$. Recalling that fitnesses and Hamiltonians have opposite signs, a comparison of Eqs.~\ref{Hop} and~\ref{HIs_b} shows that $\vec{\Delta}$ plays the part of a repulsive pattern in the two-body coupling terms, with the exact correspondence given by $\vec{\xi'}=\vec{\Delta}\,\sqrt{\kappa L}/2$. Note that in our model the local fields are proportional to the components of $\vec{\Delta}$.

Ref.~\cite{Cocco11} proposed a method to infer attractive and repulsive patterns from data generated using a generalized Hopfield model Eq.~\ref{Hop}. Introducing the correlation matrix $G$, which is related to the covariance matrix $C$ through 
\begin{equation}
G_{ll'}=\frac{C_{ll'}}{\tilde{\sigma}_l\tilde{\sigma}_{l'}}\,,
\label{correlmat}
\end{equation}
where $\tilde{\sigma}_l^2=\langle s_l^2\rangle_* -\langle s_l\rangle_*^2=1 -\langle s_l\rangle_*^2$.
Ref.~\cite{Cocco11} found, to lowest order, the following approximation for a single repulsive pattern $\vec{\xi}'$ (see Eq.~9 in Ref.~\cite{Cocco11}):
\begin{equation}
\xi'_l\approx\sqrt{L\left(\frac{1}{\lambda^{(L)}}-1\right)}\,\,\frac{\nu_l^{(L)}}{\tilde{\sigma}_l},
\label{eq:Cocco0}
\end{equation}
where $\lambda^{(L)}$ is the smallest (last) eigenvalue of the correlation matrix $G$ and $\nu_l^{(L)}$ is the associated eigenvector. This yields
\begin{equation}
\Delta_l\propto\frac{\nu_l^{(L)}}{\tilde{\sigma}_l}.
\label{eq:Cocco}
\end{equation}Inference of $\vec{\Delta}$ based on Eq.~\ref{eq:Cocco} is referred to as GHI (for Generalized Hopfield Inference) below.

GHI performs very well for the sequence ensembles from the elastic network model used in Fig.~1 and Fig.~2 (Fig.~\ref{fig:Cocco_performance}). Importantly, just as for simple PCA and for ICOD (see main text), the top Recovery is obtained for the (modified) bottom eigenvector of the covariance matrix, consistently with $\vec{\Delta}$ being a repulsive pattern, but the large-eigenvalue modes also contain some information about $\vec{\Delta}$ (Fig.~\ref{fig:Cocco_performance}).  

\begin{figure}[h!]
	\centering
	\includegraphics[width=9cm]{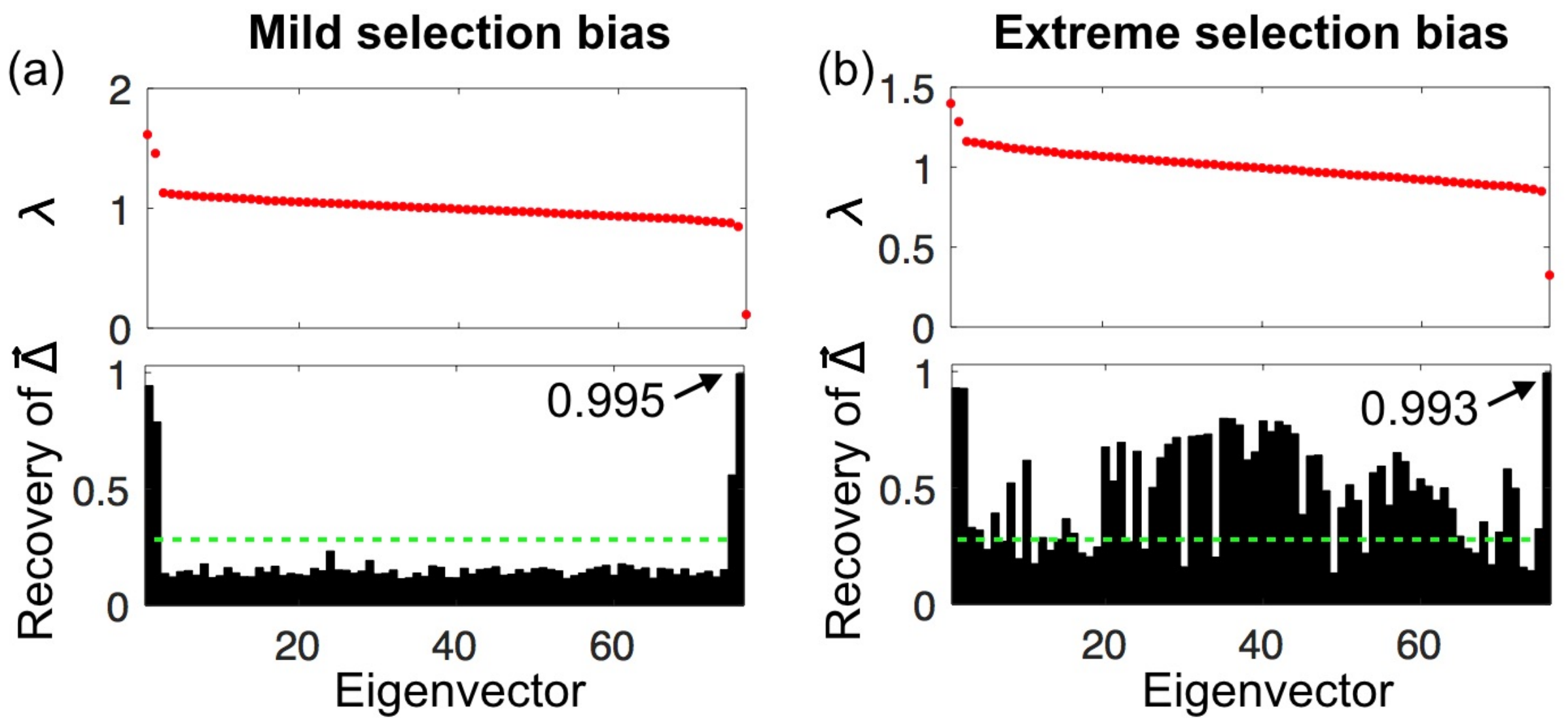}
	\vspace{0.2cm}
	\caption{  {\bf Performance of GHI on sequence ensembles generated with our elastic-network $\vec{\Delta}$.} (a)  Eigenvalues of $G$ and Recovery under mild selection bias, as in Fig.~1 in the main text.  (b) Eigenvalues of $G$ and Recovery under extreme selection bias, as in Fig.~2 in the main text. The green dashed lines in (a,b) indicate the random expectation of Recovery (Eq.~\ref{eq:random-Recovery}).  	} 
	\label{fig:Cocco_performance}
\end{figure}

In Fig.~\ref{fig:Cocco_sector_size}, we systematically compare all methods discussed in our work to recover $\vec{\Delta}$ from sequence data under various selection biases, using different sector sizes, for selectively weighted ensembles of 50,000 random sequences. We focus on the case of a single selection and compare Recovery of $\vec{\Delta}$ according to:
\setlist{nolistsep}
\begin{itemize}[noitemsep]
\item ICOD, using the first eigenvector of the modified inverse covariance matrix $\tilde{C}^{-1}$ (see main text, Eq.~\ref{eq:ICODmatrix})
\item PCA, using the last principal component of the data (last eigenvector of the covariance matrix, see main text)
\item SCA, using the first eigenvector of the absolute value of a conservation-weighted covariance matrix, $\tilde{C}_{ll'}^{\mathrm{(SCA)}}=|\phi_l C_{ll'}\phi_{l'}|$ (see main text and Ref.~\cite{halabi2009protein})
\item GHI, using the reweighted last eigenvector of the correlation matrix (see Eqs.~\ref{correlmat} and~\ref{eq:Cocco}).
\end{itemize} 
\noindent
Overall, ICOD and GHI perform best. For small selection biases, all methods perform accurately, except SCA, which fails when selection bias vanishes, as explained above. When the sector size is small compared to the sequence length $L$ (Fig.~\ref{fig:Cocco_sector_size} (a-d)), GHI performs a little bit better than ICOD for relatively small selection biases (however Recovery remains $\gtrsim 95\%$ with ICOD). Conversely, GHI is significantly outperformed by ICOD for relatively large selection bias, and the performance of PCA and SCA falls off quite rapidly in this regime. The performances of ICOD, PCA, and GHI become similar when the sector size becomes comparable to the sequence length (Fig.~\ref{fig:Cocco_sector_size} (e, f)).

\begin{figure}[htb]
	\centering
	\includegraphics[width=12cm]{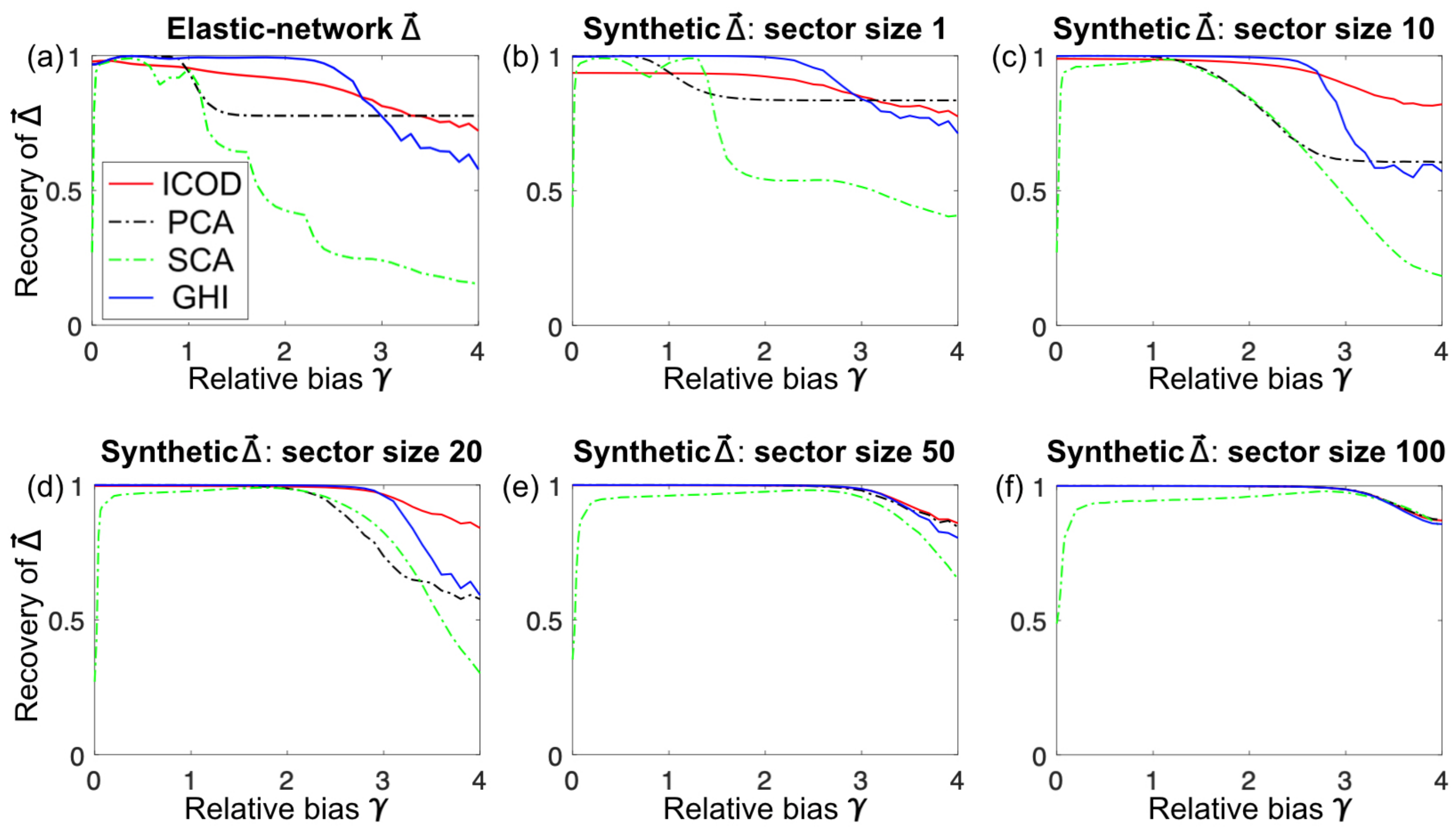}
	\vspace{0.2cm}
	\caption{ {\bf Comparing Recovery of different methods for various $\vec{\Delta}$s.}  Here, GHI refers to inference based on Eq.~\ref{eq:Cocco}.  Curves are obtained by averaging over 100 realizations, each for an ensemble of 50,000 random sequences.  For synthetic $\vec{\Delta}$s, each realization corresponds to a new $\vec{\Delta}$.  
	}
	\label{fig:Cocco_sector_size}
\end{figure}

We further find that GHI is more sensitive to the size of the sequence ensemble than ICOD, although it becomes the most accurate for very large dataset sizes (see Fig.~\ref{fig:Cocco_sequence_size}). The performance of ICOD is quite robust to dataset size. Note that PCA outperforms other methods when the data size becomes very small (Fig.~\ref{fig:Cocco_sequence_size}, number of sequences~$=500$).

\begin{figure}[htb]
\centering
\includegraphics[width=13cm]{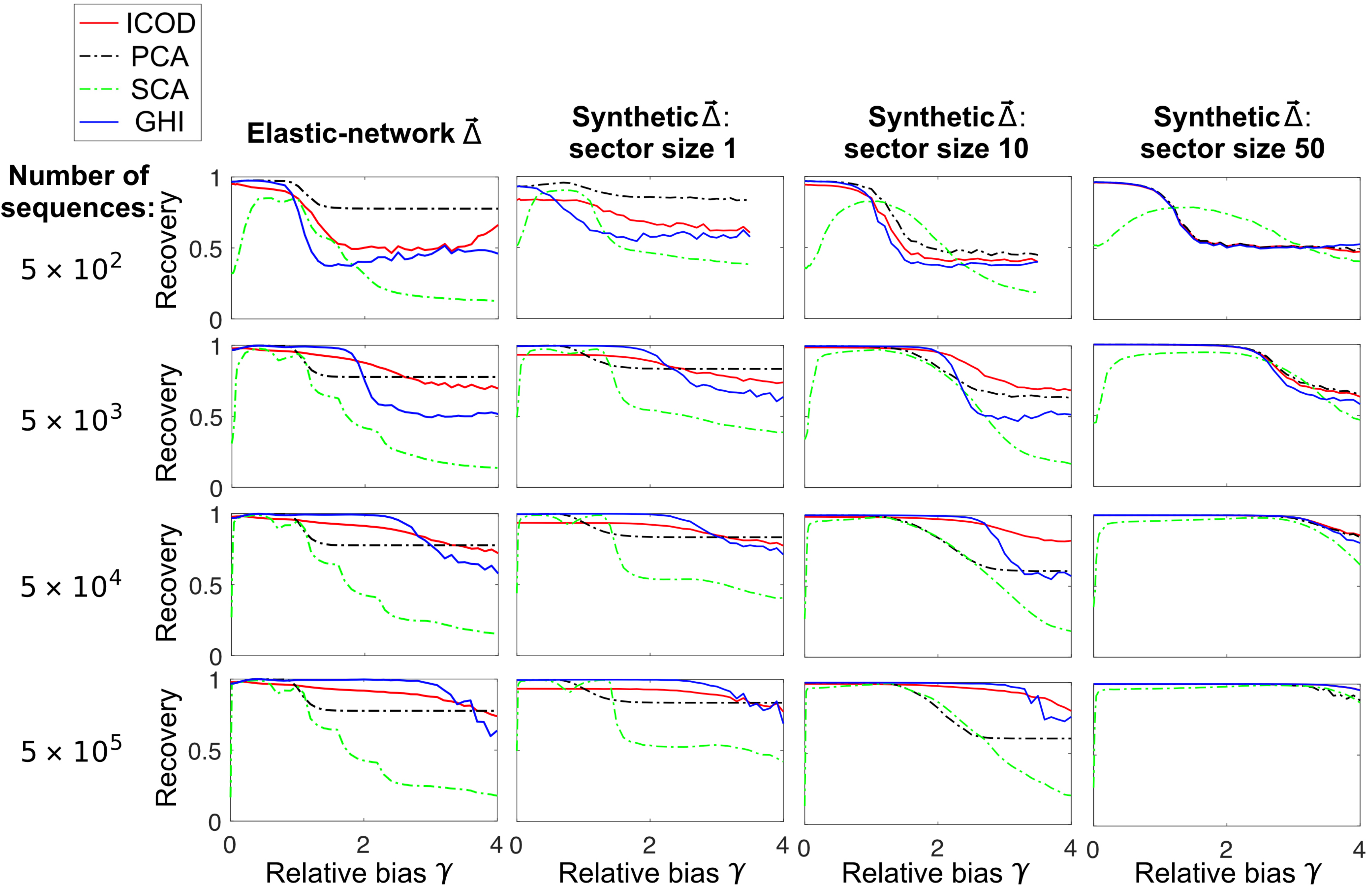}
\vspace{0.2cm}
\caption{ {\bf Effect of dataset size on Recovery of $\vec{\Delta}$.}  Selectively reweighted ensembles of $5\times 10^2$, $5\times 10^3$, $5\times 10^4$, and $5\times 10^5$ random sequences are generated for the elastic-network $\vec{\Delta}$ and synthetic $\vec{\Delta}$s with sector sizes 1, 10, and 50. All results are averaged over 100 realizations, except those using $5\times 10^5$ sequences, where only 5 realizations were used. For synthetic $\vec{\Delta}$s, each realization employs a different $\vec{\Delta}$ with the same sector size.  For the case of 500 sequences, some Recoveries were not computed at high biases due to numerical instabilities. 
}
\label{fig:Cocco_sequence_size}
\end{figure}

\clearpage

Overall, we find that GHI is very well suited to infer $\vec{\Delta}$ from very large synthetic datasets. However, ICOD is more robust to variation of dataset size and to selection bias, which should be an advantage in the application to real protein data.

\section{Application of ICOD to a multiple sequence alignment of PDZ domains}

Our general physical model for sectors provides insights into the statistical signatures of sectors in sequence data. In particular, we have found that the primary signature of physical sectors lies in the modes associated with the smallest eigenvalues of the covariance matrix, even though there is often additional signal from these sectors in the large eigenvalue modes, as studied more conventionally, e.g. in SCA. The success of ICOD on synthetic data demonstrates that information about sectors can indeed be extracted from the small eigenvalue modes of the covariance matrix.

How well does ICOD perform on real sequence data? Here, we apply ICOD to an actual alignment of sequences of PDZ domains from the Pfam database (\texttt{https://pfam.xfam.org/}) containing 24,934 sequences of length $L=79$ (corresponding to sites 313-391 in the numbering in Fig.~2 of Ref.~\cite{mclaughlin2012spatial}). In Ref.~\cite{mclaughlin2012spatial}, sites important for the specific binding of PDZ to peptide ligands were identified experimentally  via complete single-site mutagenesis. In particular, 20 sites showing particularly high mutational effects were deemed functionally significant~\cite{mclaughlin2012spatial}. It was further shown that 15 among the 20 sector amino acids found by SCA (i.e. 75\%) were also functionally significant sites.

In order to compute the empirical covariance matrix of the data, we first removed sites with more than { $15\%$ gaps (11 sites out of 79). } To eliminate the confounding effects of very rare residues at particular sites, we used a pseudocount weight { $\Lambda=0.02$.}

Next, we performed both SCA and ICOD using this empirical covariance matrix:
\begin{itemize}
	\item For SCA, we computed the conservation reweighting factors as in Refs.~\cite{halabi2009protein,Rivoire16}, using the background frequency values from Ref.~\cite{halabi2009protein}. We compressed the conservation-reweighted covariance matrix using the Frobenius norm, and we focused on the first eigenvector of this reweighted and compressed covariance matrix in order to predict sector sites.  { Finally, we took the square root of each component of this eigenvector to predict the mutational effect at each site} (see above, Section~\ref{SCAsub}, and Ref.~\cite{Rivoire16}). 
	\item For ICOD, we inverted the covariance matrix and set its diagonal blocks to zero, thus obtaining the ICOD-modified inverse covariance matrix (see Eq.~\ref{ICOD_Potts}). Next, we computed the Frobenius norm of each $20\times 20$ block associated to each pair of sites $(i,j)$ according to Eq.~\ref{Frob}. The magnitude of the $l$-th component of the first eigenvector $\vec{\nu}^{(1)}$ of this compressed $L\times L$ matrix, denoted by $||\nu_l^{(1)}||$, is the ICOD prediction of the overall mutational effect at site $l$  (see above, Section~\ref{secMulti}, especially Fig.~\ref{fig:protein-data-with-pseudocount}). Since mutational effects were experimentally measured with respect to the wild-type residues~\cite{mclaughlin2012spatial}, we used as reference the wild-type sequence of the PDZ domain employed in Fig.~\ref{Fig1} and retained this reference-sequence gauge to perform ICOD, thus allowing direct comparison to experiments. 
\end{itemize}

{
We then assessed the ability of SCA and ICOD to predict experimentally-measured mutational effects~\cite{mclaughlin2012spatial}. Specifically, we compared SCA and ICOD predictions to the overall mutational effects corresponding to the Frobenius norm of the experimentally-measured residue-specific mutational effects $\Delta_l(\alpha)$ with $\alpha\in\{1,\dots,20\}$ : 
\begin{equation}
||\Delta_l||= \sqrt{\sum_{\alpha=1}^q \left(\Delta_{l}(\alpha)\right)^2},
\label{eq:fn_Delta_2}
\end{equation}
which is the counterpart in the reference-sequence gauge of the ``site significance'' introduced in the zero-sum gauge in Eq.~\ref{eq:fn_Delta}.  
The ability of SCA and ICOD to identify the sites with the experimentally most important mutational effects is shown in Fig.~5 in the main text.  Here, we discuss the impact of parameters on these results. 
Fig.~\ref{fig:Supp_Effect_Cut_gap} shows the effect of varying the cutoff for removal of sites with a large proportion of gaps.  As illustrated in panel (a), sites with a fraction of gaps larger than a cutoff are discarded. Many of these sites are on the edges of the PDZ domain, and tend to be less conserved. Fig.~\ref{fig:Supp_Effect_Cut_gap}(b) shows that ICOD performance is robust to variations of this cutoff within a reasonable range. We have chosen a cutoff of 15\% in the rest of this analysis.

\begin{figure}[htb]
\centering
\includegraphics[width=8cm]{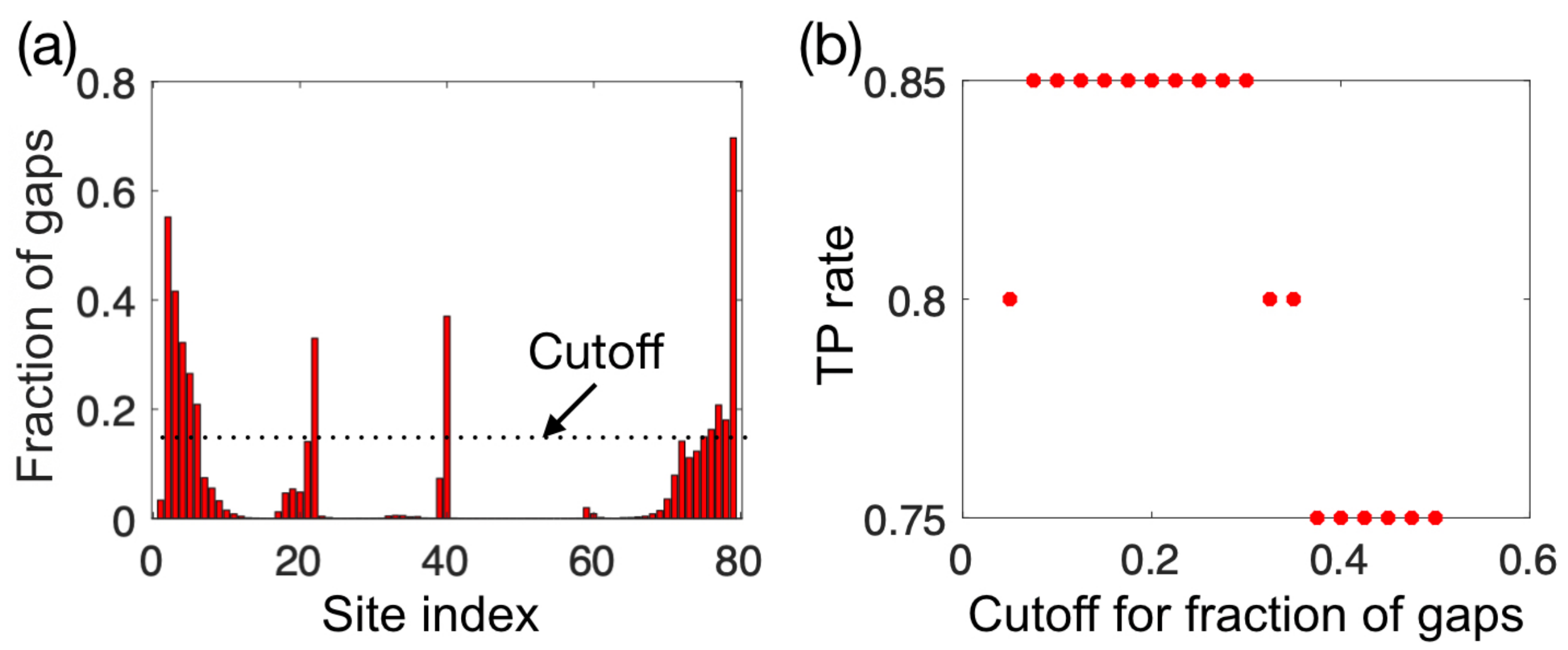}
\vspace{0.2cm}
\caption{{\bf Impact of varying the cutoff for removal of sites with a large fraction of gaps.} (a) Fraction of sequences that have a gap at each site. Sites with a fraction of gaps larger than the cutoff shown by the dashed line are discarded in the rest of our analysis.  (b) Impact of varying the gap-fraction cutoff on the performance of ICOD. A pseudocount weight of $\Lambda=0.02$ is used. }
\label{fig:Supp_Effect_Cut_gap}
\end{figure}

\newpage

Fig.~\ref{fig:Supp_PseudoCount_ICOD_SCA} shows the effect of varying the pseudocount weight, both for ICOD and for SCA. Panels (a) and (b) show the TP rate, defined as the fraction of the top 20 predicted sites that are among the 20 sites with the largest experimentally-determined mutational effects. Panels (c) and (d) show the Pearson correlation between ICOD or SCA predictions of mutational effects and the corresponding experimental measurements. Both ICOD and SCA identify experimentally-important sites significantly better than random expectation over the whole range of pseudocounts shown. However, ICOD performs best with small but nonzero pseudocount weights (panels (a) and (c)), while the performance of SCA is more robust to changing the pseudocount weight (panels (b) and (d)).

\begin{figure}[htb]
\centering
\includegraphics[width=8cm]{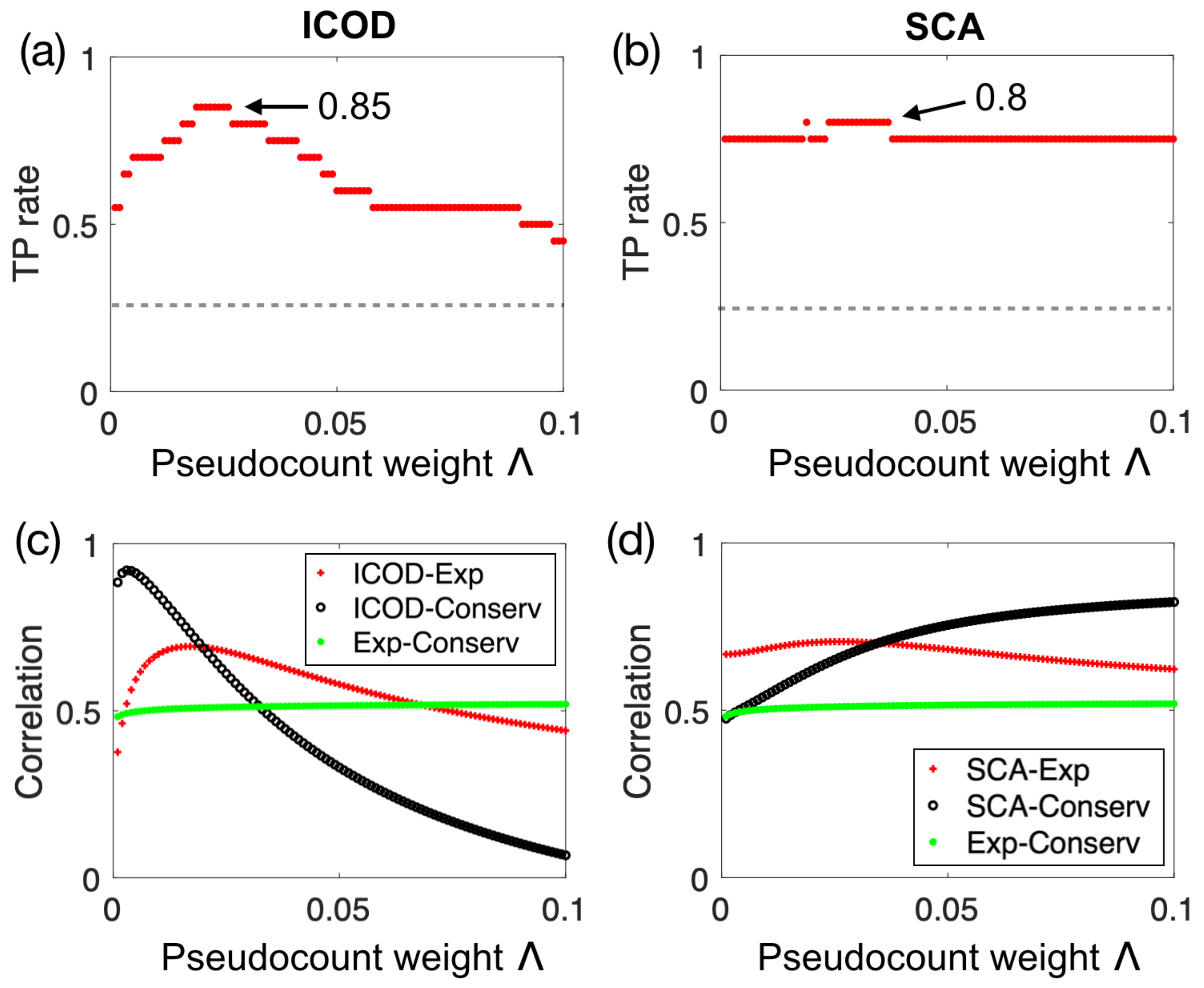}
\vspace{0.2cm}
\caption{ {\bf Impact of the pseudocount weight $\Lambda$ on the performance of ICOD and SCA}. (a) Fraction of the 20 top sites predicted by ICOD that are among the 20 sites with the largest experimentally-determined mutational effects (``TP rate'') versus pseudocount weight. The TP rate definition is the same as that shown in Fig.~\ref{Fig5}. Gray dashed line: random expectation for the TP rate, namely 20/68=0.29 (68 sites are left after removing those with a gap fraction larger than the cutoff). (b) Counterpart of (a) for SCA. (c) Pearson correlation: between mutational effects predicted by ICOD and those measured experimentally at each site (ICOD-Exp; data from Fig.~\ref{Fig5}(c) for $\Lambda=0.02$); between mutational effects predicted by ICOD and conservation scores $\phi_l$ (ICOD-Conserv; see Ref.~\cite{Rivoire16} and Eq.~\ref{phil} in the binary case); and between experimentally measured mutational effects and conservation scores $\phi_l$ (Exp-Conserv). (d) Counterpart of (b) for SCA. In all panels, a gap-fraction cutoff of $15\%$ is used.}
\label{fig:Supp_PseudoCount_ICOD_SCA}
\end{figure}

Since residue conservation plays a very important part in the PDZ sector~\cite{Tesileanu15}, we compared prediction based simply on conservation to those of SCA and ICOD. We employed the conservation scores $\phi_l$ used in SCA~\cite{Rivoire16}, which are a generalization of Eq.~\ref{phil} to 21 states. Conservation alone identifies 70\% of the 20 sites with largest experimentally-determined mutational effects, versus 85\% for ICOD (for $\Lambda=0.02$) and 75\% for SCA (see Fig.~\ref{Fig5}).  In addition, the Pearson correlation between conservation scores and experimentally-measured mutational effects is significant (see Fig.~\ref{fig:Supp_Similarity_Conservation_ICOD_SCA}(c)), even though it is smaller than between ICOD or SCA scores and experimentally-measured mutational effects (see Fig.~\ref{fig:Supp_Similarity_Conservation_ICOD_SCA}(a-b)). In fact, both ICOD and SCA scores are significantly correlated with conservation scores (see Fig.~\ref{fig:Supp_Similarity_Conservation_ICOD_SCA} (d-e)). In the case of SCA, this is not surprising given that conservation scores are explicitly used to weight the covariance matrix. Interestingly, ICOD naturally identifies these conserved sites as being important. This correlation between ICOD and conservation highlights the ability of ICOD to identify functionally important amino acids in a principled way that only relies on covariance.

}

\begin{figure}[htb]
\centering
\includegraphics[width=14cm]{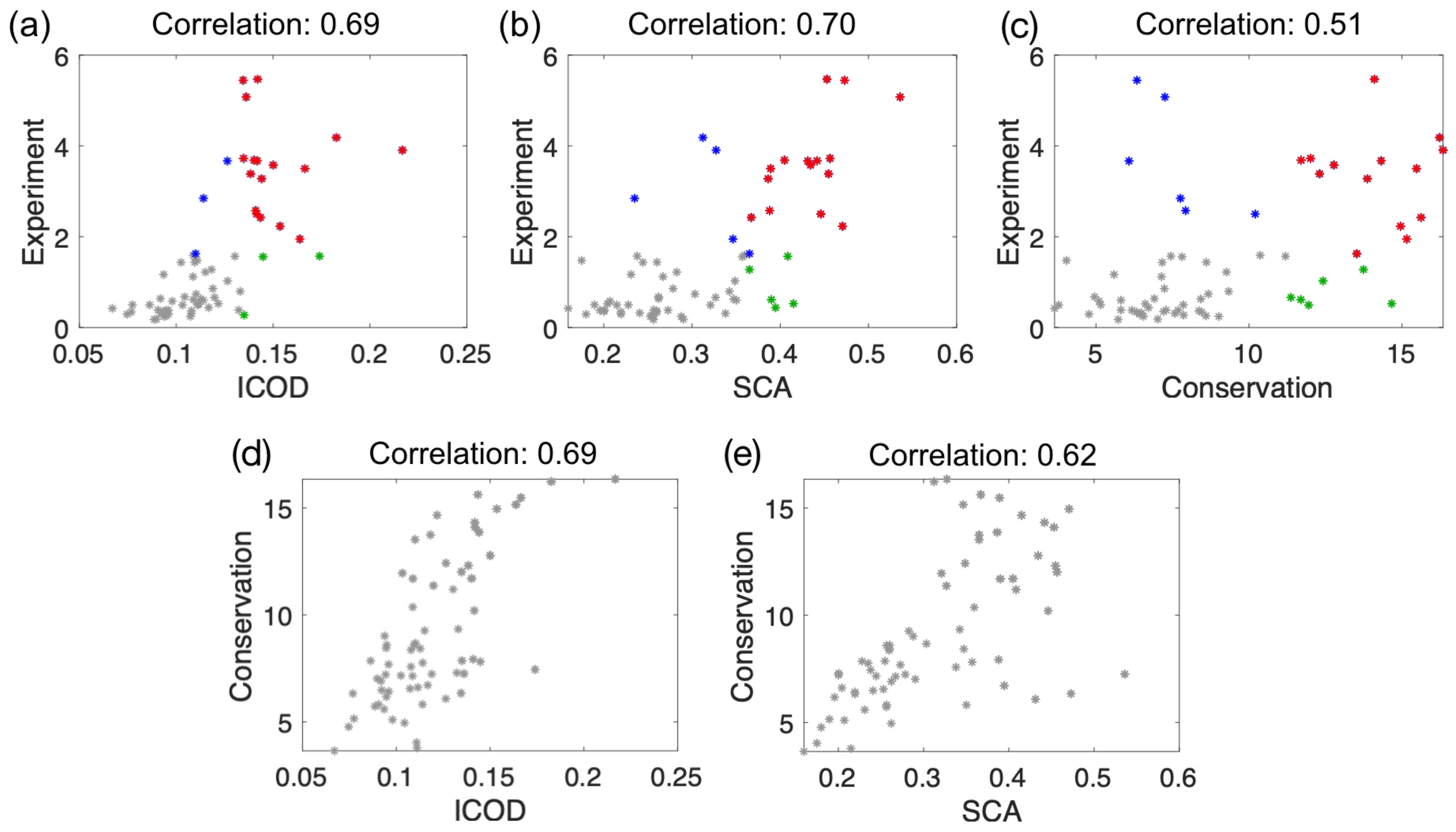}
\vspace{0.2cm}
\caption{ {\bf Predicting experimentally-measured mutational effects using ICOD, SCA, or conservation.}\protect\\
	(a) Experimentally-measured mutational effect versus mutational effect predicted by ICOD for each site of the PDZ sequence. (b) Experimentally-measured mutational effect versus mutational effect predicted by SCA for each site of the PDZ sequence. (c) Experimentally-measured mutational effect versus Conservation score $\phi_l$ for each site of the PDZ sequence. In panels (a, b, c), to highlight the matches between the top 20 predictions and the top 20 experimentally important sites~\cite{mclaughlin2012spatial}, correct hits are shown in red, false negatives in blue, and false positives in green. (d) Conservation score $\phi_l$ versus mutational effect predicted by ICOD for each site of the PDZ sequence. (e) Conservation score $\phi_l$ versus mutational effect predicted by SCA for each site of the PDZ sequence. In all panels, a pseudocount weight $\Lambda=0.02$ and a gap-fraction cutoff of $15\%$ were used. }
\label{fig:Supp_Similarity_Conservation_ICOD_SCA}
\end{figure}


\end{document}